\documentclass[usenatbib]{mn2e}
\usepackage{amssymb}
\usepackage{psfig}
\usepackage{amsmath}

\newcommand\nuddd{\ifmmode\stackrel{\bf \,...}{\textstyle \nu}\else$\stackrel{\,...}{\textstyle \nu}$\fi}
\def\lsim{~\rlap{$<$}{\lower 1.0ex\hbox{$\sim$}}}
\def\gsim{~\rlap{$>$}{\lower 1.0ex\hbox{$\sim$}}}

\title{Physical and chemical properties of Red MSX Sources in the southern sky: HII regions }

\author[Naiping Yu, Jun-Jie Wang and Nan Li]{Naiping Yu$^{1,2}$\thanks{E-mail: yunaiping09@mails.gucas.ac.cn},
Jun-Jie Wang$^{1,2}$ and Nan Li$^{1,2}$\footnotemark[1]\thanks{}\\
$^{1}$National Astronomical Observatories, Chinese Academy of
Sciences, Beijing 100012, China\\
$^{2}$NAOC-TU Joint Center for Astrophysics, Lhasa 850000, China}

\begin{document}
\maketitle
\pagestyle{plain}

\begin{abstract}
We have studied the physical and chemical properties of 18 southern
Red MSX Sources (RMSs), using archival data taken from the APEX
Telescope Large Area Survey of the Galaxy (ATLASGAL), the Australia
Telescope Compact Array (ATCA), and the Millimeter Astronomy Legacy
Team Survey at 90 GHz (MALT90). Most of our sources have simple
cometary/unresolved radio emissions at 4.8GHz and/or 8.6GHz. The
large number of Lyman continuum fluxes ($N_L$) indicates they are
probably massive O or early B type star formation regions. Archival
IRAS infrared data is used to estimate the dust temperature, which
is about 30 K of our sources. Then the H$_2$ column densities and
the volume-averaged H$_2$ number densities are estimated using the
870 $\mu$m dust emissions. Large scale infall and ionized accretions
may be occurring in G345.4881+00.3148. We also attempt to
characterize the chemical properties of these RMSs through molecular
line (N$_2$H$^+$ (1-0) and HCO$^+$ (1-0)) observations. Most of the
detected N$_2$H$^+$ and HCO$^+$ emissions match well with the dust
emission, implying a close link to their chemical evolution in the
RMSs. We found the abundance of N$_2$H$^+$ is one order of magnitude
lower than that in other surveys of IRDCs, and a positive
correlation between the abundances of N$_2$H$^+$ and HCO$^+$. The
fractional abundance of N$_2$H$^+$ with respect to H$_2$ seems to
decrease as a function of $N_L$. These observed trends could be
interpreted as an indication of enhanced destruction of N$_2$H$^+$,
either by CO or through dissociative recombination with electrons
produced by central UV photons.

\end{abstract}

\begin{keywords}
stars: formation - ISM: HII regions - ISM: abundances - ISM:
molecules - ratio lines
\end{keywords}

\section{Introduction}
One of the key challenges in stellar astrophysics is to understand
the formation of massive stars and their role in the evolution of
the universe. A lot of research has been done in astrophysics in the
last two decades and considerable progress has been seen in the
understanding of massive star formation (Zinnecker $\&$ Yorke 2007).
Even though their short lives and rare sources, massive stars play a
major role in determining physical, chemical and morphological
structures of galaxies through ionizing radiation, heating of dust
and expansions of their HII regions. These processes may also
trigger next generation of star formation by compressing neighboring
molecular clouds to the point of gravitational instability. It is
generally accepted that massive stars evolve through starless cores
in infrared dark clouds (IRDCs) to hot molecular cores (HMCs; e.g.
Kurtz et al. 2000). Then substantial UV photons and ionized stellar
winds rapidly ionize the surrounding hydrogen, forming a
hyper-compact HII region (HCHII) or ultra-compact HII region
(UCHII). The final stages are compact and classical HII regions.
However, many questions are still unclear: whether accretion could
be halted by the strong outward radiation and thermal pressure? Does
it continue in an ionized form? Does it continue through a molecular
or ionized disk? Chemical composition of molecular gas is thought to
evolve due to the physical changes that occur during the star
formation process. As material collapses and gets ionized by central
young stars, densities and temperature rise, leading to the
production and destruction of different molecular species.
Multi-wavelength observations are vital to our understanding of the
physical and chemical conditions where massive star formation
occurs.

IRDCs provide us with the possibility of investigating the
subsequent early stages of high mass star formation, and a lot of
research has also been done to investigate their physical and
chemical properties (e.g. Vasyunina et al. 2009; Vasyunina et al.
2011; Hoq et al. 2013; Liu et al. 2013; Miettinen 2014). In order to
characterize the different evolutionary stages of IRDCs, Chambers et
al. (2009) proposed an evolutionary sequence in which ``quiescent''
clumps evolve into ``intermediate'', ``active'' and ``red'' clumps,
based on the Spitzer observations. Sanhueza et al. (2012) and Hoq et
al. (2013) characterize the chemical properties of a sample of IRDCs
using the data of MALT90. Correlations and anti-correlations between
molecular abundances and the evolutionary stages have been found.
These studies suggest chemical processes in high-mass star
formations may differ from their low-mass counterparts, as their
temperature, densities and UV fluxes are very different from each
other. Establishing the chemical properties of massive star
formation requires that large surveys be conducted.

\begin{table*}
\begin{minipage}{13cm}
 \caption{\label{tab:test}List of our sources.}
 \begin{tabular}{lclclclclclclcl}
  \hline
  \hline
 RMS  & RA  & Dec.  & V$_{lsr}$ &$D$  & $S$$_{4.8GHz}$$^a$ & log($N_L$) &  \\

 name    & (J2000) & (J2000) &(km s$^{-1}$) & (kpc) & (mJy)  & (S$^{-1}$) &  \\
  \hline
G330.2838+00.4933  & 16:03:43.28 &-51:51:45.6     & -94.2 &5.32$\pm$0.06 & 44.9 &47.05$\pm$0.05 \\
G331.4181$-$00.3546  & 16:12:50.24 &-51:43:28.5   & -63.2 &3.90$\pm$0.05 & 83.9 &46.78$\pm$0.06 \\
G332.2944$-$00.0962  & 16:15:45.83 &-50:56:02.4   & -48.7 &3.64$\pm$0.03 & 175.6 &47.30$\pm$0.04\\
G332.5438$-$00.1277  & 16:17:02.33 &-50:47:03.3   & -47.1 &3.64$\pm$0.06 & 42.2 &46.69$\pm$0.07\\
G332.8256$-$00.5498A & 16:20:11.06 &-50:53:16.2   & -58.1 &3.64$\pm$0.05 & 1259.0 &48.16$\pm$0.07\\
G333.3401$-$00.1273  & 16:20:37.00 &-50:13:32.7   & -59.7 &3.64$\pm$0.03 & 3.3 &45.58$\pm$0.04\\
G337.0047+00.3226  & 16:34:04.73 &-47:16:29.2     & -62.4 &11.43$\pm$0.05 & 249.3 &48.46$\pm$0.02\\
G337.4032$-$00.4037  & 16:38:50.44 &-47:28:03.0   & -40.0 &3.14$\pm$0.04 & 90.3 &46.89$\pm$0.06\\
G337.7091+00.0932A & 16:37:51.80 &-46:54:33.4     & -75.9 &10.98$\pm$0.04 & 144.9 &48.19$\pm$0.02\\
G337.8442$-$00.3748  & 16:40:26.68 &-47:07:13.1   & -39.9 &3.06$\pm$0.06 & 11.1 &45.95$\pm$0.09\\
G339.1052+00.1490  & 16:42:59.58 &-45:49:43.6     & -78.2 &4.73$\pm$0.06 & 42.9 &46.93$\pm$0.03\\
G340.0708+00.9267  & 16:43:15.69 &-44:35:16.0     & -73.5 &4.69$\pm$0.04 & 93.8 &46.96$\pm$0.04\\
G343.5024$-$00.0145  & 16:59:20.78 &-42:32:37.5   & -27.9 &2.72$\pm$0.08 & 176.9 &47.06$\pm$0.12\\
G345.0034$-$00.2240A & 17:05:11.19 &-41:29:06.3   & -27.3 &2.87$\pm$0.08 & 140.0 &46.99$\pm$0.12\\
G345.4881+00.3148  & 17:04:28.04 &-40:46:23.3     & -17.9 &2.12$\pm$0.10 & 1980.0 &47.89$\pm$0.21\\
G346.0774$-$00.0562  & 17:07:53.91 &-40:31:34.2   & -83.9 &10.92$\pm$0.03 & 8.3 &46.90$\pm$0.01\\
G347.8707+00.0146  & 17:13:08.77 &-39:02:28.2     & -30.6 &3.42$\pm$0.04 & 206.1 &47.33$\pm$0.05\\
G348.8922$-$00.1787  & 17:17:00.10 &-38:19:26.4   & 8.13   &11.20$\pm$0.05 & 185.7 &48.32$\pm$0.02\\

  \hline
 \end{tabular}
 \label{tb:rotn}

Notes. Columns are (1)source name, (2)right ascension,
(3)declination, (4)local standard of rest velocity, (5)distance,
(6)flux density at 4.8 GHz $S_{4.8}$, (7)number of UV ionizing
photons

 a: Urquhart et al. (2007a)
\end{minipage}
\end{table*}

In this paper, we describe our physical and chemical studies of RMSs
(mainly HII regions) and compare our results with previous IRDCs
observations. The RMS survey was conceived at Leeds to
systematically search the entire Galaxy for MYSOs, by comparing the
colors of sources from the Midcourse Space Experiment (MSX) and
Two-Micron All-Sky Survey (2MASS) point sources (Lumsden et al.
2002). The RMS survey is an ongoing multi-wavelength observational
programme (e.g. Urquhart et al. 2007a; Urquhart et al. 2007b) and
will provide us with the largest MYSO sample for statistical
studies. We have analyzed 18 RMSs in the southern sky using archival
data from ATLASGAL, ATCA and MALT90. N$_2$H$^+$ and HCO$^+$ are both
good tracers of dense gases, as their critical density $n_{crit}$
$>$ $10^5$ cm$^{-3}$. N$_2$H$^+$ is more resistent to freeze-out on
grains than the carbon-bearing species (Hirota et al. 1998). So it
is an excellent tracer of cold and dense molecular clouds. In star
formation regions, it is primarily formed through the gas-phase
reaction H$_3$$^+$ + N$_2$ $\rightarrow$ N$_2$H$^+$ + H$_2$, and
destroyed by CO molecules in the gas-phase through N$_2$H$^+$ + CO
$\rightarrow$ HCO$^+$ + N$_2$. It can also be destroyed in the
electron recombination: N$_2$H$^+$ + e$^-$ $\rightarrow$ N$_2$ + H
or NH + N (e.g. Dislaire et al. 2012; Vigren et al. 2012). In dense
molecular clouds, HCO$^+$ is mainly formed through the gas-phase
ion-neutral reaction H$_3$$^+$ + CO $\rightarrow$ HCO$^+$ + H$_2$
(e.g. Herbst $\&$ Klemperer 1973). The HCO$^+$ abundance can also be
increased in regions where shocks are generated. In chemical models
of low mass star formation, a relative enhancement of N$_2$H$^+$ and
a depletion of HCO$^+$ abundances are expected in the cold
prestellar phase, as CO is thought to be depleted in starless cores
(e.g. Lee et al. 2003, 2004; Bergin \& Tafalla 2007). When the
central star evolves, the gas gets warmer and CO should be
evaporated from the dust grains if the dust temperature exceeds
about $\sim$ 20 K (Tobin et al. 2013). Thus the HCO$^+$ abundance
should increase while the N$_2$H$^+$ abundance drops.

\begin{table*}
\begin{minipage}{13cm}
 \caption{\label{tab:test}Properties of the RMSs.}
 \begin{tabular}{lclclclclclclcl}
  \hline
  \hline
 RMS  & $S_{peak}$/$\Omega$$^a$  & $S_{870}$$^a$  & $T_d$$^b$& R$_{eff}$$^a$ & $M_{tot}$  & $N_{H_2}$$^c$ & $<n_{H_2}>$ &  \\

 name    & (Jy/beam) & (Jy) &(K) & ($^\prime$$^\prime$) & $(M_\odot$) & ($\times$ 10$^{22}$ cm$^{-2}$)  & ($\times$ 10$^{4}$ cm$^{-3}$) &  \\
  \hline
G330.2838+00.4933    & 2.00 & 7.50   & 30.0 &30 & 1995$\pm$46 &2.2& 1.5$\pm$0.1 \\
G331.4181$-$00.3546  & 3.48 &39.81   & 18.9 &75 & 9099$\pm$238 &6.0& 0.3$\pm$0.1 \\
G332.2944$-$00.0962  & 7.04 &34.52   & 27.6 &62 & 4603$\pm$77 &8.2& 1.3$\pm$0.1 \\
G332.5438$-$00.1277  & 2.03 &26.30   & 30.5 &69 & 3174$\pm$97 &2.1& 0.6$\pm$0.1 \\
G332.8256$-$00.5498A &32.86 &139.92  & 31.7 &79 &16246$\pm$460 &33.5& 2.1$\pm$0.1\\
G333.3401$-$00.1273  & 1.01 & 3.43   & 40.5 &28 & 311$\pm$5  &0.8& 0.9$\pm$0.1 \\
G337.0047+00.3226    & 2.28 & 6.67   & 33.9 &25 & 6341$\pm$56 &2.2& 0.8$\pm$0.1 \\
G337.4032$-$00.4037  &18.35 &58.81   & 29.0 &47 & 5535$\pm$143 &20.4&5.4$\pm$0.2 \\
G337.7091+00.0932A   & 9.73 &48.94   & 30.0(A) &66 &55045$\pm$424 &10.5&0.4$\pm$0.1 \\
G337.8442$-$00.3748  & 3.32 & 9.90   & 32.8 &31 & 771$\pm$30  &3.3 &2.9$\pm$0.1 \\
G339.1052+00.1490    & 1.98 &8.64    & 29.6 &34 & 1831$\pm$23 &2.2 &1.4$\pm$0.1 \\
G340.0708+00.9267    & 2.69 &15.10   & 33.4 &40 & 2716$\pm$48 &2.6 &1.3$\pm$0.1 \\
G343.5024$-$00.0145  & 6.21 &45.12   & 24.8 &71 & 3766$\pm$209 &8.1 &1.6$\pm$0.1 \\
G345.0034$-$00.2240A &18.51 &137.14  & 31.0 &113& 9850$\pm$528 &19.3&0.9$\pm$0.1 \\
G345.4881+00.3148    &17.27 &143.50  & 26.5 &77 & 6782$\pm$646 &21.0&4.8$\pm$0.4 \\
G346.0774$-$00.0562  & 2.38 &11.57   & 30.0(A) &43 &13013$\pm$76 &2.6 &0.4$\pm$0.1 \\
G347.8707+00.0146    & 3.16 &17.56   & 30.0(A) &45 & 1922$\pm$45 &3.4 &1.6$\pm$0.1 \\
G348.8922$-$00.1787  & 3.02 &7.80    & 30.0(A) &17 & 4177$\pm$19 &1.8 &1.8$\pm$0.1 \\

  \hline
 \end{tabular}
 \label{tb:rotn}
Notes. Columns are (1)source name, (2)peak intensities $S_{peak}$,
(3)flux densities $S_{870}$, (4)dust temperature, (5)effective
radii, (6)clump mass, (7)H$_2$ column density, (8)volume-averaged
H$_2$ number density

 a: Contreras et al. (2013)

 b: The $T_{d}$ values marked with ``A'' represent the average value derived for other sources.

 c: The data has been converted to the Mopra telescope beam size.
\end{minipage}
\end{table*}

\section{data sets and source selections}
The MALT90 is a large international project aimed at characterizing
the sites within our Galaxy where high-mass stars will form. The
target clumps are selected from the 870 $\mu$m ATLASGAL to host the
early stages of high-mass star formation and include pre-stellar
clumps, protostellar clumps, and HII regions. Exploiting the unique
broad frequency range and fast-mapping capabilities of the Mopra
22-m telescope, MALT90 maps 16 emission lines simultaneously at
frequencies near 90 GHz. These molecular lines will probe the
physical, chemical and evolutionary states of dense high-mass
star-forming cores. The survey covers a Galactic longitude range  of
$\sim$ -60 to $\sim$ 15$^{\circ}$ and Galactic latitude range of -1
to +1 $^{\circ}$. The observations were carried out with the newly
upgraded Mopra Spectrometer (MOPS). The full 8 GHz bandwidth of MOPS
was split into 16 zoom bands of 138 MHz each providing a velocity
resolution of $\sim$ 0.11 km s$^{-1}$. The angular resolution of
Mopra is about 38 arcsec, with beam efficiency between 0.49 at 86
GHz and 0.42 at 115 GHz (Ladd et al. 2005). The mapping sizes are
3$^{\prime}$.4 $\times$ 3$^{\prime}$.4 with 9$^{\prime\prime}$
spacing between adjacent rows. The MALT90 data includes (\emph{l},
\emph{b}, \emph{v}) data cubes and (\emph{l}, \emph{b}) moment and
error maps, and is publicly available from the MALT90 Home
Page\footnote{See
  http://atoa.atnf.csiro.au/MALT90}. More information
about this survey can be found in Foster et al. (2011) and Hoq et
al. (2013). The data processing was conducted using CLASS (Continuum
and Line Analysis Single-Disk Software) and GreG (Grenoble Graphic)
software packages.

The ATLASGAL is the first systematic survey of the inner Galactic
plane in the sub-millimeter. The observations were carried out with
the Large APEX Bolometer Camera (LABOCA), an array of 295 bolometers
observing at 870 $\mu$m (Contreras et al. 2013). The aim of this
survey is to obtain a complete census of cold dusty clumps in the
Galaxy, and to study their distributions across the Galaxy. The
half-power beam width (HPBW) of the telescope is
18.$^\prime$$^\prime$6 ($\sim$ 0.22 pc at 2.4 kpc) at the frequency
used. The total field of view (FoV) of LABOCA is 11.$^\prime$4. The
instrument and its observing modes are described in Siringo et al.
(2009). And the data reduction is described in Contreras et al.
(2013).

In order to choose a genuine sample of MYSOs from the RMSs, Urquhart
et al. (2007a) have completed the 5-GHz observations of 892 RMS
sources in the southern sky using the ATCA. According to the
observations, they divided these sources into three groups: real
MYSO candidates, HII regions (UCHII and HCHII) and others such as
evolved stars and planetary nebulae (PNe). The radio continuum
emissions allow us to derive the expected spectral type of the
exciting stars. Assuming an electron of $T_e$ = 10$^4$ K, the number
of UV ionizing photons needed to keep an HII region ionized is given
as (Chaisson 1976):
\begin{equation}
N_L = 7.6 \times 10^{46} s^{-1}
(\frac{S_\nu}{Jy})(\frac{D}{kpc})^2(\frac{\nu}{GHz})^{0.1}(\frac{T_e}{10^4
K})^{-0.45}
\end{equation}
The large number of Lyman continuum fluxes (listed in Table 1,
column 7) indicates they are probably massive O or early B type star
formation regions.

In order to study physical and chemical properties in massive star
formation regions, we selected 18 sources from the RMS survey by
applying the following criteria: (i)According to the observations of
ATCA (Urquhart et al. 2007a), sources should have simple
spherical/unresolved/shell-like/cometary radio emissions. (ii)The
N$_2$H$^+$ and HCO$^+$ emissions of our sources should be detected
by MALT90, and the local standard of rest (LSR) velocities of
HCO$^+$ and N$_2$H$^+$ should be similar to those of Urquhart et al.
(2007b, 2008). (iii)The center locations of RMSs and 870 $\mu$m peak
emissions are within 18$^\prime$$^\prime$, which is about the
angular resolution of MSX at 8.0 $\mu$m. (iv)Sources should not be
on the edge of known HII or supernova regions, considering the large
beam of the 22-m Mopra telescope. The information of our selected
sources is listed in Table 1. The detection of radio emissions in
our sources indicates they are much more evolved, at least be
similar to the ``active''/``red'' stages of IRDCs. However, sources
of Hoq et al. (2013) and Miettinen (2014) range from prestellar
(``quiescent'') to HII regions with bright IR emissions (``red'').
We suppose to find different chemical properties of our RMSs using
the above archival data.

\begin{table*}
\begin{minipage}{17cm}
 \caption{\label{tab:test}Parameters for molecular lines N$_2$H$^+$ and HCO$^+$.}
 \begin{tabular}{lclclclclclclcl}
  \hline
  \hline
 RMS  & Molecular   & Width  & $T_{ex}$$^a$ & $\tau$ & $\int$T$_{mb}$ d$v$  & $N$ & $\chi$  \\

 name    &  line &(km s$^{-1}$) & (K) &   & (K km s$^{-1}$)  & ($\times$ 10$^{12}$ cm$^{-2}$) &($\times$ 10$^{-10}$)   \\
  \hline
G330.2838+00.4933    & N$_2$H$^+$  & 2.65$\pm$ 0.16 &11.2  $\pm$0.5   & 0.14$\pm$0.06 &3.6$\pm$ 0.5& 5.9$\pm$1.7&2.7$\pm$ 0.8\\
& HCO$^+$  & 3.54$\pm$ 0.17 &12.5$\pm$2.5   & 0.53$\pm$0.18 &13.3$\pm$ 1.0& 18.3$\pm$ 1.4&8.3$\pm$ 0.6\\
G331.4181$-$00.3546  & N$_2$H$^+$  & 2.34$\pm$ 0.21 &6.2$\pm$1.8   & 1.17$\pm$ 0.17&10.6$\pm$ 0.8& 13.8$\pm$3.2&2.3$\pm$ 0.5\\
& HCO$^+$  & 4.39$\pm$ 0.45 &12.5$\pm$2.5   & 0.47$\pm$0.16 &15.5$\pm$ 2.0&  20.9$\pm$ 2.7&3.5$\pm$ 0.5\\
G332.2944$-$00.0962  & N$_2$H$^+$  & 2.88$\pm$ 0.10 &7.2$\pm$0.5   & 0.83$\pm$0.12 &12.2$\pm$ 0.4&  16.4$\pm$2.1&2.0$\pm$ 0.3\\
& HCO$^+$  & 4.51$\pm$ 0.40 &12.5$\pm$2.5   & 0.23$\pm$0.07 &9.3$\pm$ 1.7&  11.7$\pm$ 2.1&1.4$\pm$ 0.2\\
G332.5438$-$00.1277  & N$_2$H$^+$  & 1.90$\pm$ 0.18 & 7.8$\pm$1.4   & 0.41$\pm$ 0.27 &4.5$\pm$ 0.3&  6.3$\pm$ 0.3&3.0$\pm$ 0.3\\
& HCO$^+$  & 2.70$\pm$ 0.14 &12.5$\pm$2.5   & 0.64$\pm$0.23 &4.1$\pm$ 0.2& 5.8$\pm$ 0.3&2.8$\pm$ 0.1\\
G332.8256$-$00.5498A & N$_2$H$^+$  & 4.19$\pm$ 0.21 &8.9$\pm$2.3(A)   & ...&16.2$\pm$ 2.4& 23.6$\pm$ 3.9&0.8$\pm$ 0.1\\
& HCO$^+$  & 11.84$\pm$ 0.31 &12.5$\pm$2.5   & 0.49$\pm$0.16 &31.0$\pm$ 0.3& 42.2$\pm$ 0.4&1.3$\pm$ 0.1\\
G333.3401$-$00.1273  & N$_2$H$^+$  & 2.10$\pm$ 0.29 &8.9$\pm$2.3(A)   & ... &2.8$\pm$ 0.4&  4.0$\pm$1.2&5.0$\pm$ 1.5\\
& HCO$^+$  & 2.47$\pm$ 0.26 &12.5$\pm$2.5   & 0.20$\pm$0.06 &4.1$\pm$ 1.2& 5.1$\pm$ 1.5&6.4$\pm$ 1.8\\
G337.0047+00.3226    & N$_2$H$^+$  & 3.76$\pm$ 0.21 &8.9$\pm$2.3(A)   & ... &2.9$\pm$ 0.5& 4.1$\pm$ 1.0&1.5$\pm$0.5\\
& HCO$^+$  & 4.63$\pm$ 0.25 &12.5$\pm$2.5   & 0.40$\pm$0.13 &13.9$\pm$ 1.6& 18.4$\pm$ 2.1&8.4$\pm$ 1.0\\
G337.4032$-$00.4037  & N$_2$H$^+$  & 3.16$\pm$ 0.01 &10.4$\pm$1.3   & 0.47$\pm$0.05 &13.2$\pm$ 0.4& 20.7$\pm$1.2&1.0$\pm$ 0.1\\
& HCO$^+$  & 6.49$\pm$ 0.18 &12.5$\pm$2.5   & 0.51$\pm$0.17 &20.9$\pm$ 1.5& 28.6$\pm$ 2.1&1.4$\pm$ 0.1\\
G337.7091+00.0932A   & N$_2$H$^+$  & 4.02$\pm$ 0.29 & 7.5$\pm$0.6   & 0.40$\pm$ 0.27 &8.9$\pm$ 0.7& 12.3$\pm$ 2.2&1.2$\pm$ 0.2\\
& HCO$^+$  & 5.40$\pm$ 0.31 &12.5$\pm$2.5   & 0.35$\pm$0.11 &15.1$\pm$ 0.7& 19.7$\pm$ 0.9&1.9$\pm$ 0.1\\
G337.8442$-$00.3748  & N$_2$H$^+$  & 2.36$\pm$ 0.12 &8.9$\pm$2.3(A)   & ... &6.2$\pm$ 0.2& 9.1$\pm$1.2&2.8$\pm$ 0.4\\
& HCO$^+$  & 2.41$\pm$ 0.26 &12.5$\pm$2.5   & 0.25$\pm$0.08 &4.9$\pm$ 0.2& 6.2$\pm$ 0.3&1.9$\pm$ 0.1\\
G339.1052+00.1490    & N$_2$H$^+$  & 2.62$\pm$ 0.16 &12.1$\pm$0.5   & 0.2$\pm$0.07 &6.5$\pm$ 0.4& 11.1$\pm$2.8&5.0$\pm$0.3\\
& HCO$^+$  & 4.01$\pm$ 0.42 &12.5$\pm$2.5   & 0.21$\pm$0.06 &7.4$\pm$ 2.2& 9.2$\pm$ 2.7&4.2$\pm$ 1.2\\
G340.0708+00.9267    & N$_2$H$^+$  & 3.58$\pm$ 0.19 &4.1$\pm$0.5   & 2.1$\pm$0.35 &7.5$\pm$ 1.0& 10.0$\pm$1.2&3.8$\pm$ 0.5\\
& HCO$^+$  & 3.41$\pm$ 0.11 &12.5$\pm$2.5   & 0.93$\pm$0.38 &17.9$\pm$ 0.9& 27.0$\pm$ 1.4&10.4$\pm$ 0.5\\
G343.5024$-$00.0145  & N$_2$H$^+$  & 3.13$\pm$ 0.10 &9.9$\pm$0.5   & 0.76$\pm$0.01 &17.1$\pm$ 0.6& 26.2$\pm$2.5&3.2$\pm$ 0.3\\
& HCO$^+$  & 3.75$\pm$ 0.23 &12.5$\pm$2.5   & 0.55$\pm$0.19 &15.0$\pm$ 1.0& 20.7$\pm$ 1.4&2.6$\pm$ 0.2\\
G345.0034$-$00.2240A & N$_2$H$^+$  & 4.21$\pm$ 0.16 &4.4$\pm$0.5   & 2.0$\pm$0.30 &12.8$\pm$ 0.8& 16.7$\pm$1.7&0.9$\pm$ 0.1\\
& HCO$^+$  & 5.02$\pm$ 0.39 &12.5$\pm$2.5   & 0.20$\pm$0.04 &9.6$\pm$ 0.4& 12.1$\pm$ 0.5&0.6$\pm$ 0.1\\
G345.4881+00.3148    & N$_2$H$^+$  & 3.83$\pm$ 0.13 &5.6$\pm$0.5   & 1.7$\pm$0.10 &17.5$\pm$ 1.5& 22.6$\pm$3.3 &1.1$\pm$ 0.2\\
& HCO$^+$  & 5.21$\pm$ 0.19 &12.5$\pm$2.5   & 1.29$\pm$0.20 &43.2$\pm$ 2.8& 96.4$\pm$ 6.2&4.6$\pm$ 0.3\\
G346.0774$-$00.0562  & N$_2$H$^+$  & 3.12$\pm$ 0.24 & 9.5$\pm$0.5   & 0.46$\pm$ 0.19&10.6$\pm$ 0.6& 16.0$\pm$ 0.7&6.1$\pm$ 0.3\\
& HCO$^+$  & 2.59$\pm$ 0.24 &12.5$\pm$2.5   & 0.32$\pm$0.10 &6.2$\pm$ 1.1& 8.0$\pm$ 1.4&3.1$\pm$ 0.6\\
G347.8707+00.0146    & N$_2$H$^+$  & 4.40$\pm$ 0.29 &8.9$\pm$2.3(A)   & ... &8.0$\pm$ 0.5& 11.6$\pm$1.1&3.4 $\pm$ 0.3\\
& HCO$^+$  & 9.02$\pm$ 0.31 &12.5$\pm$2.5   & 0.35$\pm$0.11 &17.5$\pm$ 1.0& 22.8$\pm$ 2.0&6.7$\pm$ 0.6\\
G348.8922$-$00.1787  & N$_2$H$^+$  & 3.23$\pm$ 0.23 &8.9$\pm$2.3(A)   & ... &3.1$\pm$ 0.5& 4.4$\pm$1.6 &1.3$\pm$ 0.3\\
& HCO$^+$  & 5.70$\pm$ 0.20 &12.5$\pm$2.5   & 0.50$\pm$0.17 &20.2$\pm$ 1.4& 27.5$\pm$ 1.9&8.3$\pm$ 0.6\\
  \hline
 \end{tabular}
 \label{tb:rotn}

Notes. Columns are (1)source name, (2)Molecular line name, (3)Line
width, (4)excitation temperature, (5)optical depth, (6)integrated
intensity, (7)total column density, (8)fractional abundance with
respect to H$_2$.

 a: The $T_{ex}$ of N$_2$H$^+$ values marked with ``A'' represent the average value derived for other sources.

\end{minipage}
\end{table*}

\section{results and analysis }

\subsection{Physical parameters}
The dust temperature ($T_{d}$) is essential to study the physics and
chemistry of star formation. For those sources in our sample that
are associated with IRAS point sources, we estimated the dust
temperature to be the same as the 60/100 $\mu$m color temperature
defined by Henning et al. (1990) as
\begin{equation}
T_{d} \simeq  T_c(\frac{60}{100}) = 96 [(3+\beta)ln(\frac{100}{60})
- ln(\frac{S_{60}}{S_{100}})]^{-1}
\end{equation}
where $\beta$ is the dust emissivity index set to be 1.8 to be
consistent with the Ossenkopf $\&$ Henning (1994) dust model.
$S_\lambda$ is the flux density at the wavelength $\lambda$. The
derived values are listed in Table 2. The mean dust temperature is
about 30 K, which is much larger than those in IRDCs (e.g. Miettinen
$\&$ Harju 2010; Hoq et al. 2013), indicating our sources are more
evolved. Four of our sources are not associated with any IRAS point
source. For these sources we assume their dust temperature to be 30
K. Masses will be underestimated if the temperature is lower than
the assumed value of 30 K. For example, at 20 K the masses will be
higher by a factor of 1.7.

Assuming that the dust emission at 870 $\mu$m is optically thin, the
clump masses and H$_2$ column densities could be estimated through
the following expressions:
\begin{equation}
M_{tot} = \frac{S_{870} D^2 R}{B_{870}(T_d) \kappa_{870}}
\end{equation}
\begin{equation}
N_{H_2} = \frac{S_{peak} R}{\Omega B_{870}(T_d) \kappa_{870}
m_{H_2}}
\end{equation}
where $S_{870}$ is the integrated flux density. $S_{peak}$ denotes
the peak flux density. $D$ is the distance to the RMS. $\Omega$ is
the beam solid angle. $R$ is the gas-to-dust mass ratio which is set
to be 100. $m_{H_2}$ is the mass of one hydrogen molecule.
$B_{870}(T_d)$ is the Planck function at the dust temperature $T_d$.
We assumed that $\kappa_{870}$ = 0.17 m$^2$ kg$^{-1}$ (Miettinen
$\&$ Harju 2010). The derived masses are inverse proportional to the
assumed value of the opacity $\kappa_{870}$, which has an
uncertainty of at least a factor 1.5.

The volume-averaged H$_2$ number densities ($<n(H_2)>$) could be
calculated assuming a spherical geometry for the clumps, using these
formula
\begin{equation}
<\rho> = \frac{3 M_{tot}}{4 \pi R_{eff}^3}
\end{equation}
\begin{equation}
<n(H_2)> = \frac{<\rho>}{\mu_{H_2} m_H}
\end{equation}
The linear clump effective radii is derived from the angular radii
and kinematic distances as R$_{eff}$ (pc) = R$_{eff}$(rad) $\times$
D(pc). The value of R$_{eff}$(rad) is from Contreras et al. (2013).
We adopt a mean molecular weight per H$_2$ molecule of $\mu$ = 2.8
to include helium. The derived parameters are listed in Table 2.
Taking into account the spatial resolution effects, the derived
column densities are all beyond the column density threshold
required by theoretical considerations for massive star formation.

\subsection{Chemical parameters}
Figures 3-20 show the N$_2$H$^+$ (1-0) and HCO$^+$ (1-0) spectra on
the peak emissions of 870 $\mu$m and their integrated intensities
superimposed on the 870 $\mu$m images of the RMSs. Most of the
detected N$_2$H$^+$ and HCO$^+$ emissions match well with the dust
emission, implying a close link to their chemical evolution in the
RMSs. The HCO$^+$ emission lines of some sources are far of having a
simple Gaussian shape, presenting asymmetries, and spectral wings or
shoulders, which suggest that the molecular gas is affected by the
dynamics of these star-forming regions. Yu $\&$ Wang (2014) made a
study of 19 RMSs and find these profiles are probably caused by
infall and/or outflow activities. Some of our sources are also
included in their study. The N$_2$H$^+$ (1-0) line has 15 hyperfine
structures (HFS) out of which seven have a different frequency (e.g.
Pagani et al. 2009; Keto $\&$ Rybicki 2010). These velocity
components blend into three groups (see figure 2 of Purcell et al.
2009). We followed the procedure outlined by Purcell et al. (2009)
to estimate the optical depth of N$_2$H$^+$. Assuming the line
widths of the individual hyperfine components are all equal, the
integrated intensities of the three blended groups should be in the
ratio of 1:5:3 under optically thin conditions. The optical depth
can then be derived from the ratio of the integrated intensities
($\int T_{MB} dv$) of any groups using the following equation:

\begin{equation}
\frac{\int T_{MB,1} dv}{\int T_{MB,2} dv} = \frac{1 -
exp(-\tau_1)}{1 - exp(-a\tau_2)}
\end{equation}
where $a$ is the expected ratio of $\tau_2/\tau_1$ under optically
thin conditions. We determined the optical depth only from the
intensity ratio of group 1/group 2, as anomalous excitation of the
$F_1F$ = 10-11 and 12-12 components (in our group 3) has been
reported by Caselli et al. (1995). The optical depth of group 2 is
listed in table 3 because it provides a better estimate for the
excitation temperature when $\tau
> 0.1$. Then the excitation temperature ($T_{ex}$) for N$_2$H$^+$
could be calculated with the following formula:
\begin{equation}
T_{ex} = 4.47 / ln(1+ (\frac{T_{MB}}{4.47(1 - exp(-\tau))} +
0.236)^{-1})
\end{equation}
The N$_2$H$^+$ optical depth of our six clumps could not be derived
through this method. For these cases, we adopted the mean derived
excitation temperature. Assuming local thermodynamic equilibrium
(LTE) and using Eq. (2) in Miettinen (2014), we derived the column
densities of N$_2$H$^+$. Finally, the fractional abundance of
N$_2$H$^+$ with respect to H$_2$ could be estimated by $\chi$
(N$_2$H$^+$) = N(N$_2$H$^+$)/N(H$_2$). The derived parameters are
listed in Table 3.

Generally speaking, the optical depth and $T_{ex}$ of HCO$^+$ could
be derived by comparing the intensities of HCO$^+$ and
H$^{13}$CO$^+$. However, the H$^{13}$CO$^+$ emission was not
detected in most cases. For the sources (G332.8256-00.5498A,
G337.4032-00.4037, G345.0034-00.2240A and G345.4881+00.3148) which
show distinct H$^{13}$CO$^+$ emissions, we derive the optical depths
of HCO$^+$ from
\begin{equation}
\frac{^{12}T_{mb}}{^{13}T_{mb}} =
\frac{1-exp(-\tau_{12})}{1-exp(-\tau_{12}/X)}
\end{equation}
where $\tau_{12}$ is the optical depth of the HCO$^+$ gas and $X$ is
the isotope abundance ratio
\begin{equation}
X = \frac{N_{HCO^+}}{N_{H^{13}CO^+}} = \frac{\chi (HCO^+)}{\chi
(H^{13}CO^+)} \simeq \frac{[^{12}C]}{[^{13}C]}
\end{equation}
The Galactic gradient in the $^{12}C/^{13}C$ ratio ranges from
$\sim$20 to $\sim$70 (see Savage et al. 2002, and references
therein). And the ratio of $\frac{\chi (HCO^+)}{\chi (H^{13}CO^+)}$
may also be affected by chemistry. Thus, we assumed a constant $X$ =
50. We then can calculate the excitation temperature according to
\begin{equation}
T_{ex} = 4.28(ln(1+
4.28[\frac{T_{mb}}{f(1-exp(-\tau_\nu))}+J(T_{bg})]^{-1}))^{-1}
\end{equation}
where $f$ is the filling factor, here we assume it to be 1. The
derived excitation temperature is 6.3K, 6.4K, 4.7K and 12.3K
respectively. Our values are considerably lower than the $\sim$ 10
$-$ 15K temperature in previous work (e.g. Girart et al. 2000).
Purcell et al. (2006) regard that the emission is beam diluted in a
significant fraction. If the emission is smaller than the beam, the
filling factor $f$ will be less than 1 and the excitation
temperature will be underestimated. In our final analysis, we have
assumed an excitation temperature of 10 $-$ 15 K and the column
densities are calculated by Eq. (4) in Liu et al. (2013). The
fractional abundance of HCO$^+$ with respect to H$_2$ could be
estimated by $\chi$ (HCO$^+$) = N(HCO$^+$)/N(H$_2$). The derived
parameters are also listed in Table 3. From Table 3 we find the
optical depths of HCO$^+$ are less than 1. Considering the fact that
HCO$^+$ emission should be optically thick, the optical depths in
our results are probably underestimated.

\section{discussions}
\subsection{Ionized accretion?}
The HCO$^+$ (1-0) lines show non-gaussian profiles in our four
sources (G332.8256-00.5498A, G337.4032-00.4037, G345.0034-00.2240A,
and G345.4881+00.3148).By position-velocity (PV) diagram and the
method described by Mardones et al. (1997), Yu $\&$ Wang (2014)
found recent outflow activities in G332.8256-00.5498A,
G345.0034-00.2240A and G345.4881+00.3148. Large scale infall has
also been discovered in G345.0034-00.2240A and G345.4881+00.3148.
From the 4.8 GHz radio observations of ATCA, we can see HCHII/UCHII
regions have already formed inside. The sound speed in $\sim$ 10$^4$
K photo-ionized gas is about 10 km s$^{-1}$, larger than the escape
speeds in galactic molecular cores (see Fig. 1 in Tan et al. 2014).
This means if the gas in a massive star formation region gets
ionized, the gas pressure will drive a thermal pressure that may
chock off accretion. It is still unclear whether accretion can
continue in some form such as through an ionized accretion flow. The
high resolution observations of H66$\alpha$ taken at the VLA have
directly shown an ionized accretion in G10.62-0.38 (Keto $\&$ Wood
2006), while accretion in G5.89-0.39 seems to have halted at the
onset of the UCHII region (Klassen et al. 2006). We here discuss
whether (ionized) accretion could continue in G345.4881+00.3148.

We can determine the infall velocity using the two layer radiative
transfer model of Myers et al. (1996). Using their equation (9):
\begin{equation}
V_{in} \approx \frac{\sigma^2}{v_{red}-v_{blue}} ln(\frac{1 + e
T_{BD}/T_D}{1 + e T_{RD}/T_D})
\end{equation}
where $T_D$, $T_{BD}$ and $T_{RD}$ are the brightness temperature of
the dip, the height of the blue peak above the dip and the height of
the red peak above the dip respectively. The velocity dispersion
$\sigma$ can be obtained from the FWHM of H$^{13}$CO$^+$ (1-0). The
gaussian-fitted lines of HCO$^+$ and H$^{13}$CO$^+$ is shown in
figure 17. We find G345.4881+00.3148 has an infall velocity of about
4.2 km s$^{-1}$ in large scale. To estimate the mass infall rate, we
use Eq. (3) of Klassen $\&$ Wilson (2007) $dM/dt$ = (4/3)$\pi$
$<n_{H_2}>$ $\mu$ $m_H$ $r^2$ V$_{in}$, where $\mu$ means molecular
weight, $m_H$ is the mass of Hydrogen, $r$ is the radius of the
emitting region (here is set to be the beam radius), V$_{in}$ is the
infall velocity, and $<n_{H_2}>$ is the ambient density calculated
above. From the analysis above, we estimate mass infall rate of
G345.4881+00.3148 is about 2.1 $\times$ 10$^{-3}$ M$_\odot$
yr$^{-1}$. This value is much higher than those generally observed
in low-mass star forming regions.

Walmsley (1995) considered the accretion flow in free-fall onto a
star, and showed that the escape speed from the edge of the ionized
region will exceed the ionized gas sound speed $c_i$ if the
accretion rate satisfies
\begin{equation}
\begin{split}
&\frac{dM}{dt} > [\frac{8 \pi m_H^2 G M S}{2.2 \alpha_B
ln(v_{esc}/c_i)}]\\
&= 4 \times 10^{-5} (\frac{M}{100
M_\odot}\frac{N_L}{10^{49}s^{-1}})^{1/2}M_\odot yr^{-1}
\end{split}
\end{equation}
where $M$ is the central stellar mass, $m_H$ is the mean mass per H
nucleus, $\alpha_B$ is the case B recombination coefficient and
$v_{esc}$ is the escape speed from the stellar surface. The
numerical evaluation uses $v_{esc}$ = 1000 km s$^{-1}$ and $c_i$ =
10 km s$^{-1}$ (The result is unsensitive to these parameters). Our
calculated mass infall rate is sufficient to allow continuing
accretion onto an B0 (G345.4881+00.3148) type star. Like the case of
G10.6-0.4, accretion flow in this region may be ionized. However,
given the low angular resolution of the data (at 2.1 kpc,
38$^{\prime\prime}$ is over half a parsec), it is also possible that
a lower mass star is forming in the vicinity. High angular
resolution observations are needed to verify our speculations.

\subsection{N$_2$H$^+$ and HCO$^+$}
N$_2$H$^+$ is widely detected in low- and high-mass prestellar and
protostellar cores, owing to its resistance of depletion at low
temperature and high densities. We find that its line widths for our
RMSs vary in the range from 1.9 to 4.4 km s$^{-1}$, much larger than
those found in low-mass prestellar cores (Lee et al. 2001). This
large line width can not be purely explained by thermal motions.
Internal turbulence may dominate in massive star formation regions
(Myers $\&$ Fuller 1992; Caselli $\&$ Myers 1995). Chambers et al.
(2006) and Vasyunina et al. (2011) found that the line widths and
the integrated intensities of N$_2$H$^+$ have a trend to larger
values, from ``quiescent'', ``middle(or intermediate)'' to
``active'' clumps in IRDCs. They regarded it was because that
``active'' sources are more evolved and present further evolutionary
stages compared with ``quiescent'' sources. The mean line width of
our RMSs is 3.2 km s$^{-1}$, larger than the mean value of their
``active'' clumps. It is probably that our sources are much more
evolved. However, our mean integrated intensity of N$_2$H$^+$ is
only 10 K km s$^{-1}$, between their ``middle'' and ``active'' clump
values. In our sources, we found N$_2$H$^+$ column densities in the
range of 4.0 $\times$ 10$^{12}$ $-$ 2.6 $\times$ 10$^{13}$
cm$^{-2}$, and abundances of 0.9 $\times$ 10$^{-10}$ $-$ 6.1
$\times$ 10$^{-10}$. The average values are 1.6 $\times$ 10$^{13}$
cm$^{-2}$ and 3.0 $\times$ 10$^{-10}$, respectively. Vasyunina et
al. (2011) derived fractional N$_2$H$^+$ abundances of 1.9 $\times$
10$^{-10}$ $-$ 8.5 $\times$ 10$^{-9}$ with an average of 2.8
$\times$ 10$^{-9}$ for their IRDCs. Miettien (2014) derived similar
fractional N$_2$H$^+$ abundances of 2.8 $\times$ 10$^{-10}$ $-$ 9.8
$\times$ 10$^{-9}$ with an average of 1.6 $\times$ 10$^{-9}$. Our
value of $\chi$ (N$_2$H$^+$) is nearly one order less than these
observations. Chemical models indicate when dust temperature exceeds
20 K, CO would be evaporated from the dust grains. The enhanced CO
abundance will destroy N$_2$H$^+$ producing HCO$^+$. The mean dust
temperature of our sources is about 30 K, much larger than that in
IRDCs ($\sim$ 15 K). Previous molecular-line observations have also
shown the typical gas kinetic temperature of IRDCs clumps lies in
the range of 10 and 20 K (e.g. Sridharan et al. 2005; Sakai et al.
2008; Ragan et al. 2011). This may be the reason why our values of
$\chi$ (N$_2$H$^+$) are much smaller.

HCO$^+$ can be formed through N$_2$H$^+$ + CO $\rightarrow$ HCO$^+$
+ N$_2$ when CO is evaporated from the dust grains. The HCO$^+$
abundance can also be increased in regions where shocks are
generated (e.g. the gas through which a protostellar outflow is
passing). Our HCO$^+$ column densities range from 5.1 $\times$
10$^{12}$ cm$^{-2}$ to 9.6 $\times$ 10$^{13}$ cm$^{-2}$ (2.4
$\times$ 10$^{13}$ cm$^{-2}$ on average). These values are very
close to those of Miettien (2014), and are higher by a factor of
about five on average than those derived by Liu et al. (2013). As
shown in figure 1, there is a hint that the fractional abundance of
HCO$^+$ increases as a function of the N$_2$H$^+$ abundance. A least
squares fit to the data points yields log[$\chi$ (HCO$^+$)] =
(0.76$\pm$0.51) + (1.07$\pm$0.05)log[$\chi$ (N$_2$H$^+$)], with the
linear Pearson correlation coefficient of $r$ = 0.61. The trend is
consistent with theoretical models. Because CO is both the main
supplier of HCO$^+$ and the main destroyer of N$_2$H$^+$, this
suggests that the HCO$^+$ abundance increases with respect to
N$_2$H$^+$ as sources evolve to a warmer phase. N$_2$H$^+$ can also
be destroyed in the electron recombination: N$_2$H$^+$ + e$^-$
$\rightarrow$ N$_2$ + H or NH + N (e.g. Dislaire et al. 2012; Vigren
et al. 2012). We tried to find a relationship between the N$_2$H$^+$
fractional abundance and Lyman continuum fluxes ($N_L$). Figure 2
seems to show that $\chi$(N$_2$H$^+$) decreases as a function of
$N_L$. UV radiation field may have an influence on the chemistry of
N$_2$H$^+$ in the RMSs. Since our limited data, more studies should
be done in the near future to check out this conclusion.

\section{summary}
Using archival data taken from ATLASGAL, ATCA, and MALT90, we have
studied the physical and chemical properties of 18 southern RMSs
(HII regions), and compared our results with previous observations
of IRDCs. Our sources are much more evolved than general IRDCs. The
H$_2$ column densities and gas number densities have been derived.
Ionized accretion may be taking place in G345.4881+00.3148. Most of
the detected N$_2$H$^+$ and HCO$^+$ emissions match well with the
dust emission, implying a close link to their chemical evolution in
the RMSs. We found the abundance of N$_2$H$^+$ is one order of
magnitude lower than in other surveys of IRDCs, and a positive
correlation between the abundances of N$_2$H$^+$ and HCO$^+$. The
fractional abundance of N$_2$H$^+$ with respect to H$_2$ decreases
as a function of $N_L$. It seems that the UV radiation field has an
influence on the chemistry of N$_2$H$^+$ in the RMSs. Like the
chemical model of low-mass star formation theory, these observed
trends could be interpreted as an indication of enhanced destruction
of N$_2$H$^+$, either by CO or through dissociative recombination
with electrons produced by central UV photons.

\begin{figure}
\centerline{\psfig{file=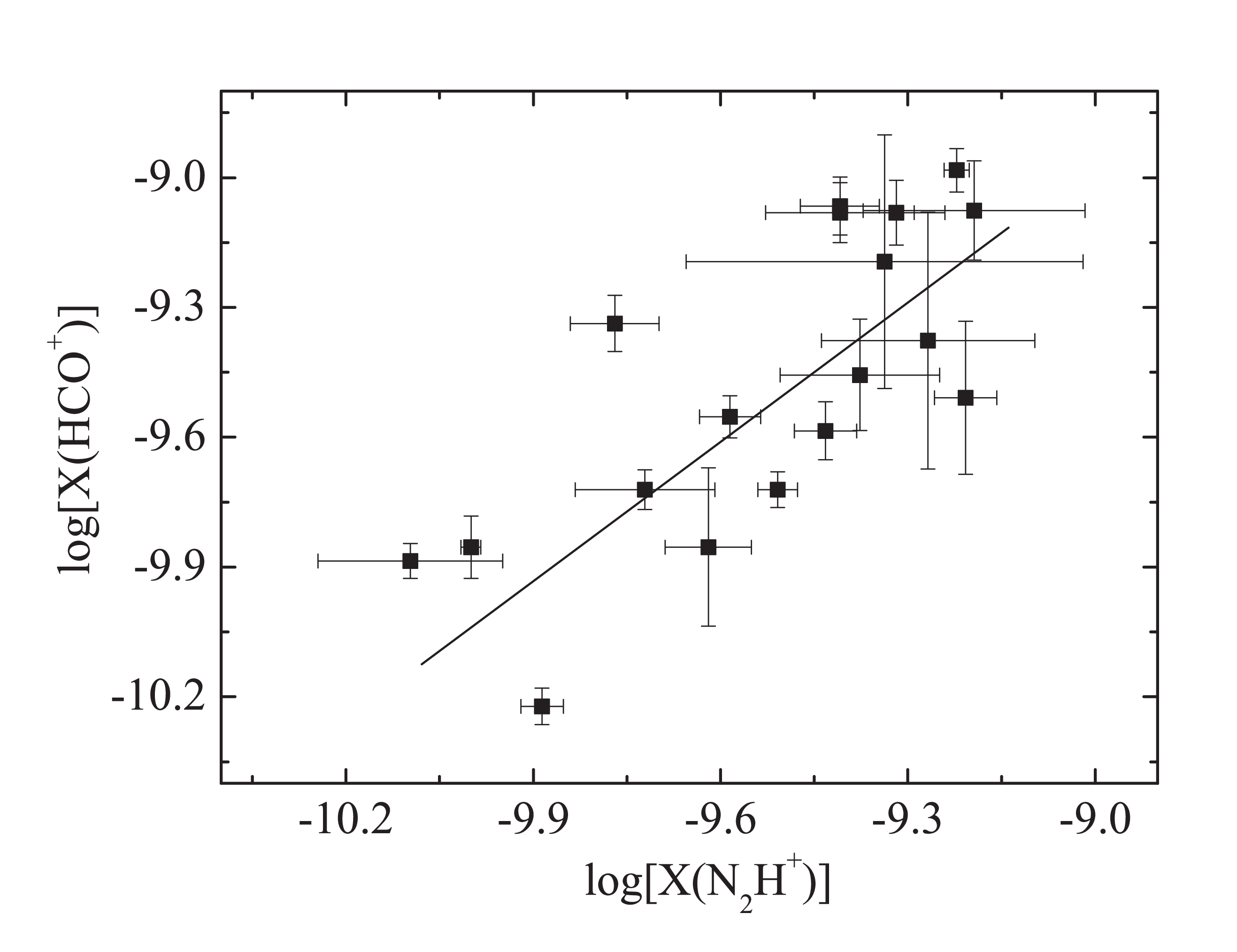,width=2.6in,height=2.0in}}
\caption{ HCO$^+$ fractional abundances vs. N$_2$H$^+$ fractional
abundances. }
\end{figure}
\begin{figure}
\centerline{\psfig{file=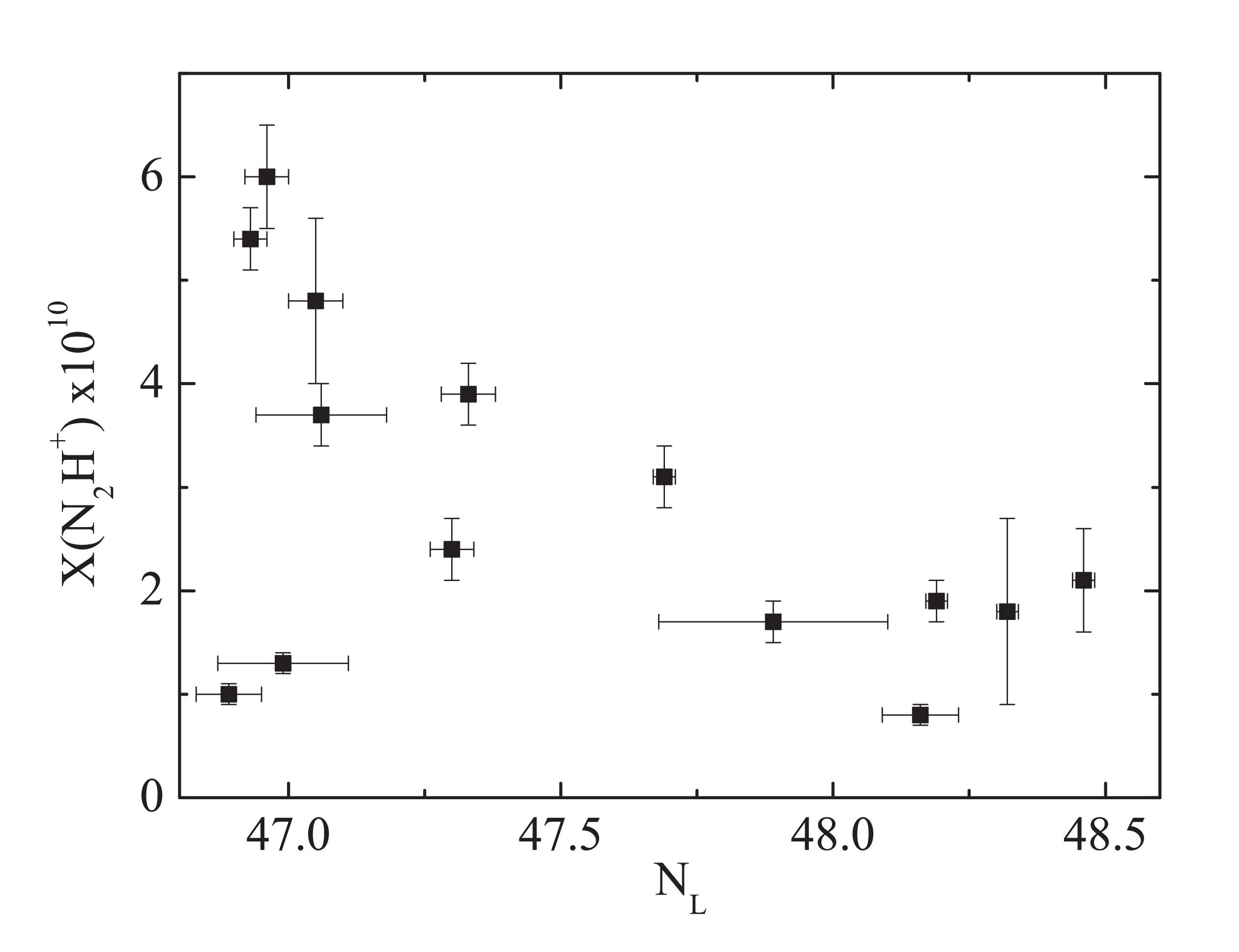,width=2.6in,height=2.0in}}
\caption{N$_2$H$^+$ fractional abundance plotted as a function of
$N_L$ in logarithmic scales.}
\end{figure}

\section*{ACKNOWLEDGEMENTS}
We thank the anonymous referee for his constructive suggestions.
This paper made use of information from the Red MSX Source (RMS)
survey database
http://rms.leeds.ac.uk/cgi-bin/public/RMS$_{-}$DATABASE.cgi which
was constructed with support from the Science and Technology
Facilities Council of the UK. This research made use of data
products from the Millimetre Astronomy Legacy Team 90 GHz (MALT90)
survey. The Mopra telescope is part of the Australia Telescope and
is funded by the Commonwealth of Australia for operation as National
Facility managed by CSIRO.

%\bibliographystyle{mn2e}
%\bibliography{journals,modrefs,psrrefs,crossrefs}

\clearpage

\begin{figure}
\centerline{\psfig{file=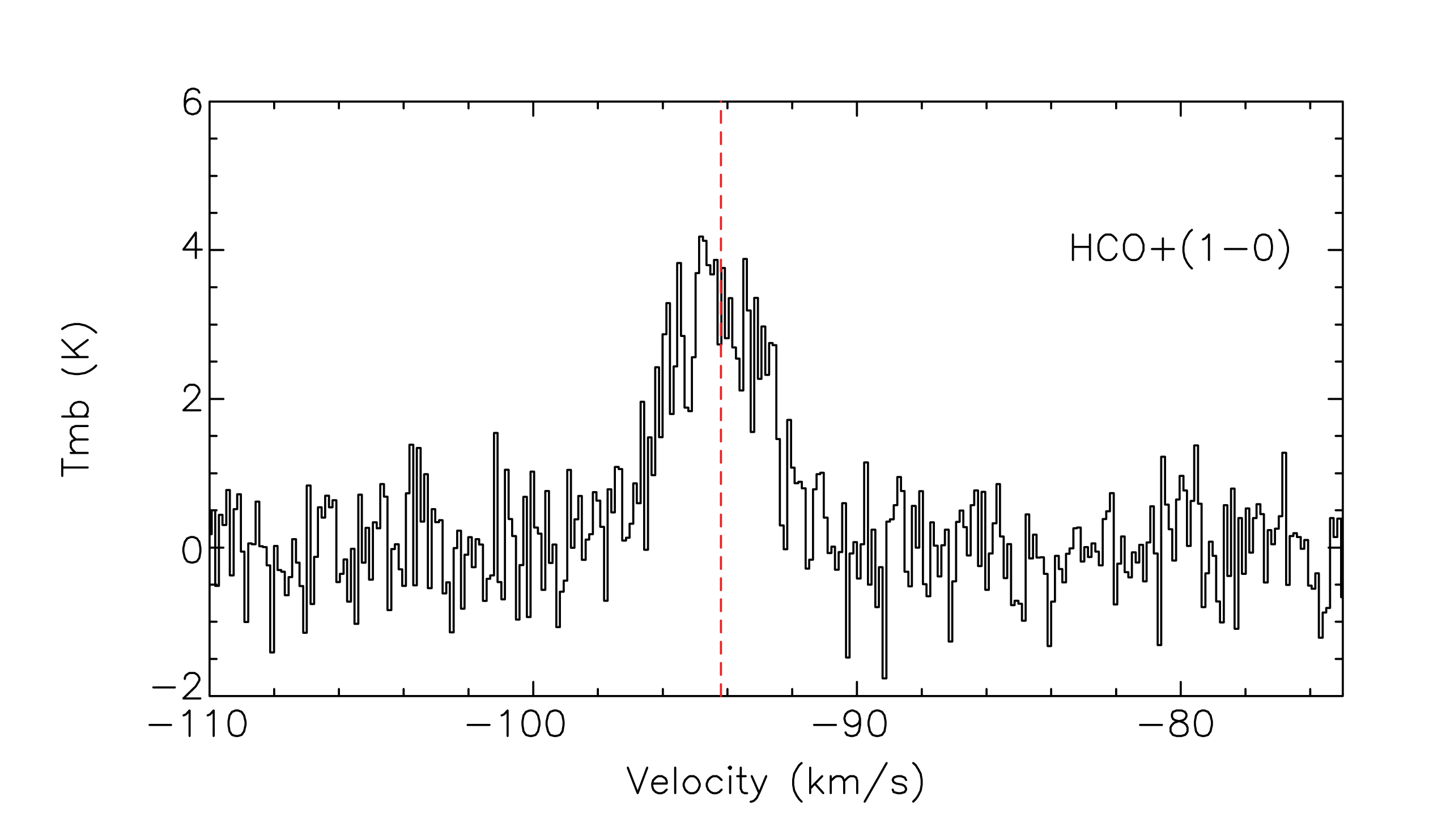,width=2.6in,height=1.8in}}
\centerline{\psfig{file=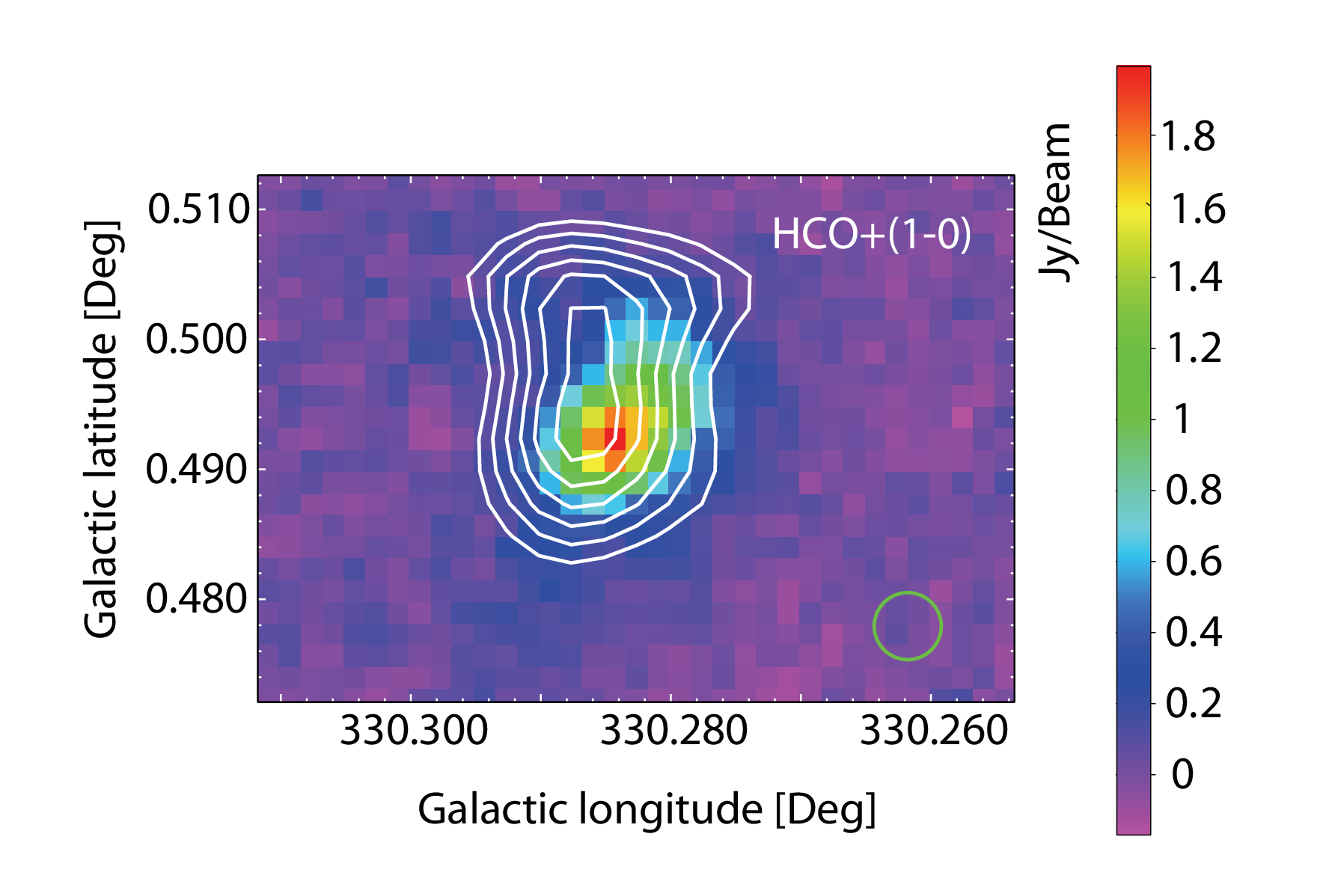,width=2.6in,height=1.8in}}
\centerline{\psfig{file=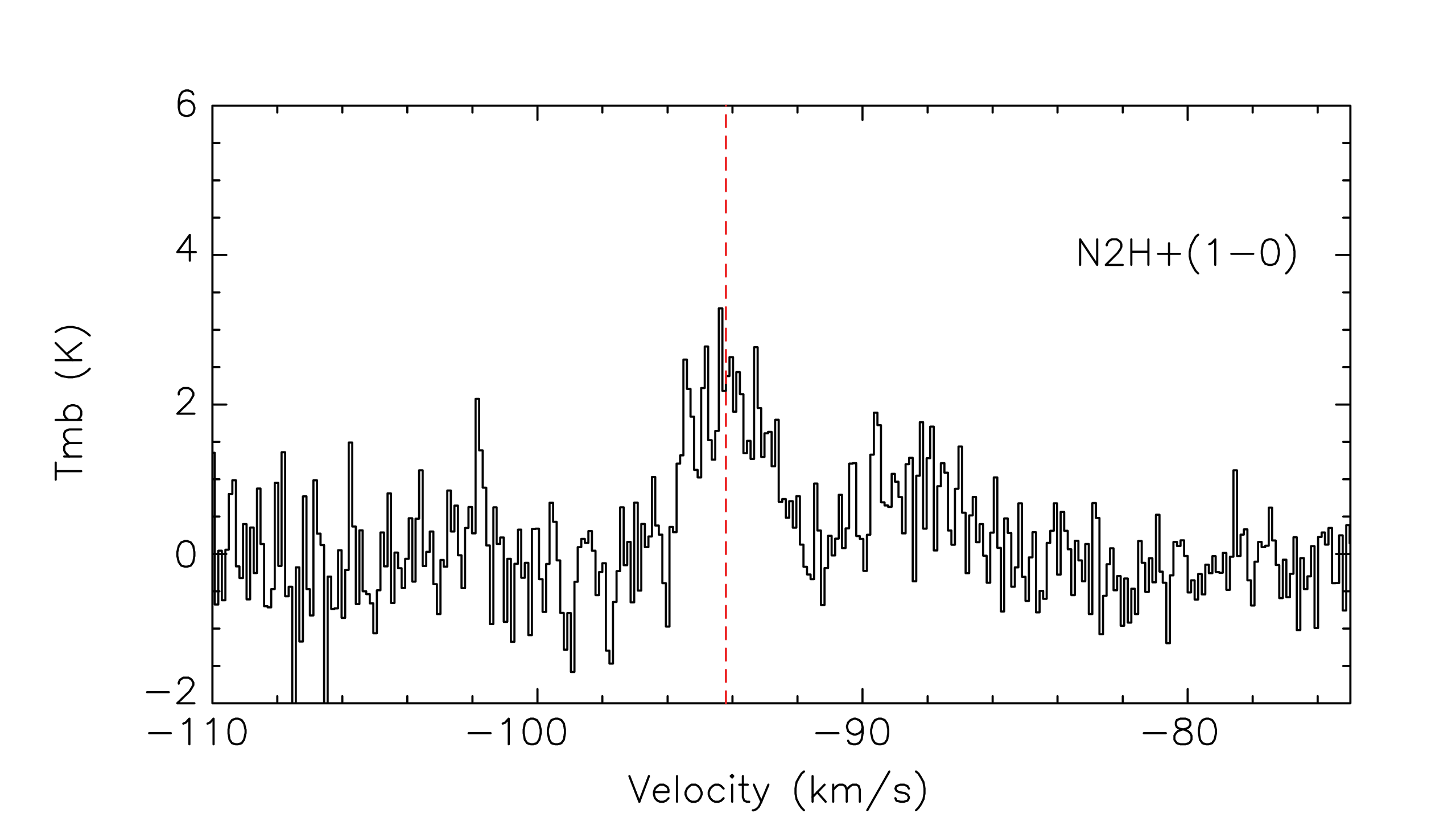,width=2.6in,height=1.8in}}
\centerline{\psfig{file=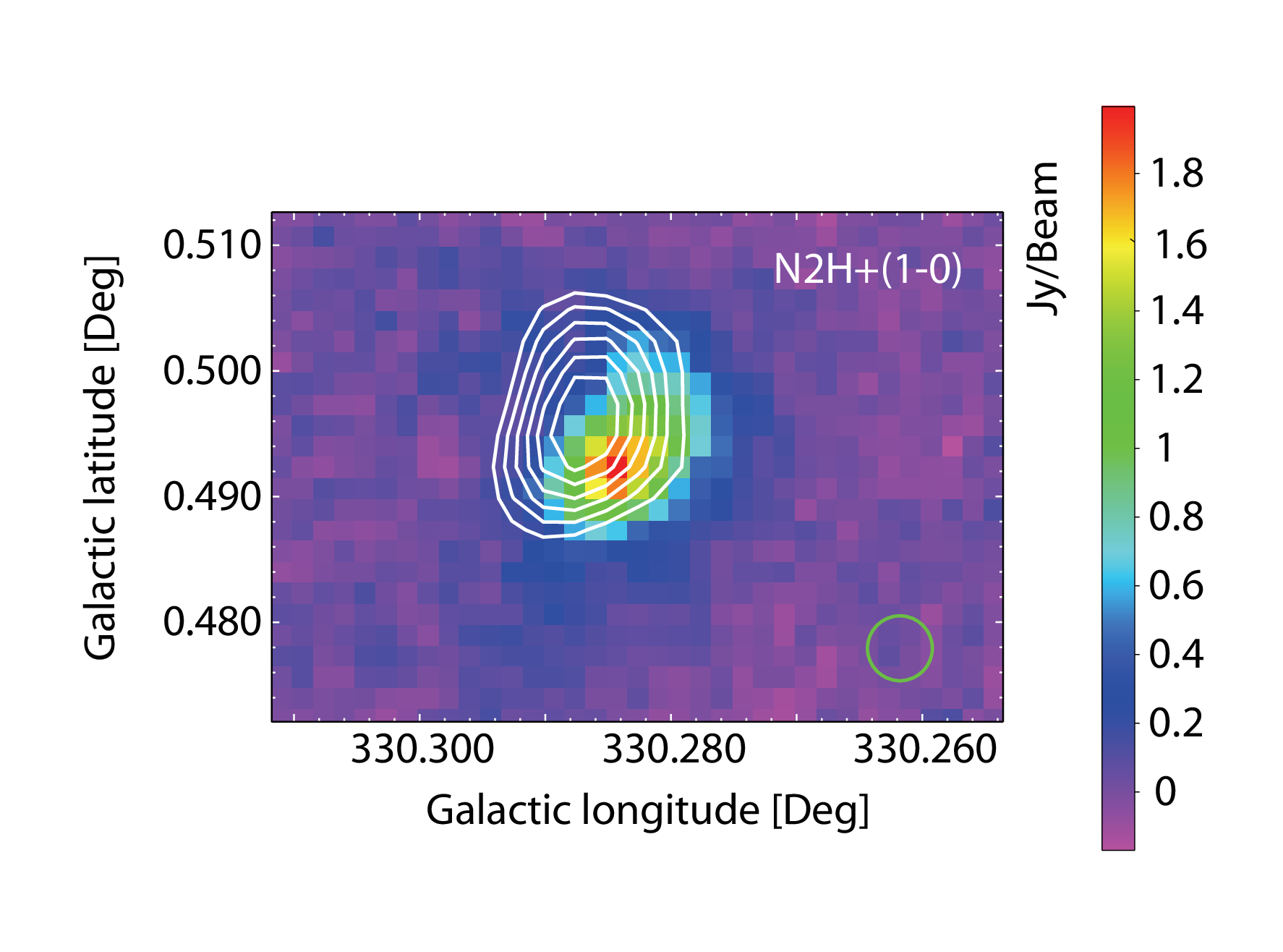,width=2.6in,height=1.8in}}
\caption{Spectra and integrated intensities superimposed on the 870
$\mu$m map in gray scale of G330.2838+00.4933. The red dash line
represents the V$_{LSR}$ of N$_2$H$^+$ line. Contour levels are
30$\%$, 40$\%$...90$\%$ of the center peak emissions. The angular
resolution of the ATLASGAL survey is indicated by the green circle
shown in the lower right corner. }
\end{figure}
\begin{figure}
\centerline{\psfig{file=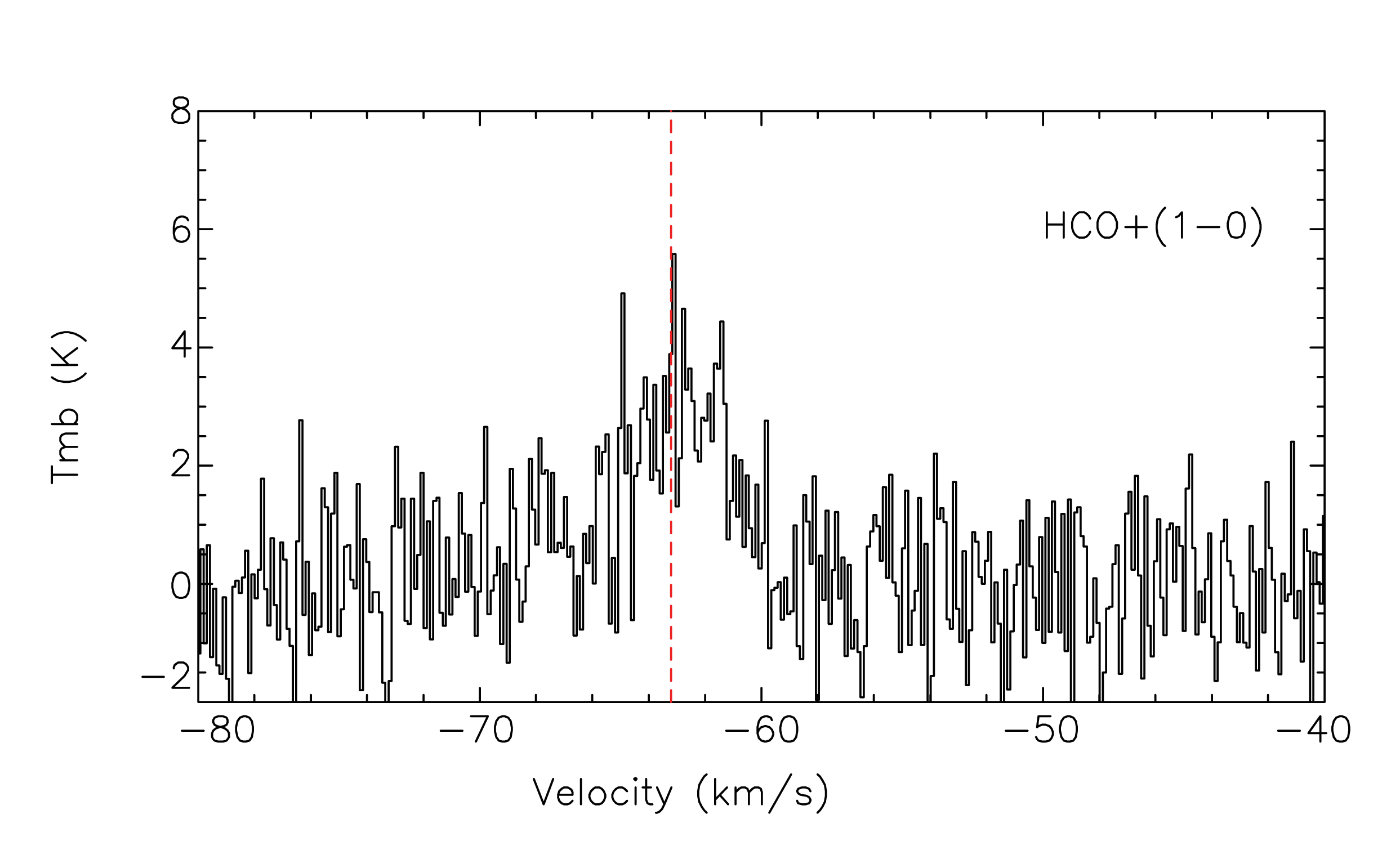,width=2.6in,height=1.8in}}
\centerline{\psfig{file=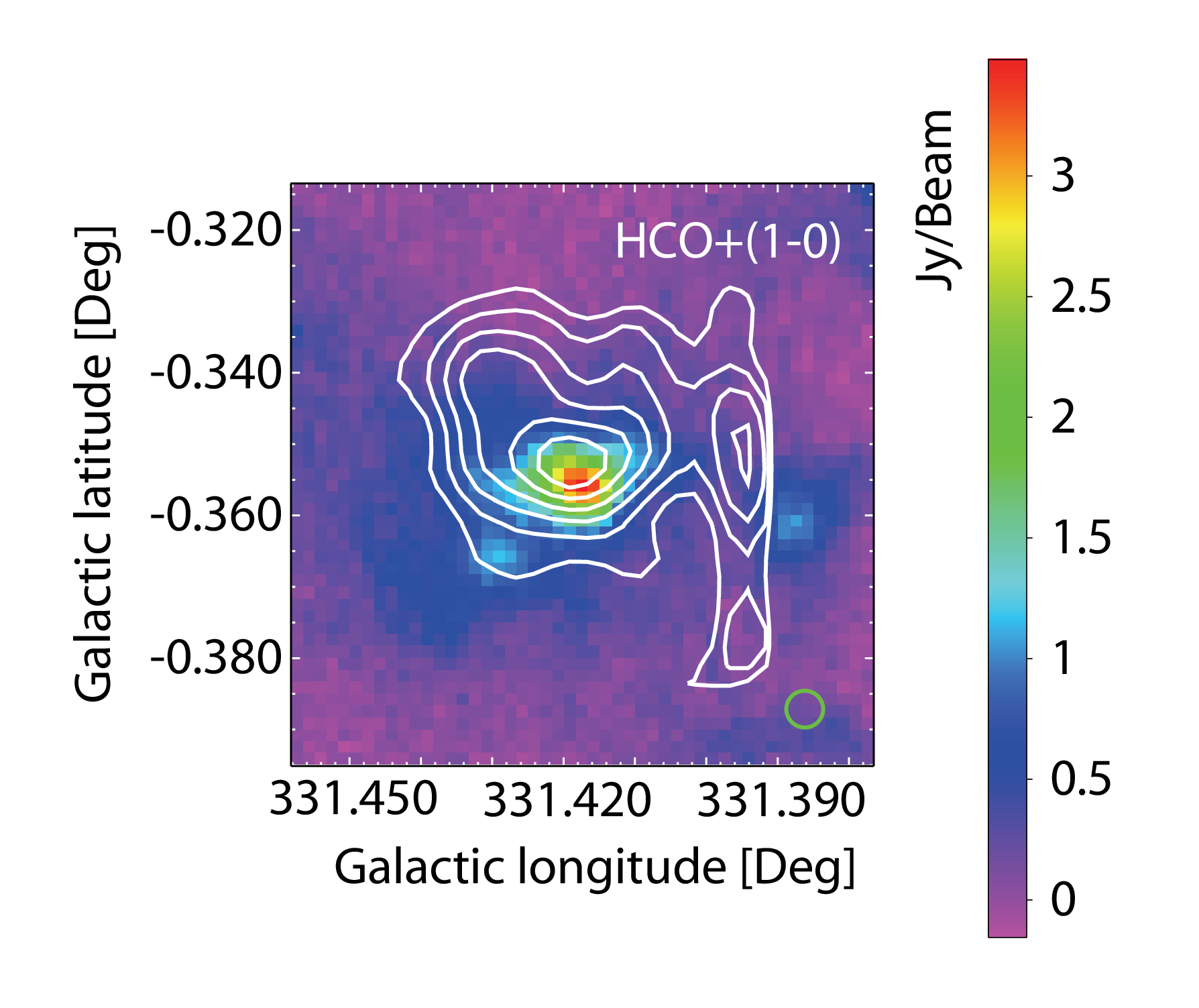,width=2.6in,height=2.3in}}
\centerline{\psfig{file=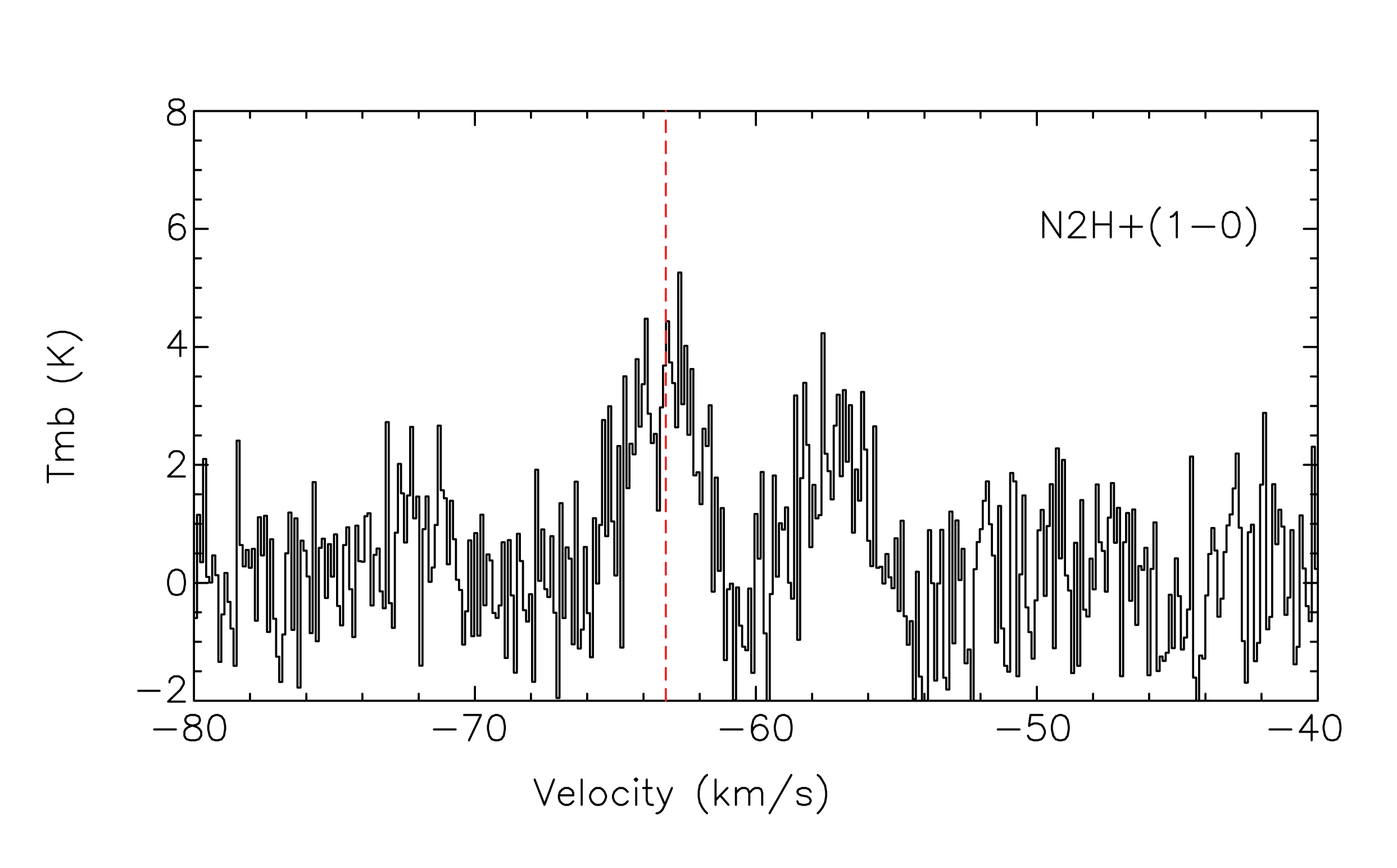,width=2.6in,height=1.8in}}
\centerline{\psfig{file=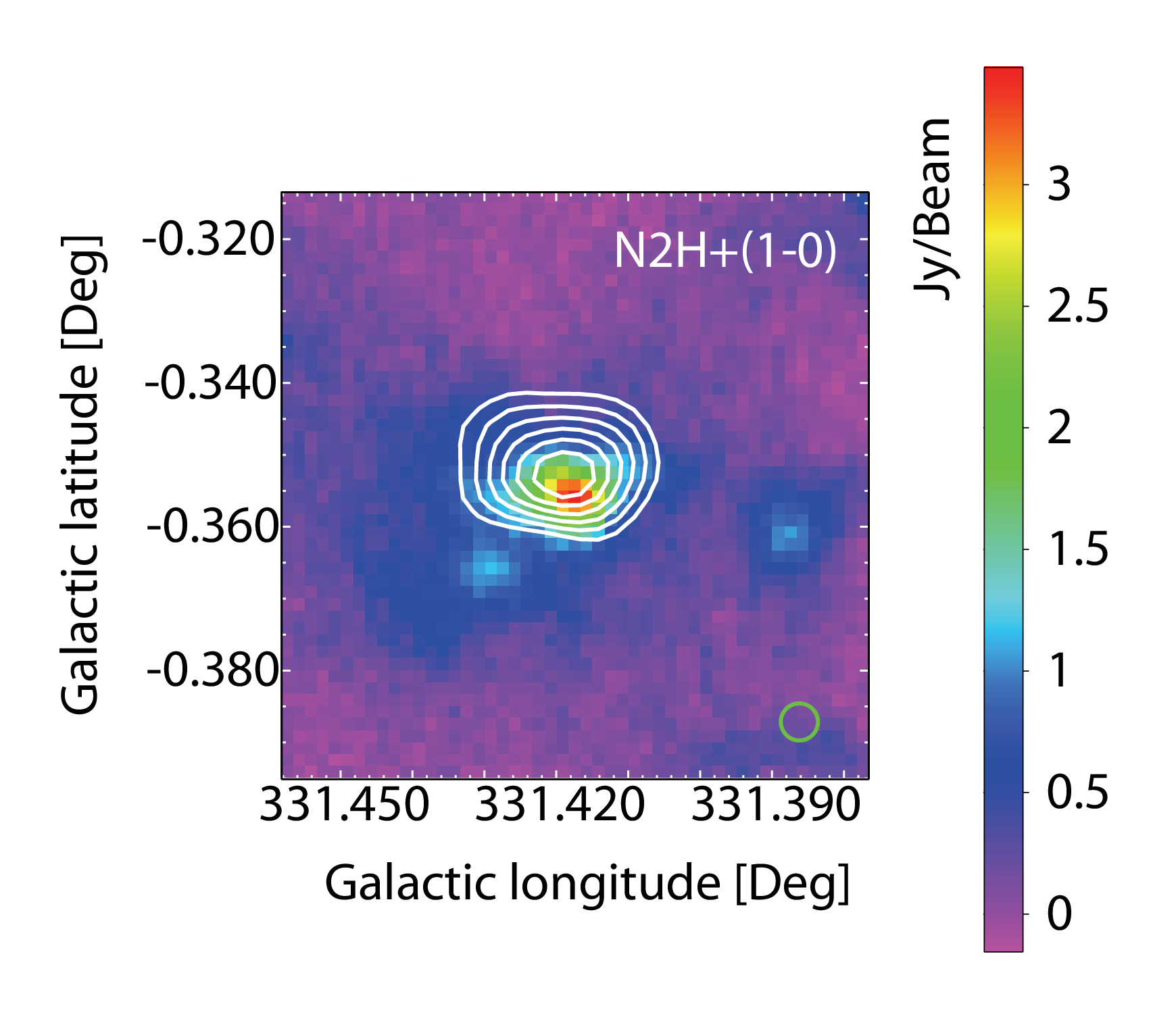,width=2.6in,height=2.3in}}
\caption{Spectra and integrated intensities superimposed on the 870
$\mu$m map in gray scale of G331.4181-00.3546. The red dash line
represents the V$_{LSR}$ of N$_2$H$^+$ line. Contour levels are
30$\%$, 40$\%$...90$\%$ of the center peak emissions. The angular
resolution of the ATLASGAL survey is indicated by the green circle
shown in the lower right corner. }
\end{figure}
\begin{figure}
\centerline{\psfig{file=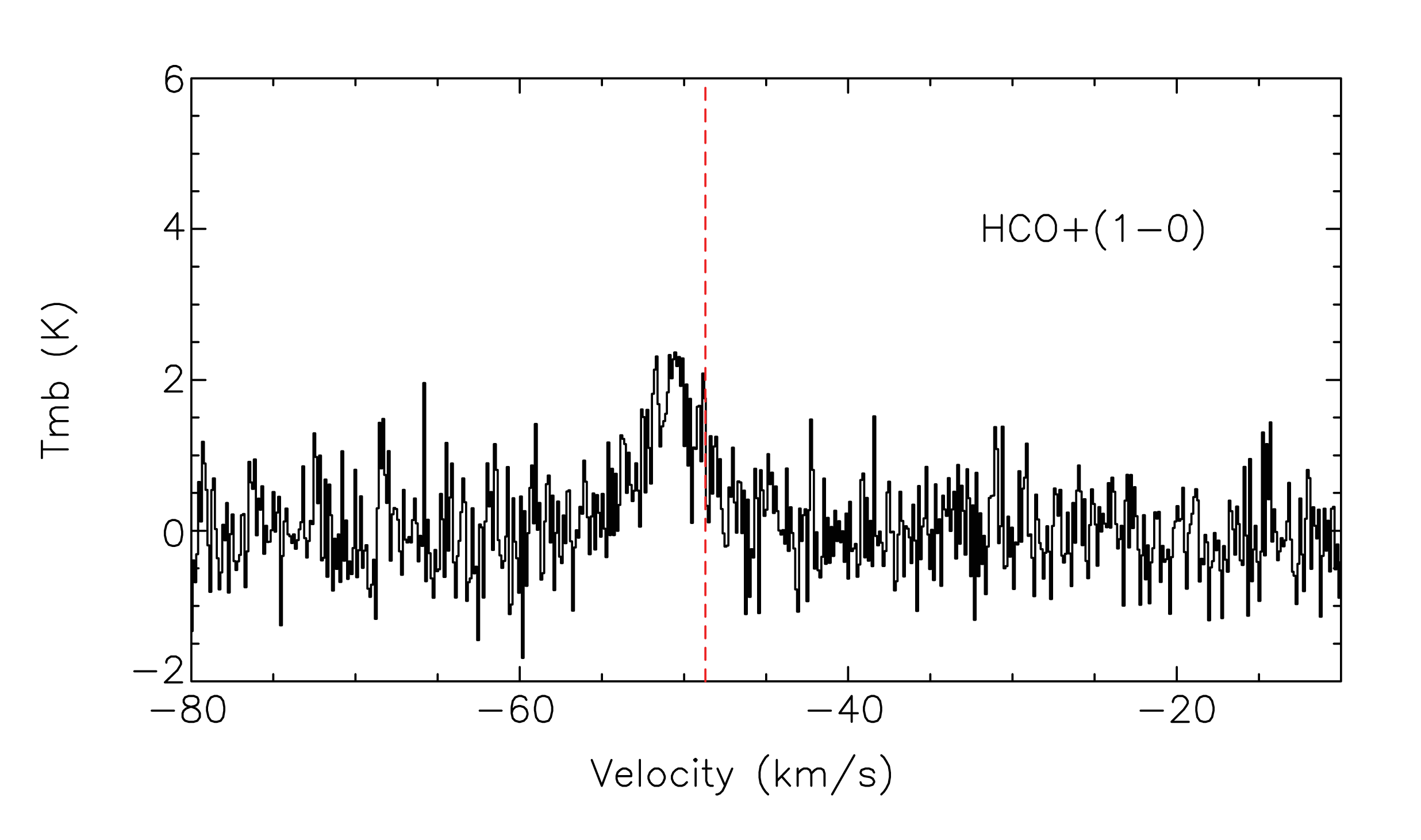,width=2.6in,height=1.8in}}
\centerline{\psfig{file=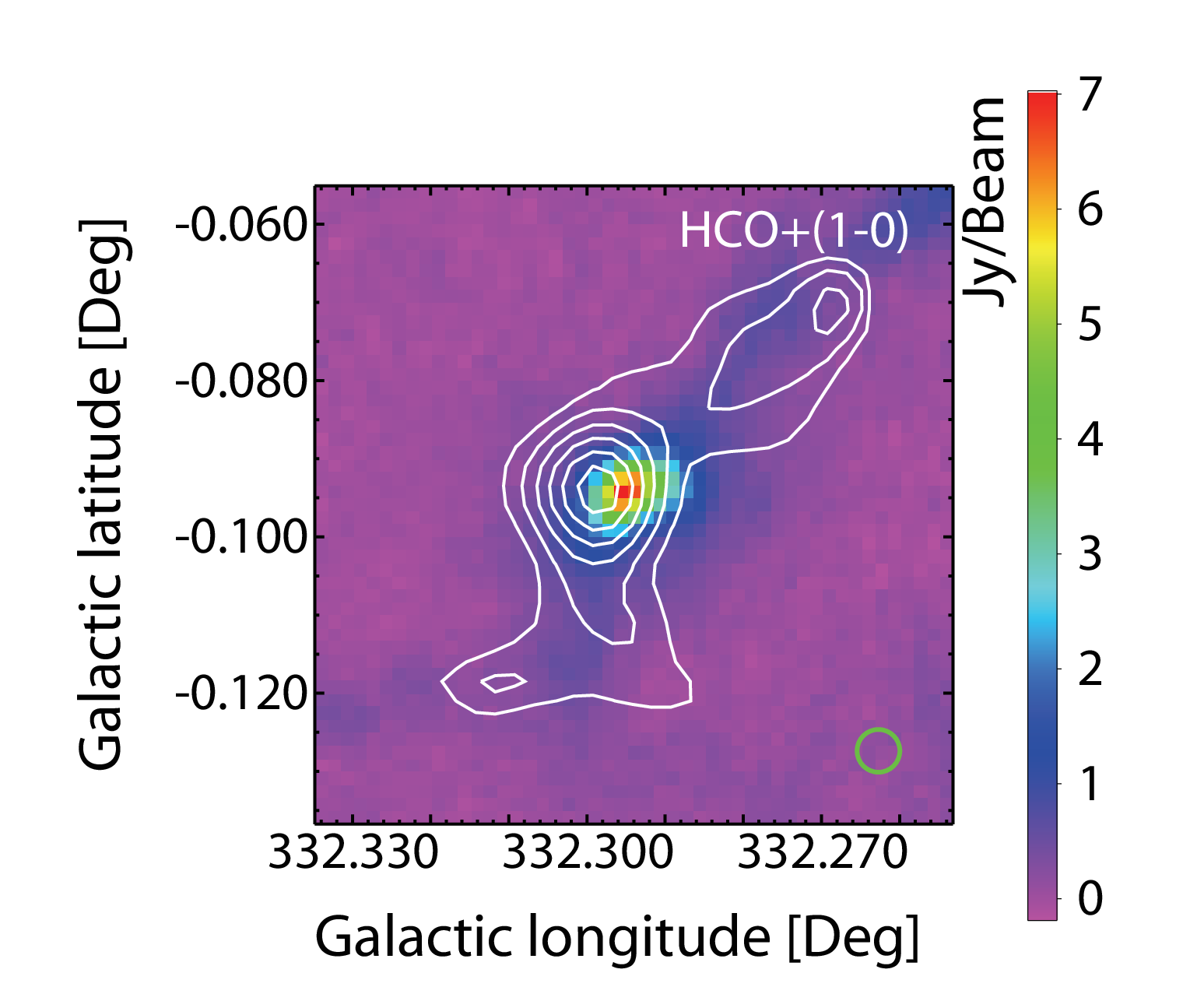,width=2.6in,height=2.3in}}
\centerline{\psfig{file=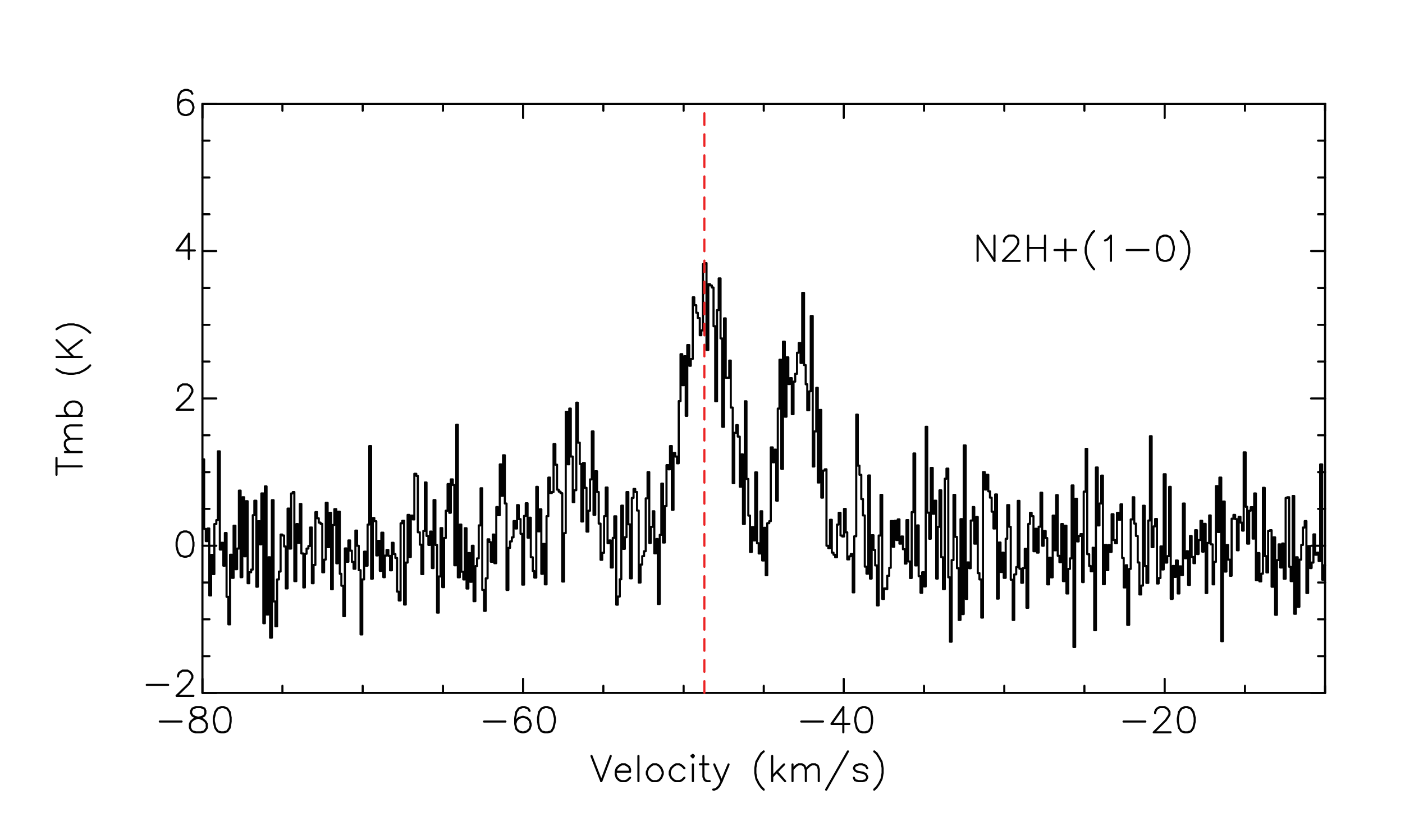,width=2.6in,height=1.8in}}
\centerline{\psfig{file=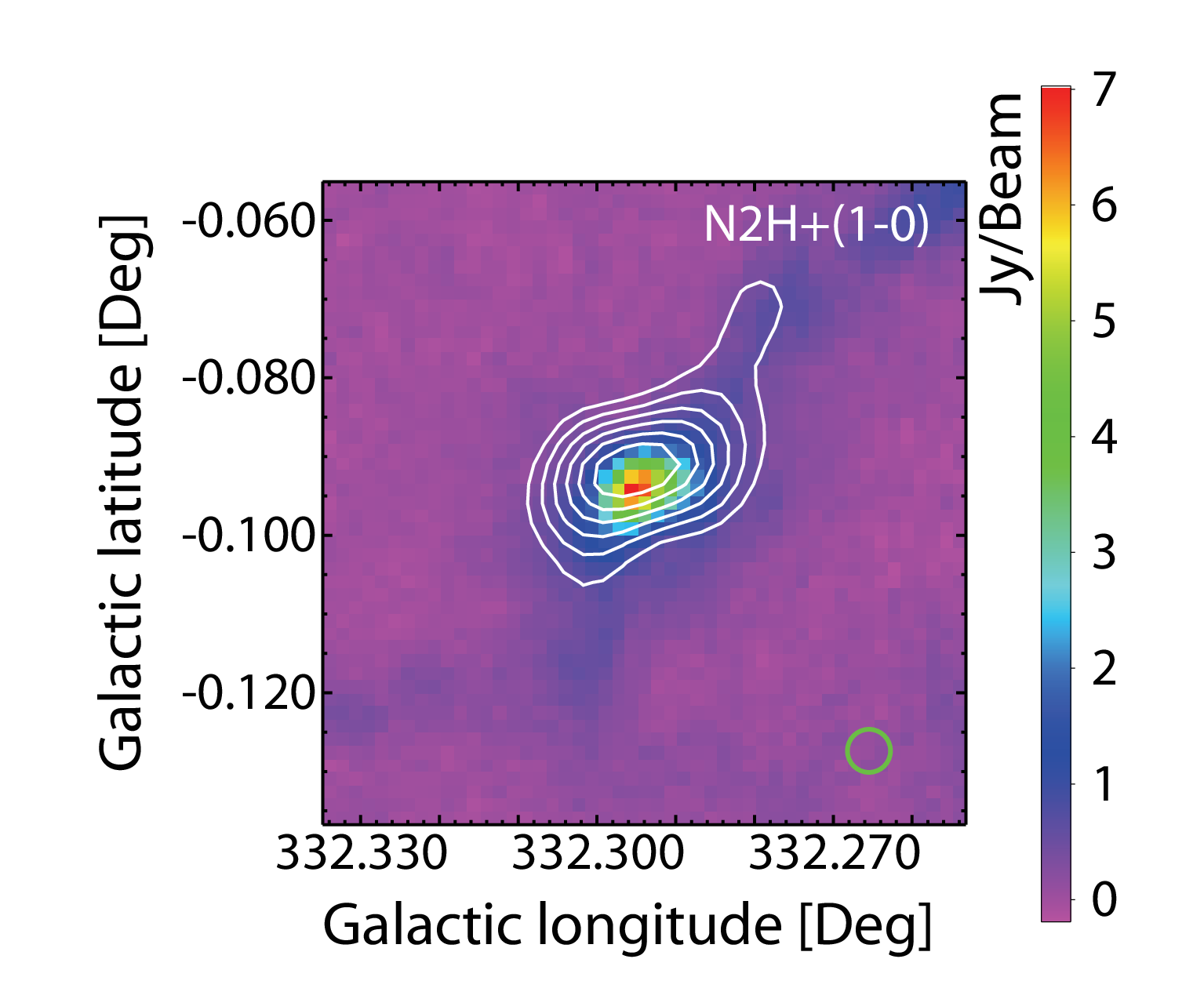,width=2.6in,height=2.3in}}
\caption{Spectra and integrated intensities superimposed on the 870
$\mu$m map in gray scale of G332.2944-00.0962. The red dash line
represents the V$_{LSR}$ of N$_2$H$^+$ line. Contour levels are
30$\%$, 40$\%$...90$\%$ of the center peak emissions. The angular
resolution of the ATLASGAL survey is indicated by the green circle
shown in the lower right corner.  }
\end{figure}
\begin{figure}
\centerline{\psfig{file=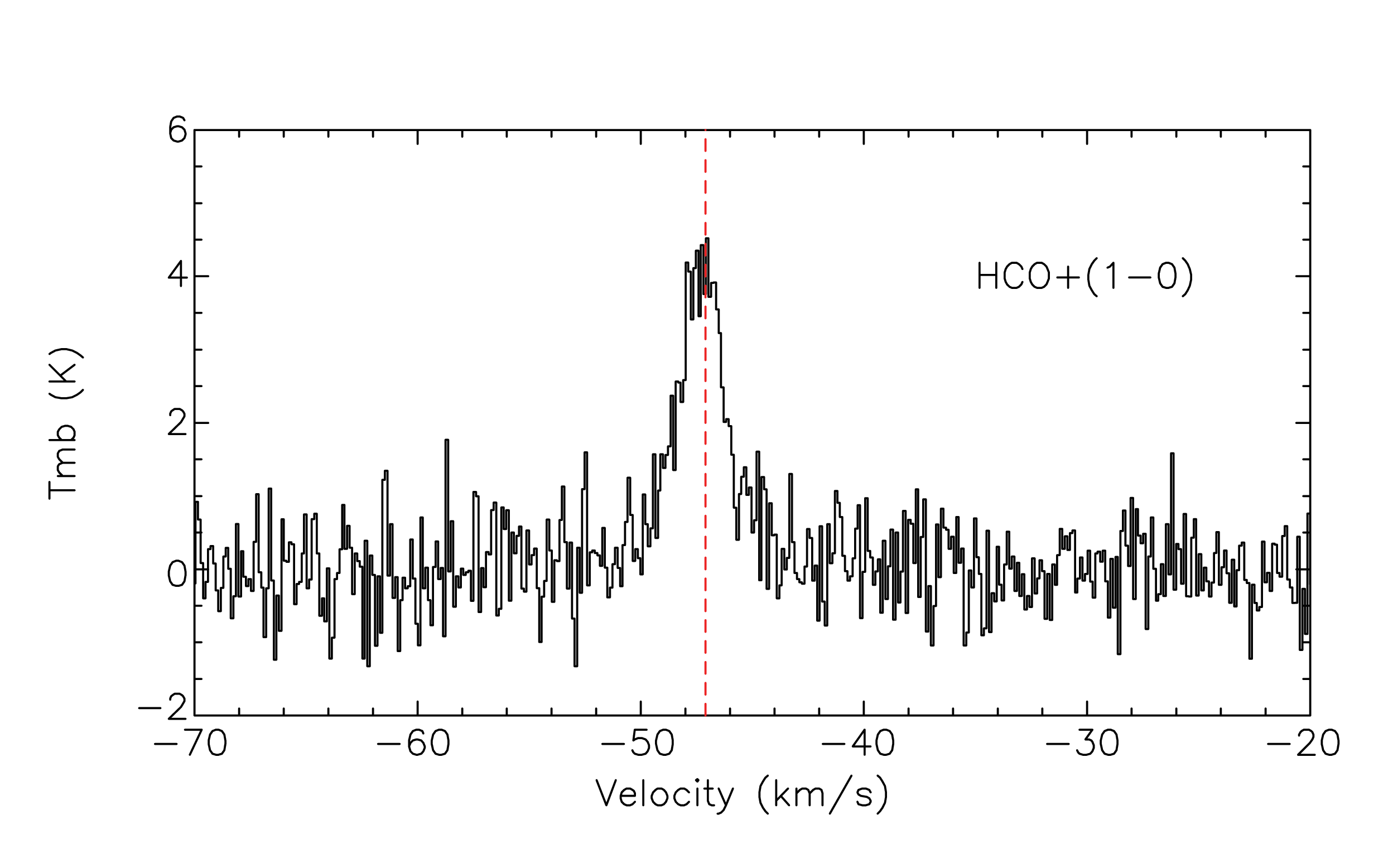,width=2.6in,height=1.8in}}
\centerline{\psfig{file=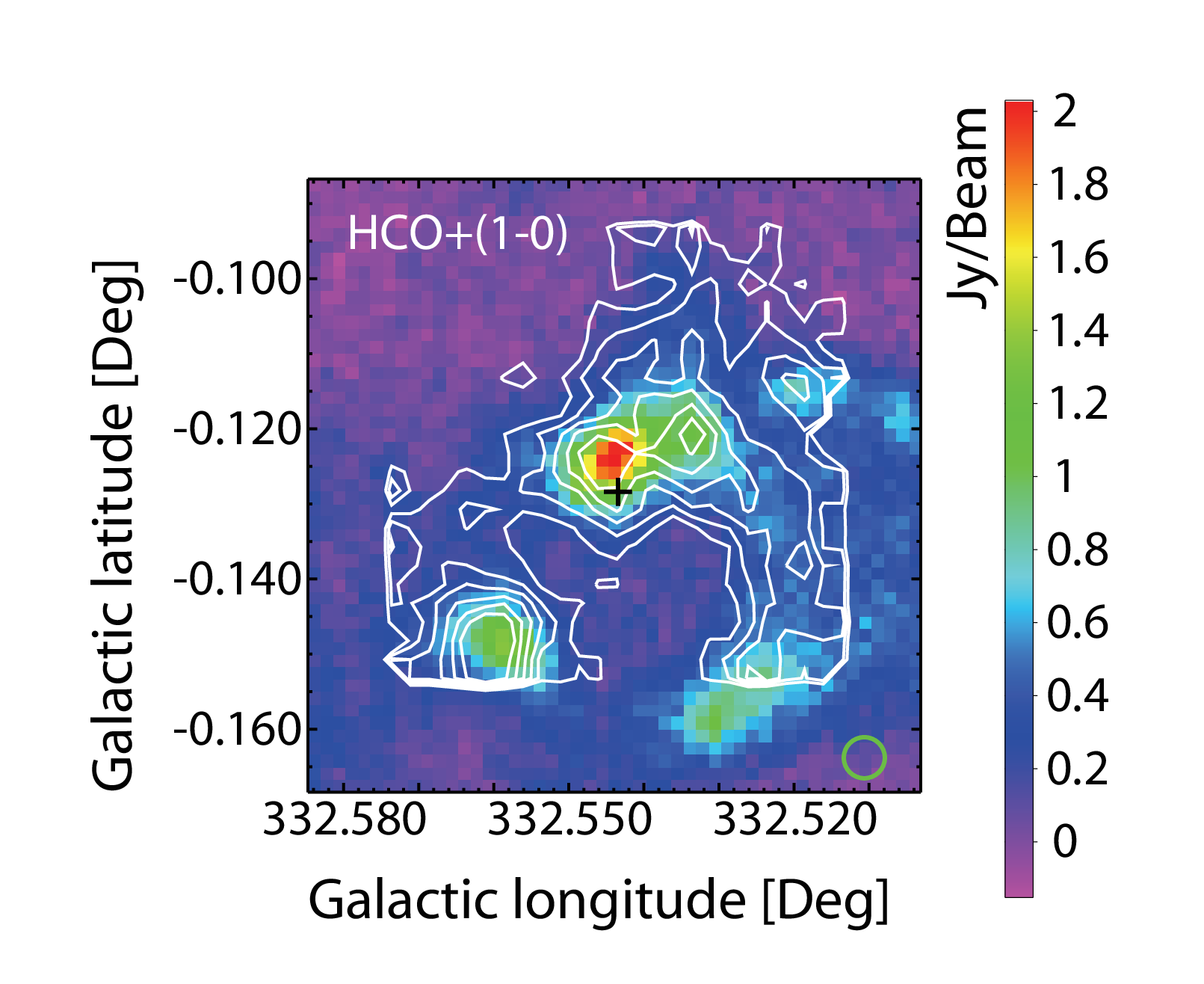,width=2.6in,height=2.3in}}
\centerline{\psfig{file=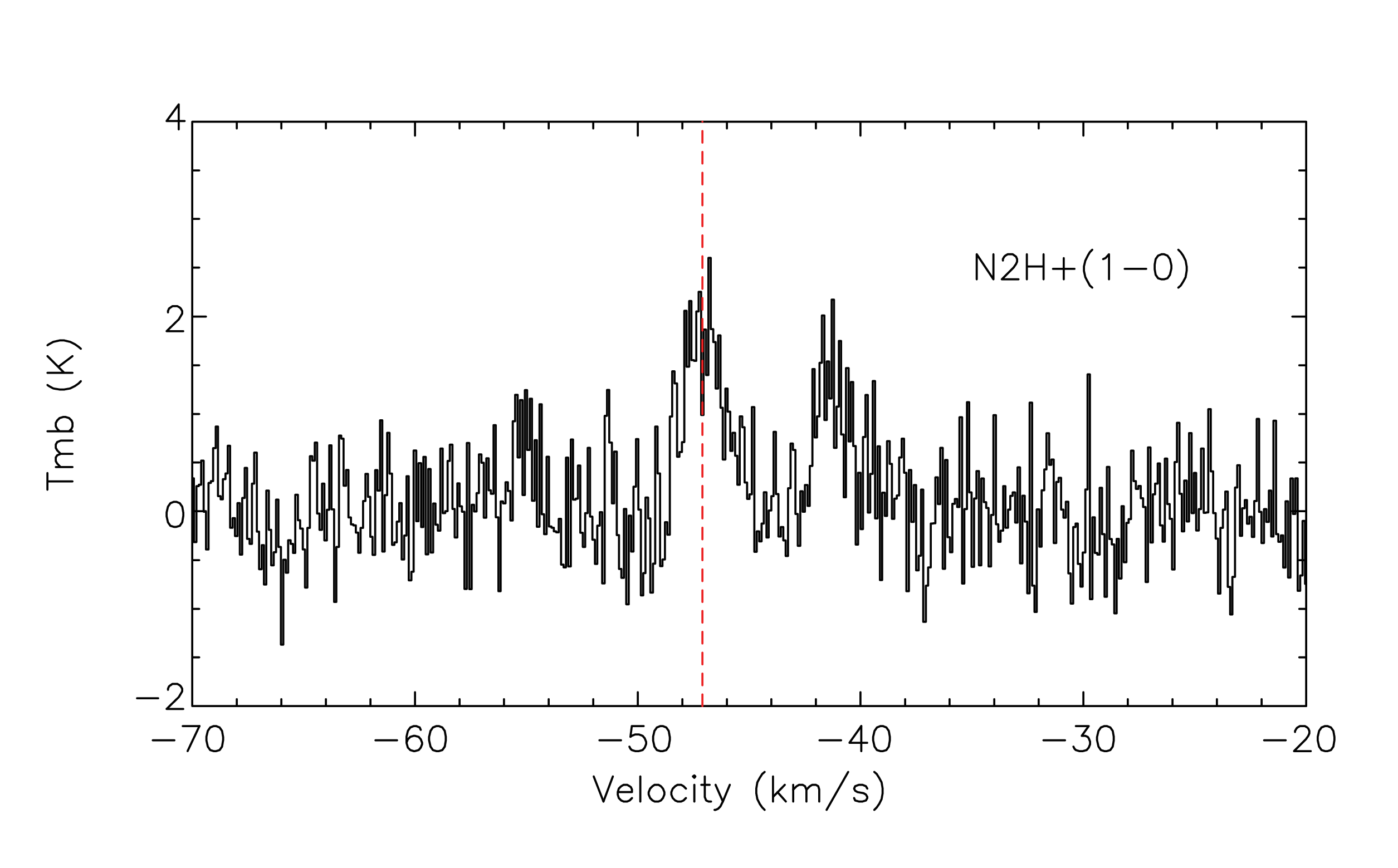,width=2.6in,height=1.8in}}
\centerline{\psfig{file=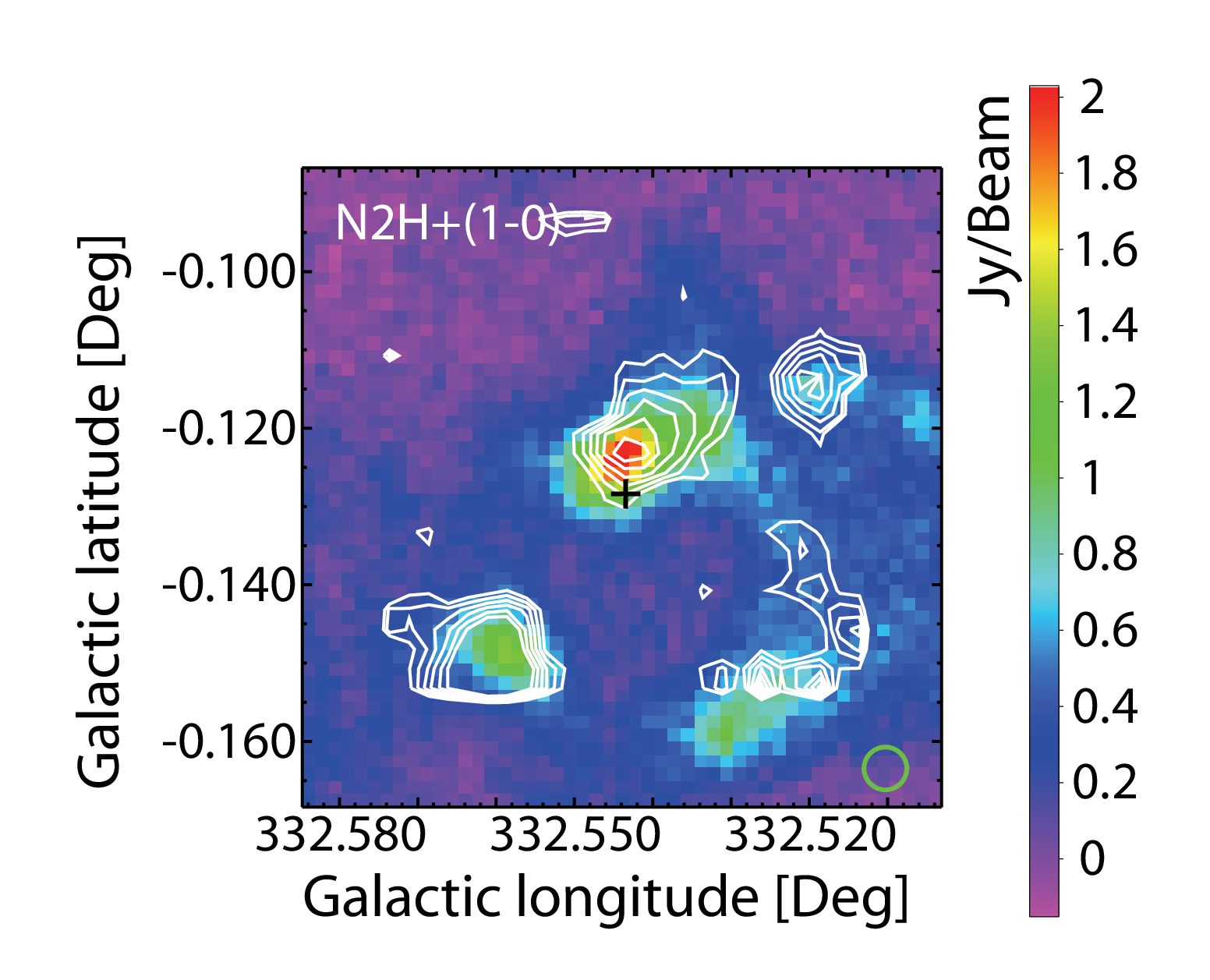,width=2.6in,height=2.3in}}
\caption{Spectra and integrated intensities superimposed on the 870
$\mu$m map in gray scale of G332.5438-00.1277. The red dash line
represents the V$_{LSR}$ of N$_2$H$^+$ line. Contour levels are
30$\%$, 40$\%$...90$\%$ of the center peak emissions. The black
cross marks the location of G332.5438-00.1277. The angular
resolution of the ATLASGAL survey is indicated by the green circle
shown in the lower right corner. }
\end{figure}
\begin{figure}
\centerline{\psfig{file=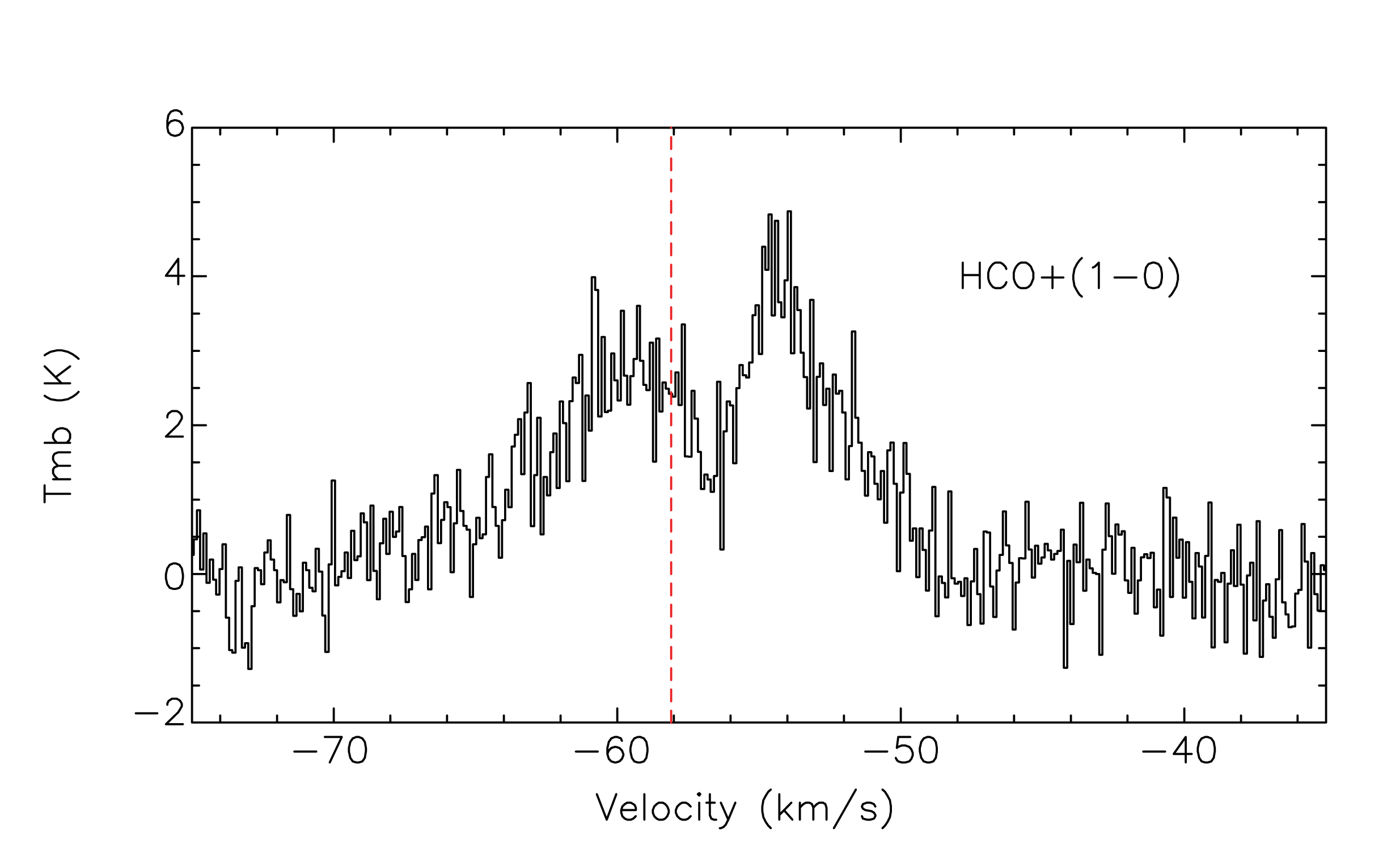,width=2.6in,height=1.8in}}
\centerline{\psfig{file=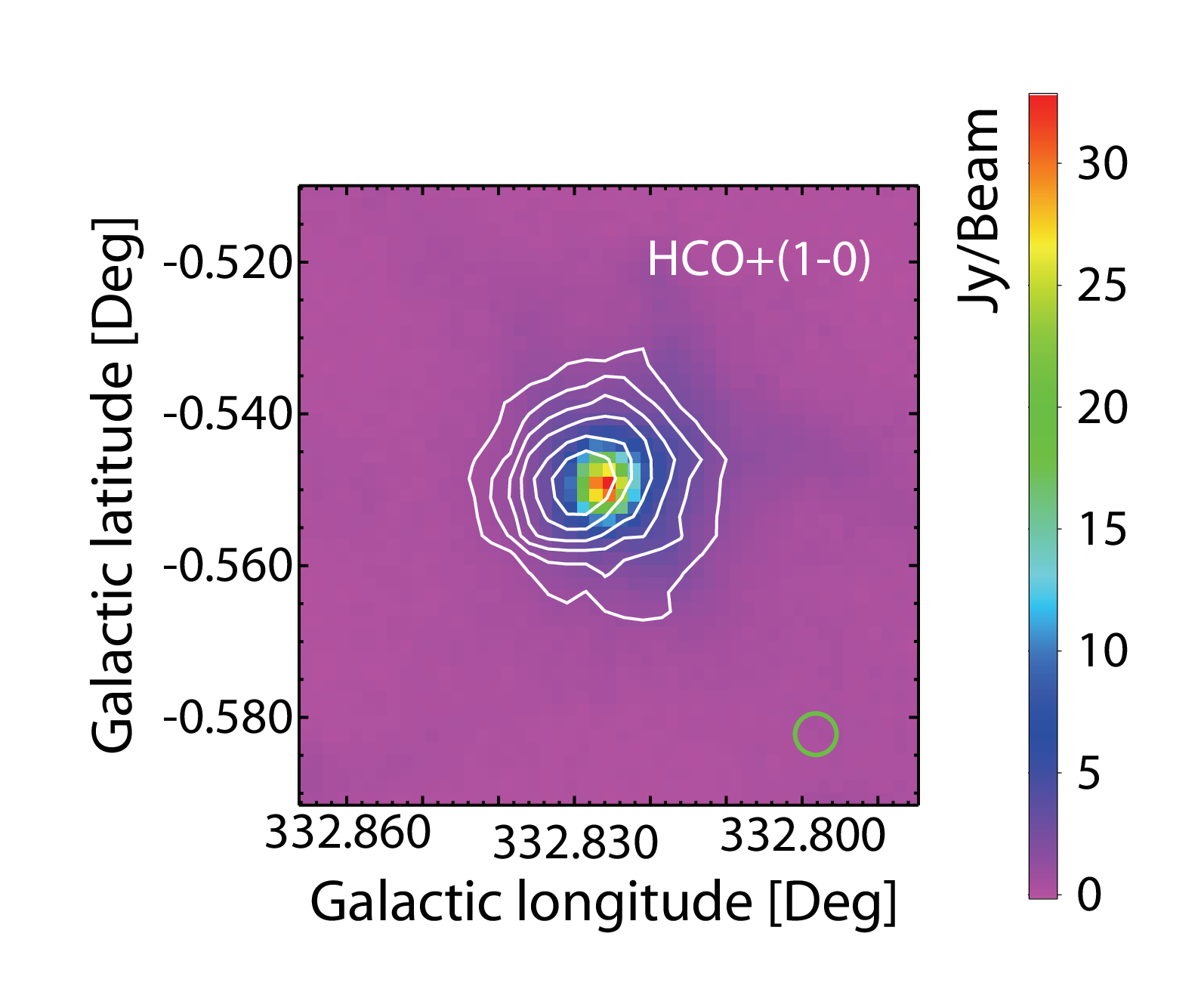,width=2.6in,height=2.3in}}
\centerline{\psfig{file=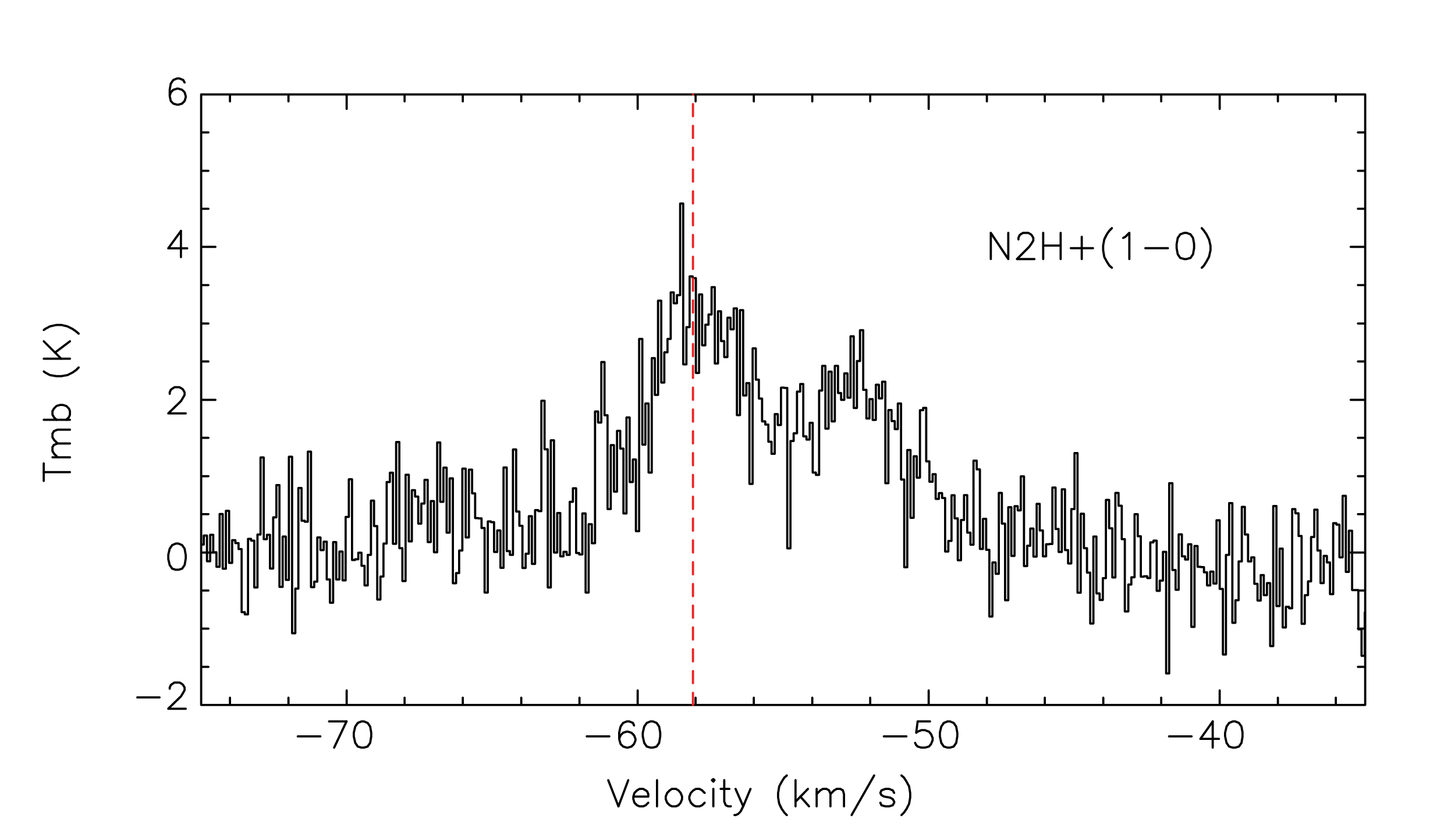,width=2.6in,height=1.8in}}
\centerline{\psfig{file=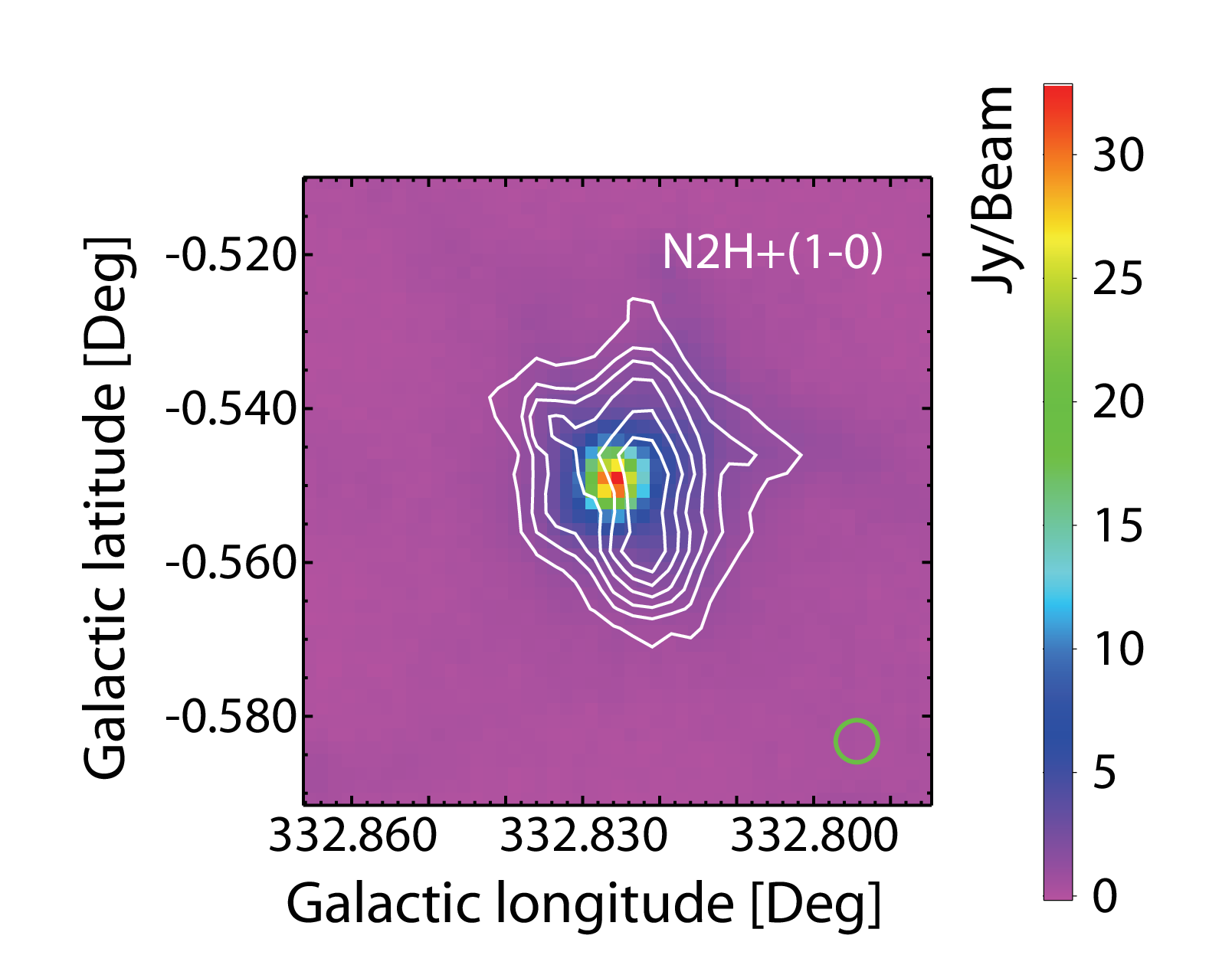,width=2.6in,height=2.3in}}
\caption{Spectra and integrated intensities superimposed on the 870
$\mu$m map in gray scale of G332.8256-00.5498A. The red dash line
represents the V$_{LSR}$ of N$_2$H$^+$ line. Contour levels are
30$\%$, 40$\%$...90$\%$ of the center peak emissions. The angular
resolution of the ATLASGAL survey is indicated by the green circle
shown in the lower right corner. }
\end{figure}
\begin{figure}
\centerline{\psfig{file=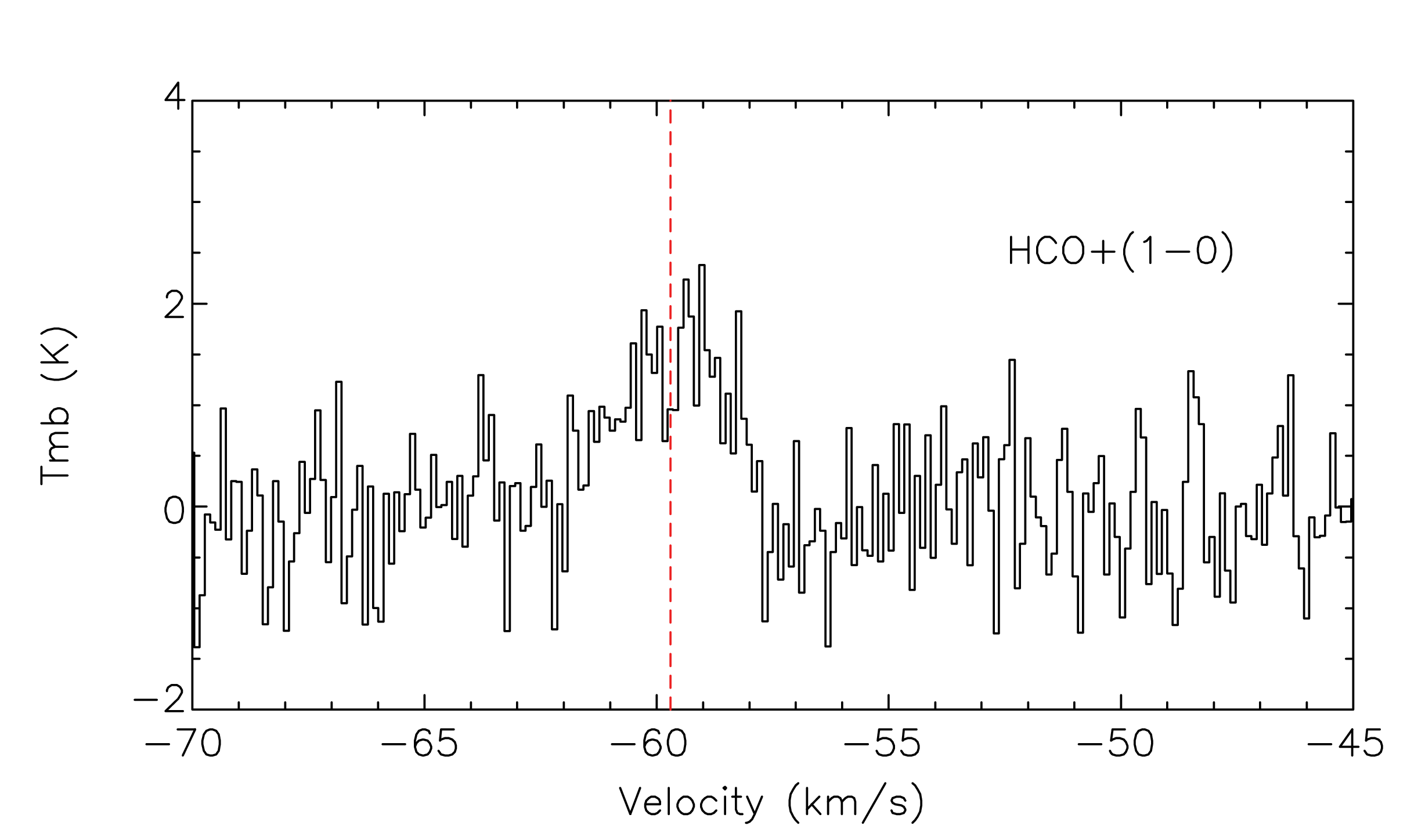,width=2.6in,height=1.8in}}
\centerline{\psfig{file=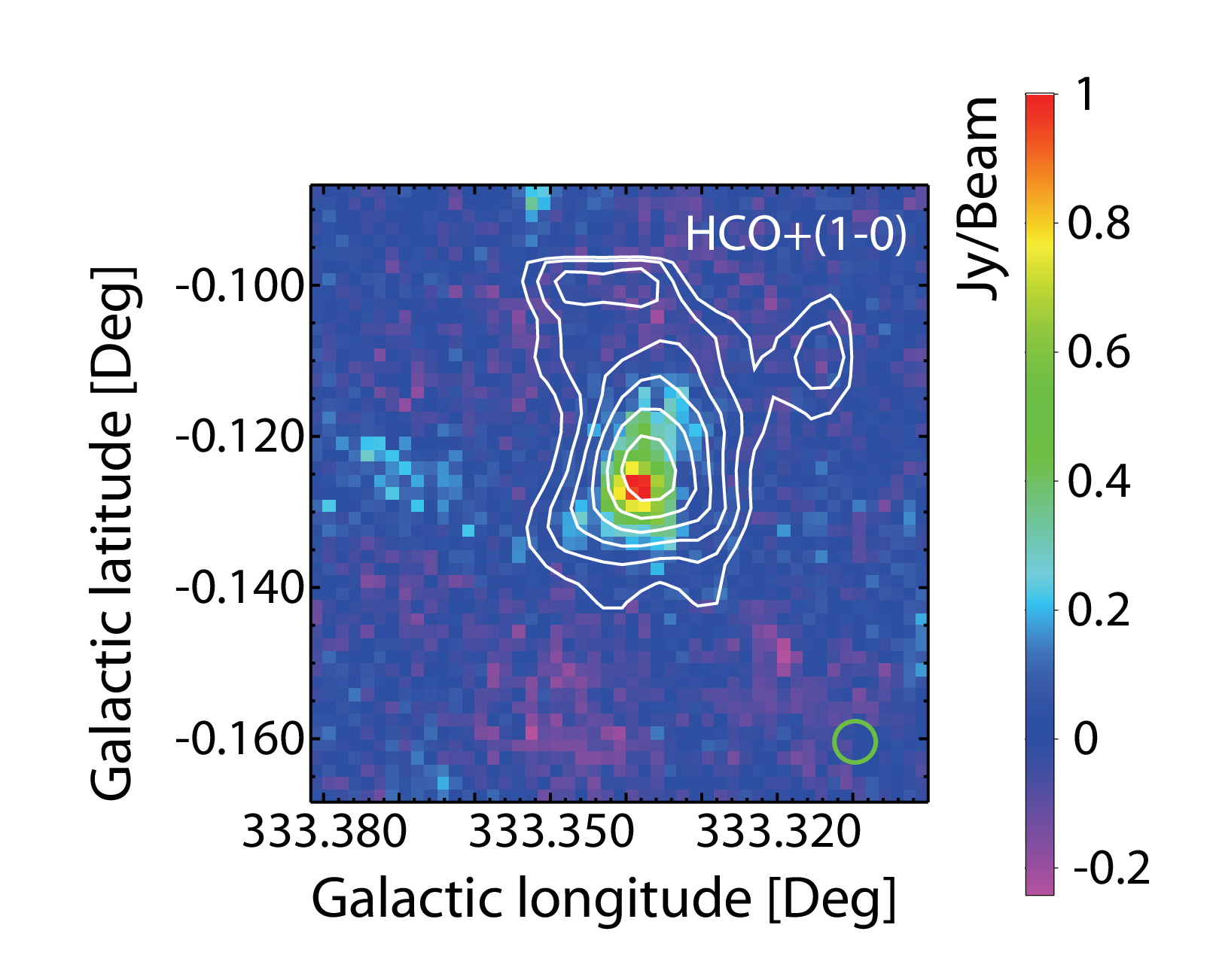,width=2.6in,height=2.3in}}
\centerline{\psfig{file=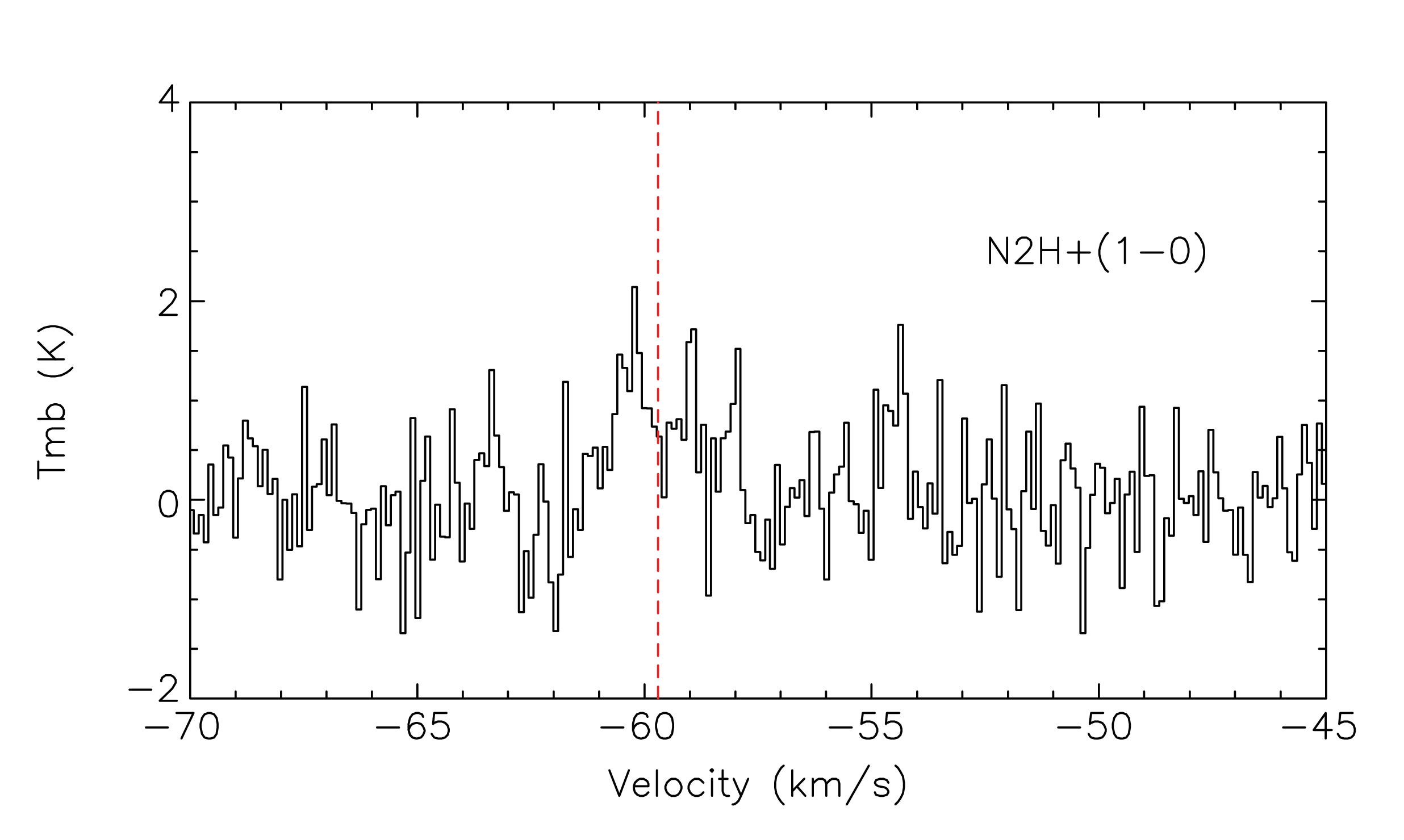,width=2.6in,height=1.8in}}
\centerline{\psfig{file=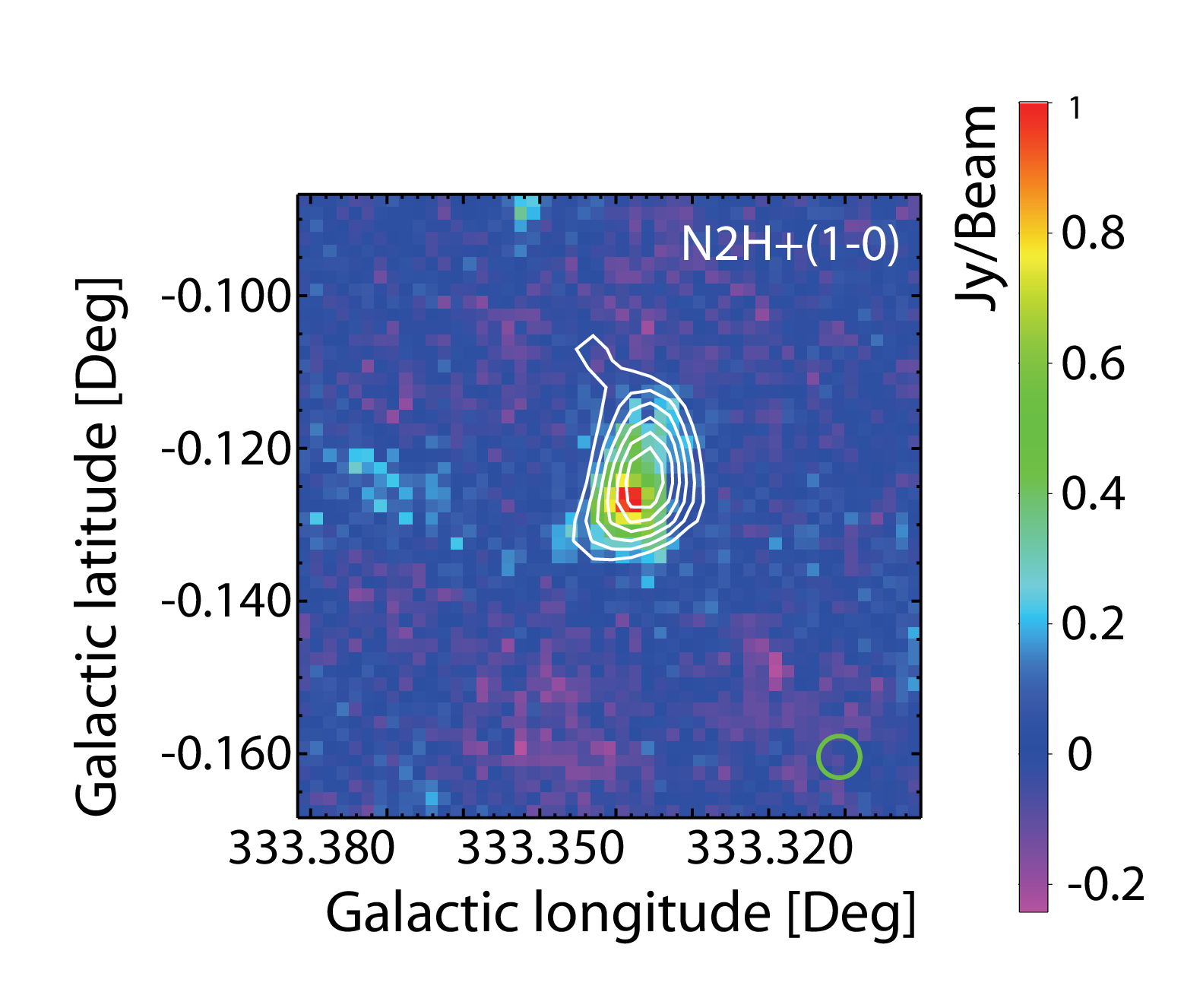,width=2.6in,height=2.3in}}
\caption{Spectra and integrated intensities superimposed on the 870
$\mu$m map in gray scale of G333.3401-00.1273. The red dash line
represents the V$_{LSR}$ of N$_2$H$^+$ line. Contour levels are
30$\%$, 40$\%$...90$\%$ of the center peak emissions. The angular
resolution of the ATLASGAL survey is indicated by the green circle
shown in the lower right corner. }
\end{figure}
\begin{figure}
\centerline{\psfig{file=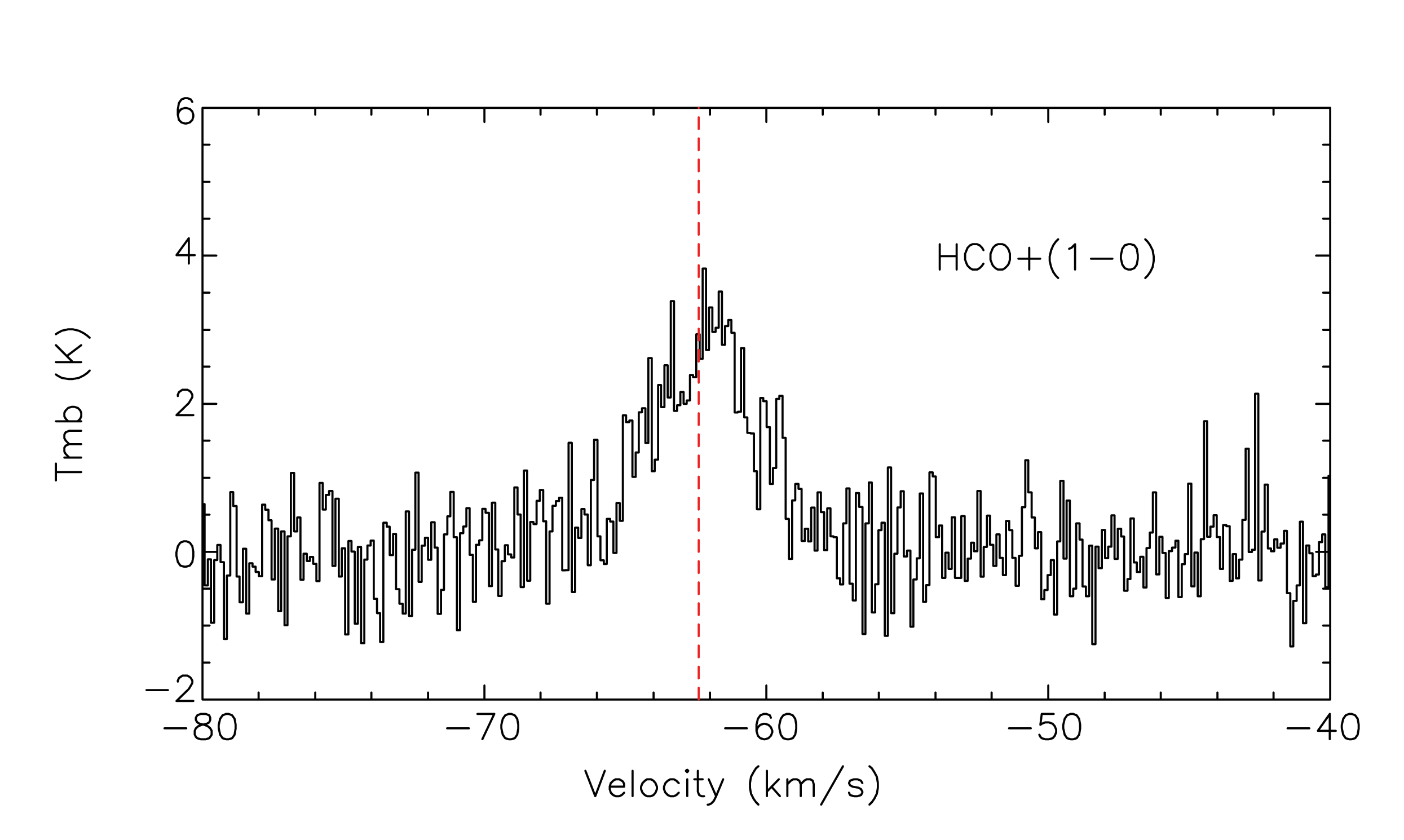,width=2.6in,height=1.8in}}
\centerline{\psfig{file=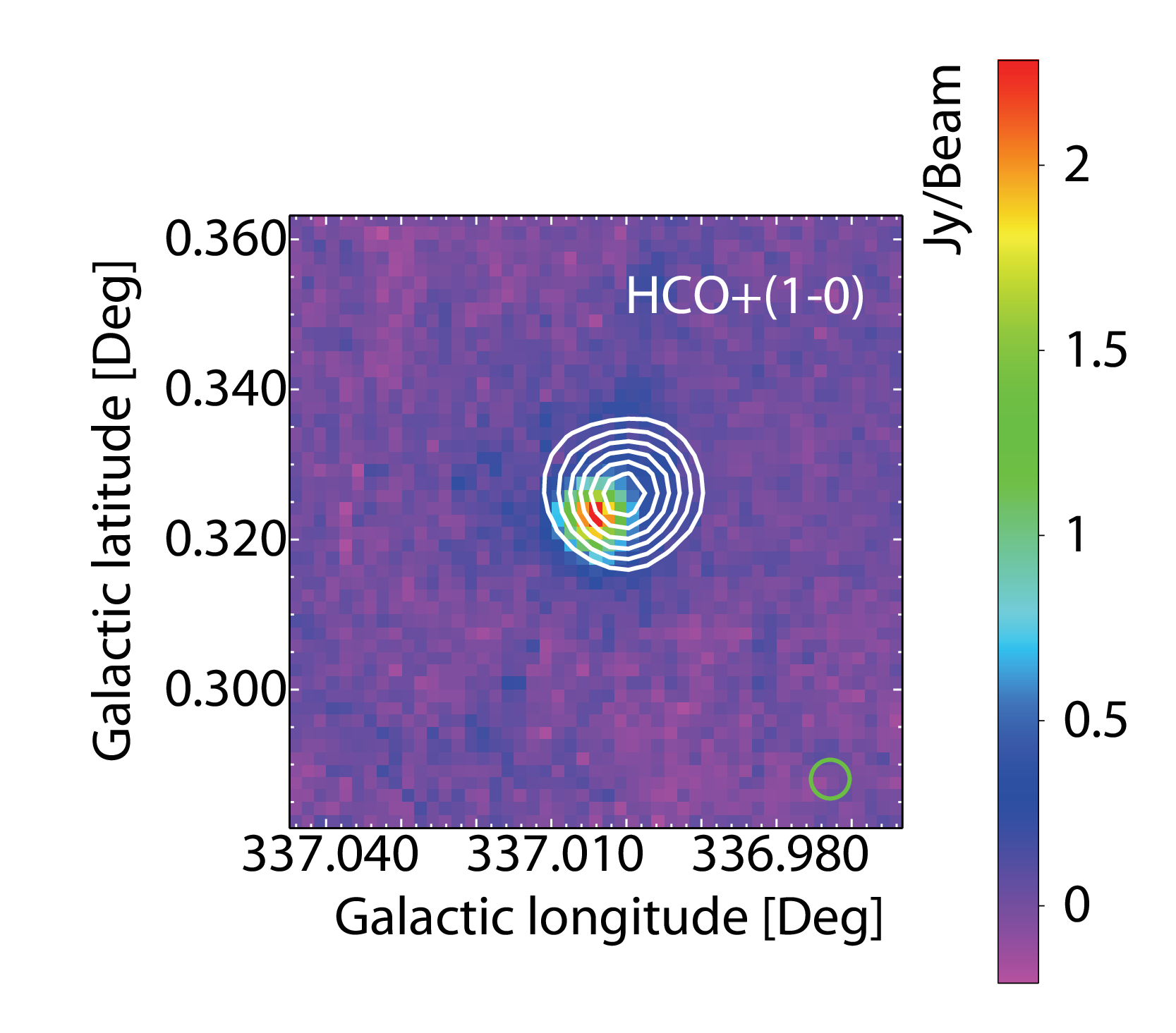,width=2.6in,height=2.3in}}
\centerline{\psfig{file=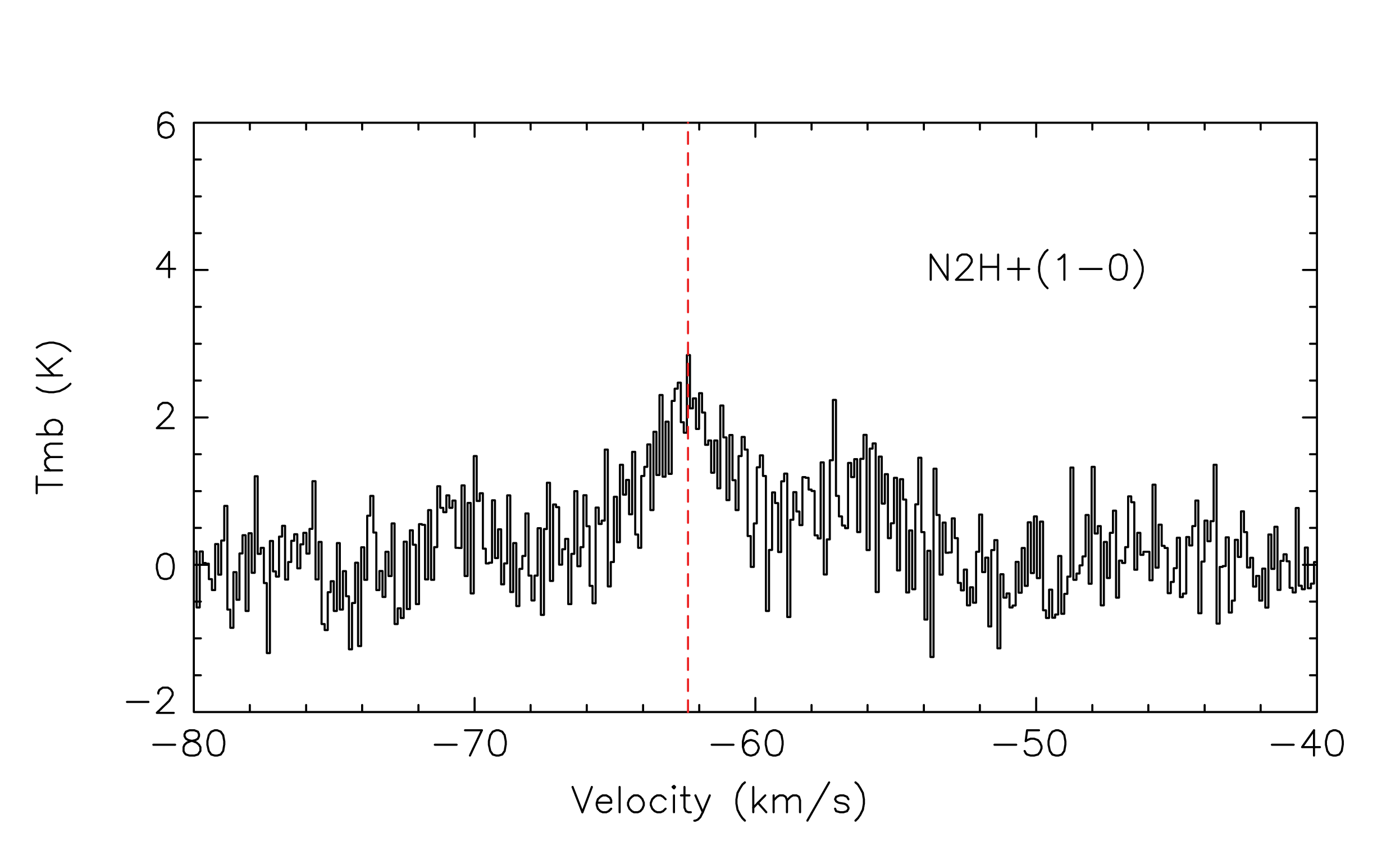,width=2.6in,height=1.8in}}
\centerline{\psfig{file=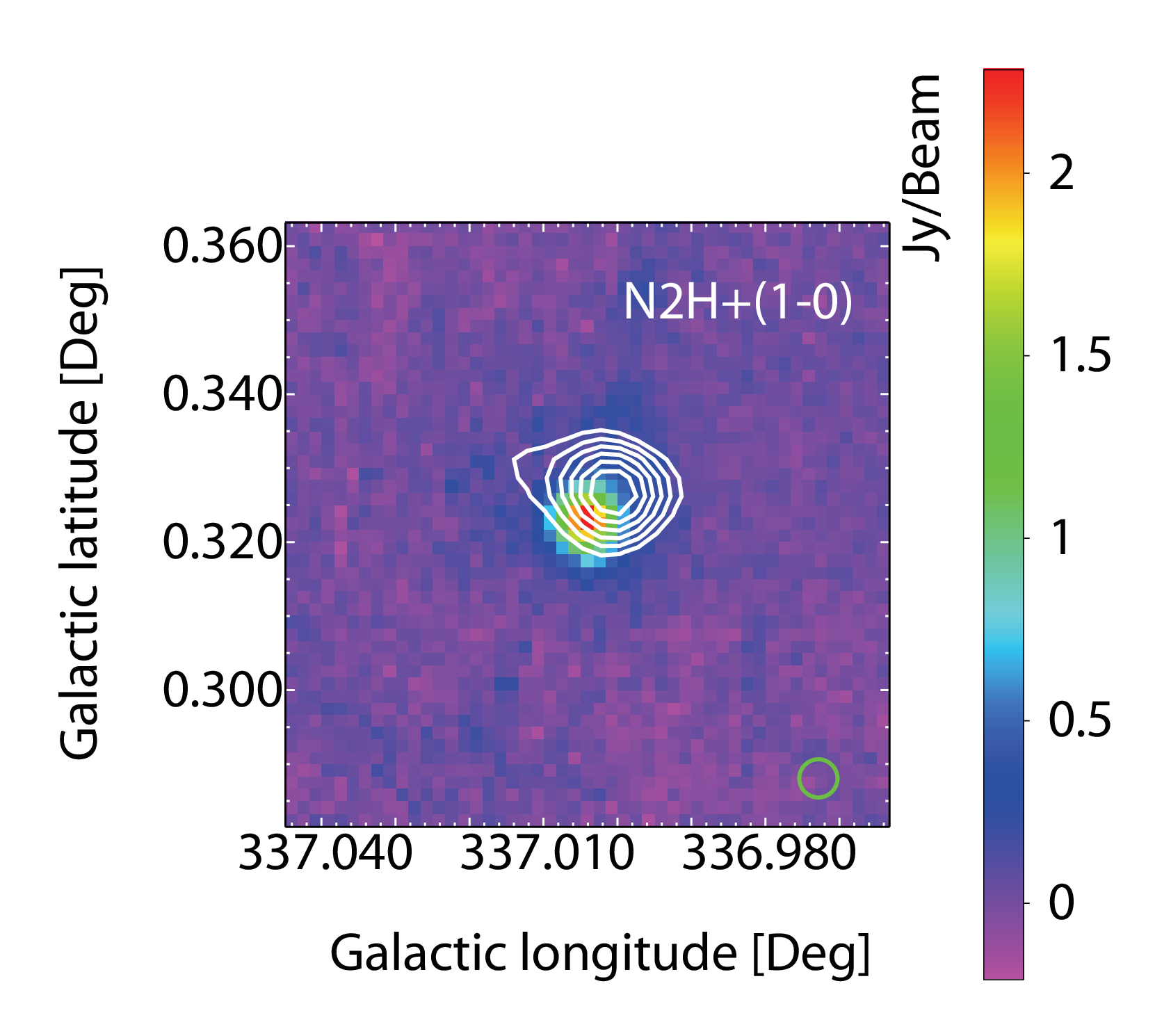,width=2.6in,height=2.3in}}
\caption{Spectra and integrated intensities superimposed on the 870
$\mu$m map in gray scale of G337.0047+00.3226. The red dash line
represents the V$_{LSR}$ of N$_2$H$^+$ line. Contour levels are
30$\%$, 40$\%$...90$\%$ of the center peak emissions. The angular
resolution of the ATLASGAL survey is indicated by the green circle
shown in the lower right corner. }
\end{figure}
\begin{figure}
\centerline{\psfig{file=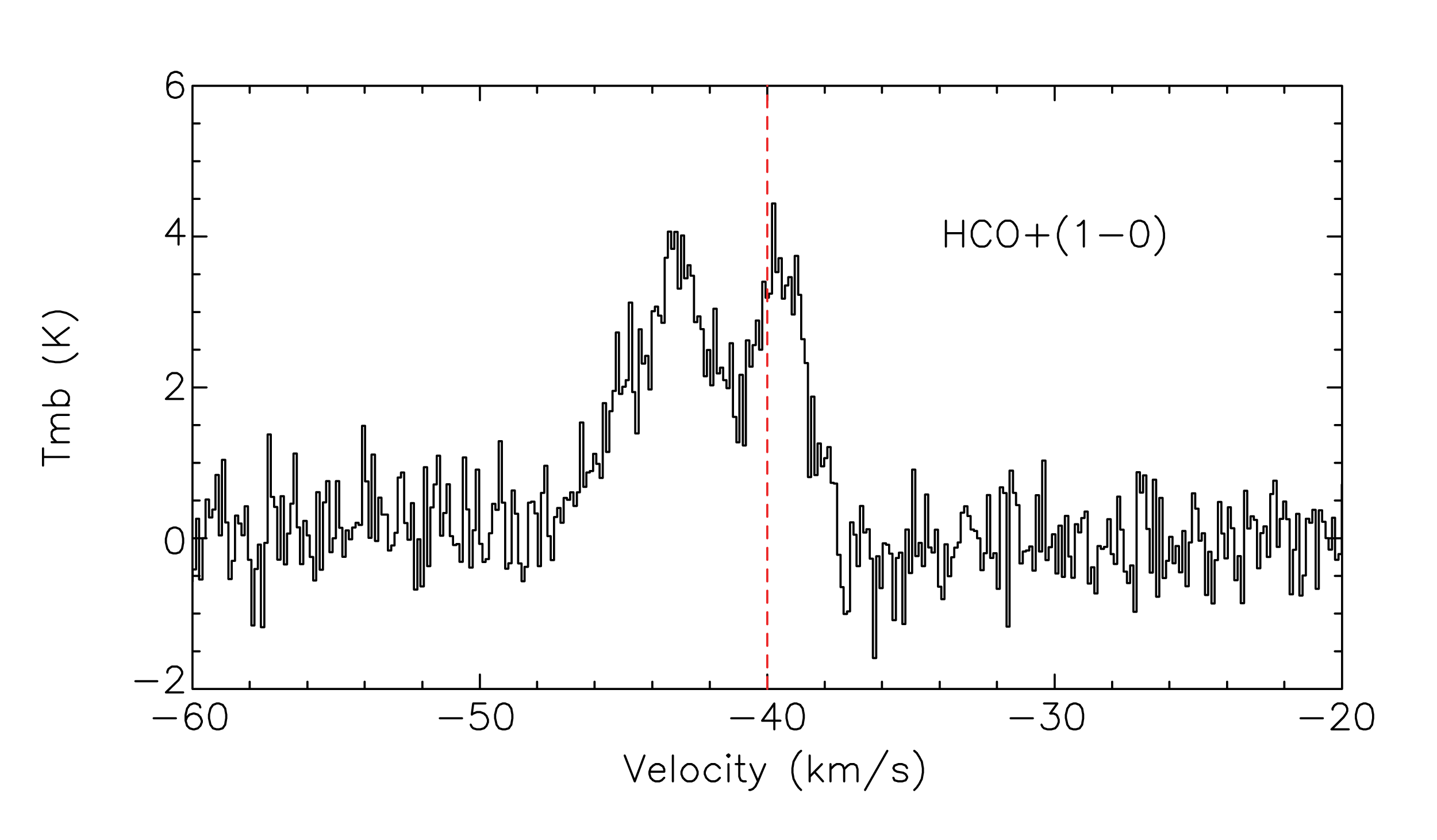,width=2.6in,height=1.8in}}
\centerline{\psfig{file=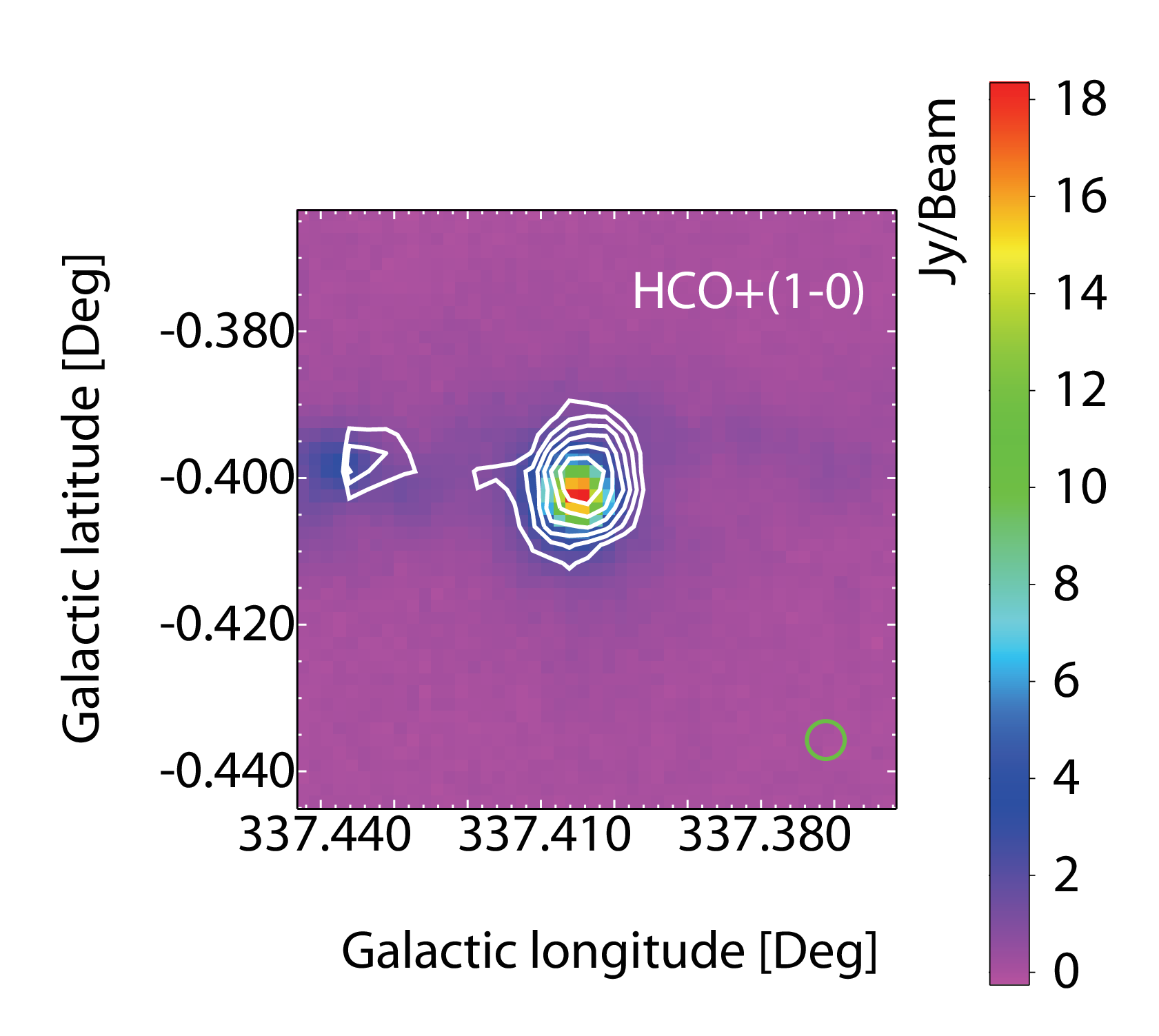,width=2.6in,height=2.3in}}
\centerline{\psfig{file=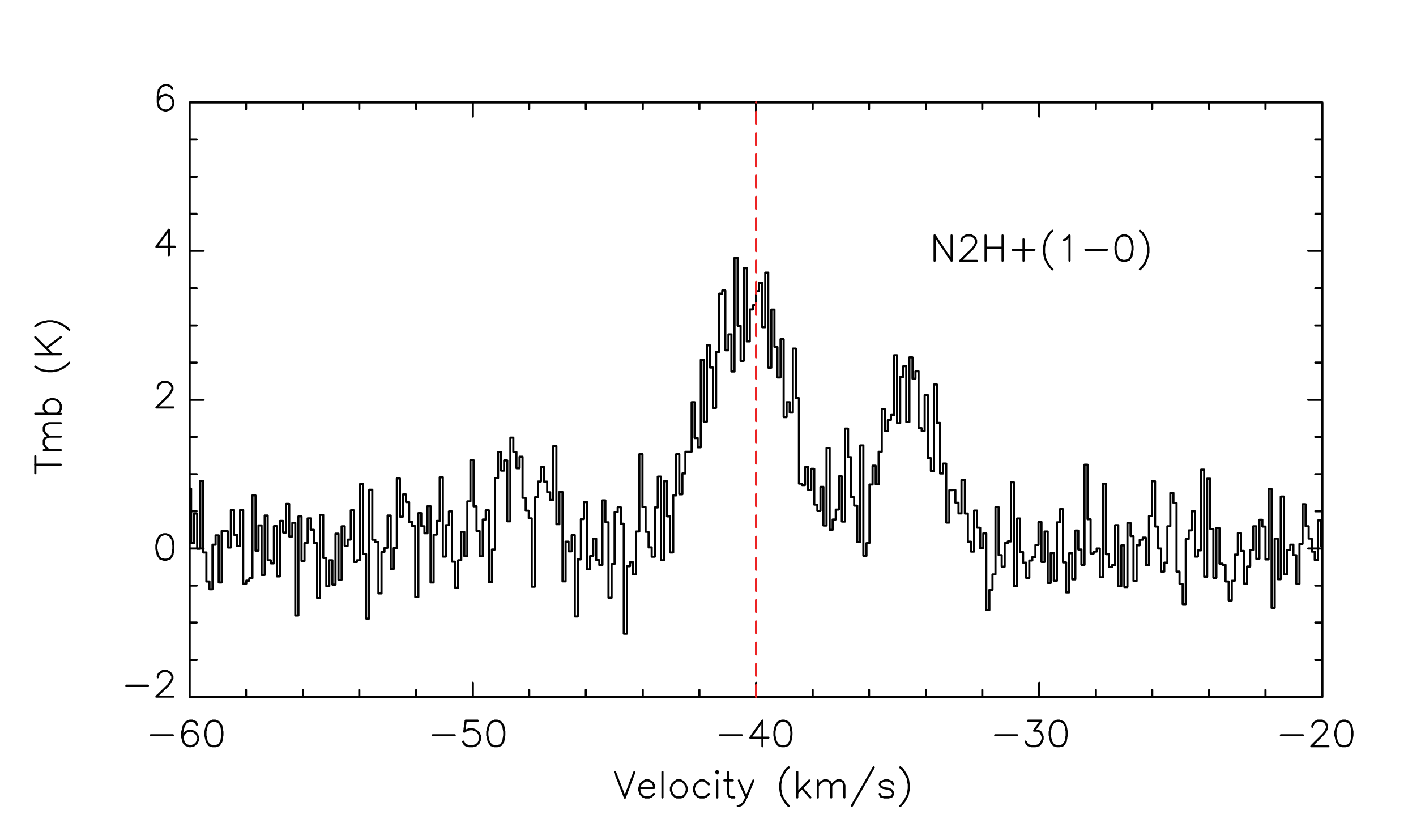,width=2.6in,height=1.8in}}
\centerline{\psfig{file=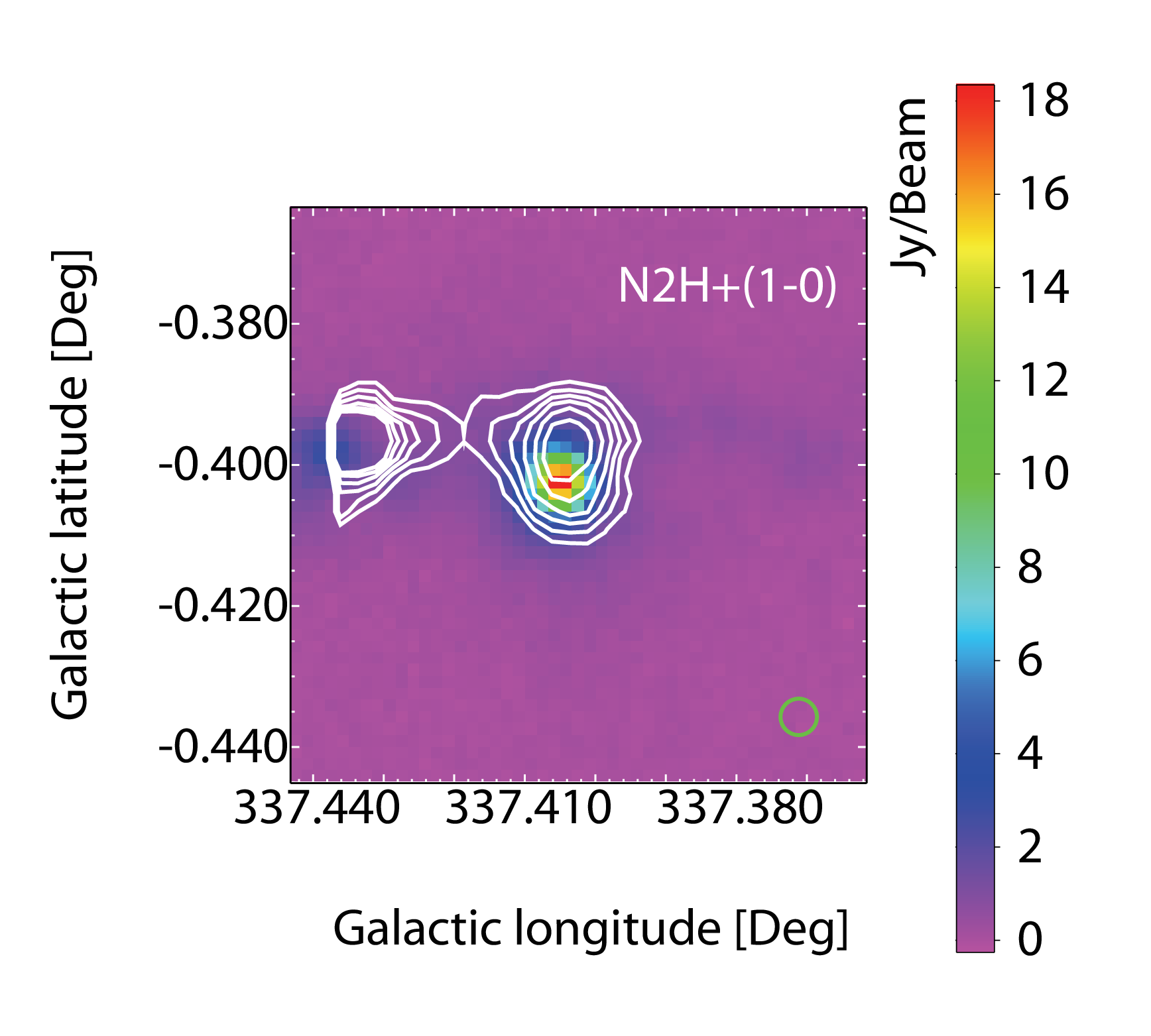,width=2.6in,height=2.3in}}
\caption{Spectra and integrated intensities superimposed on the 870
$\mu$m map in gray scale of G337.4032-00.4037. The red dash line
represents the V$_{LSR}$ of N$_2$H$^+$ line. Contour levels are
30$\%$, 40$\%$...90$\%$ of the center peak emissions. The angular
resolution of the ATLASGAL survey is indicated by the green circle
shown in the lower right corner. }
\end{figure}
\begin{figure}
\centerline{\psfig{file=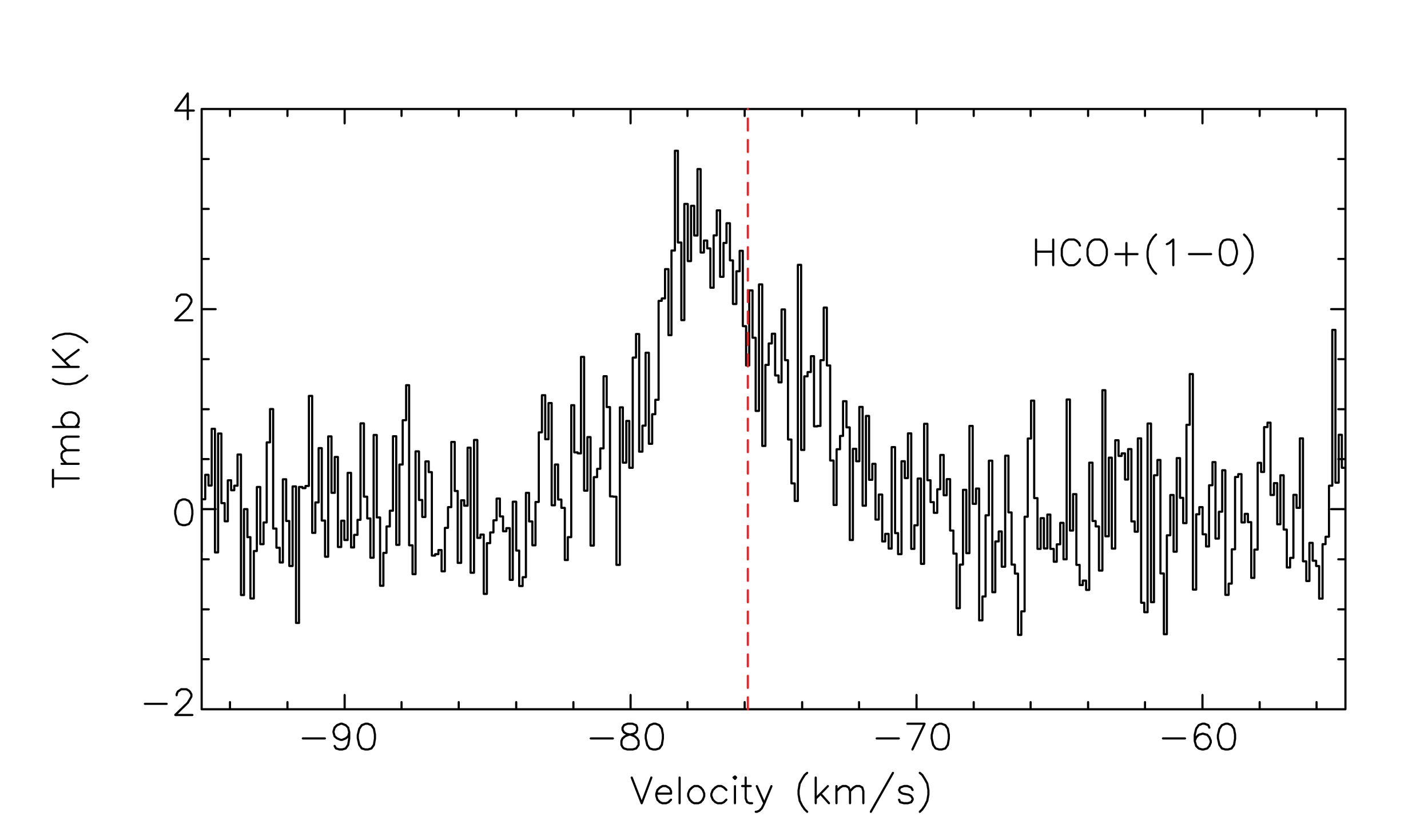,width=2.6in,height=1.8in}}
\centerline{\psfig{file=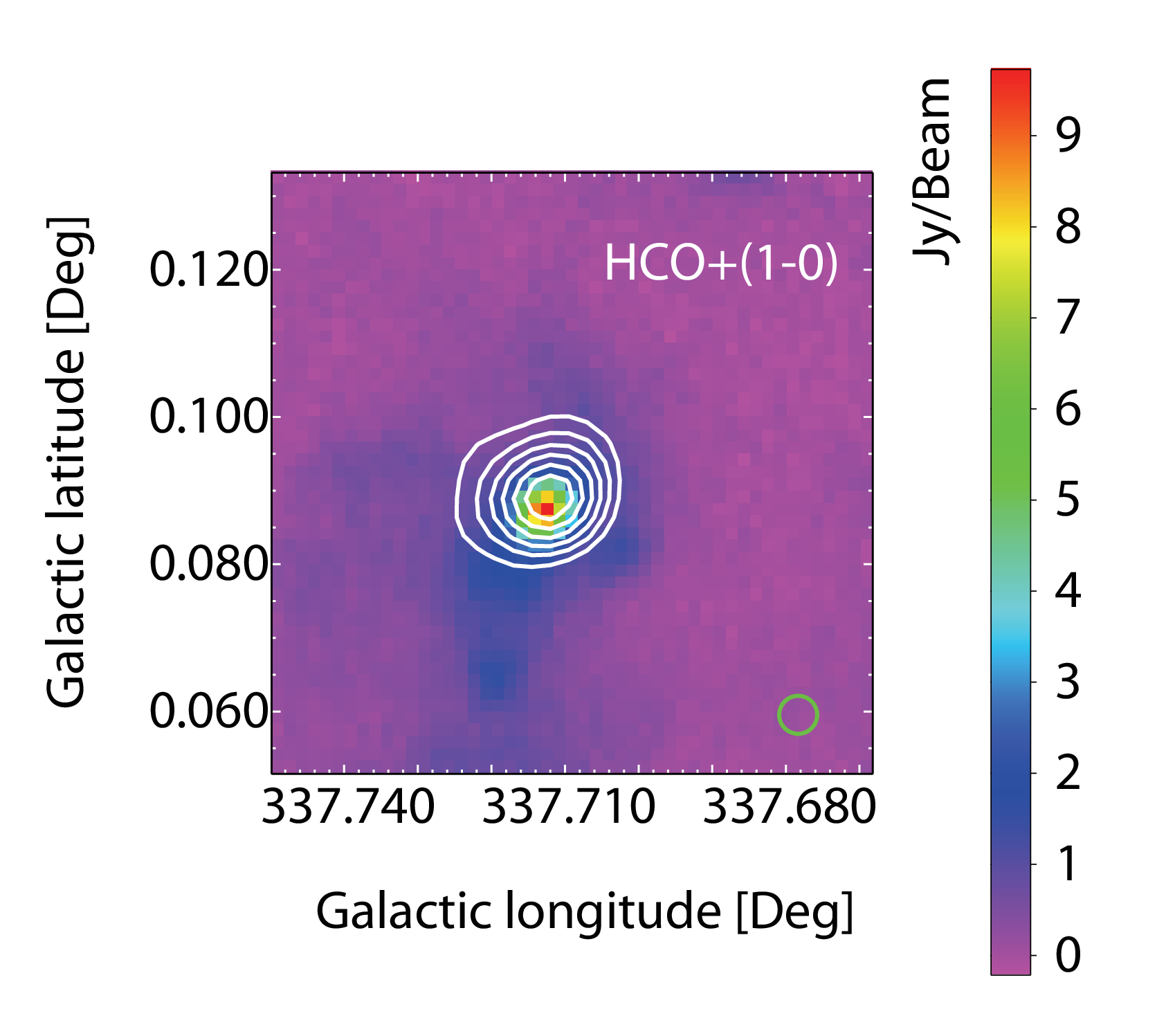,width=2.6in,height=2.3in}}
\centerline{\psfig{file=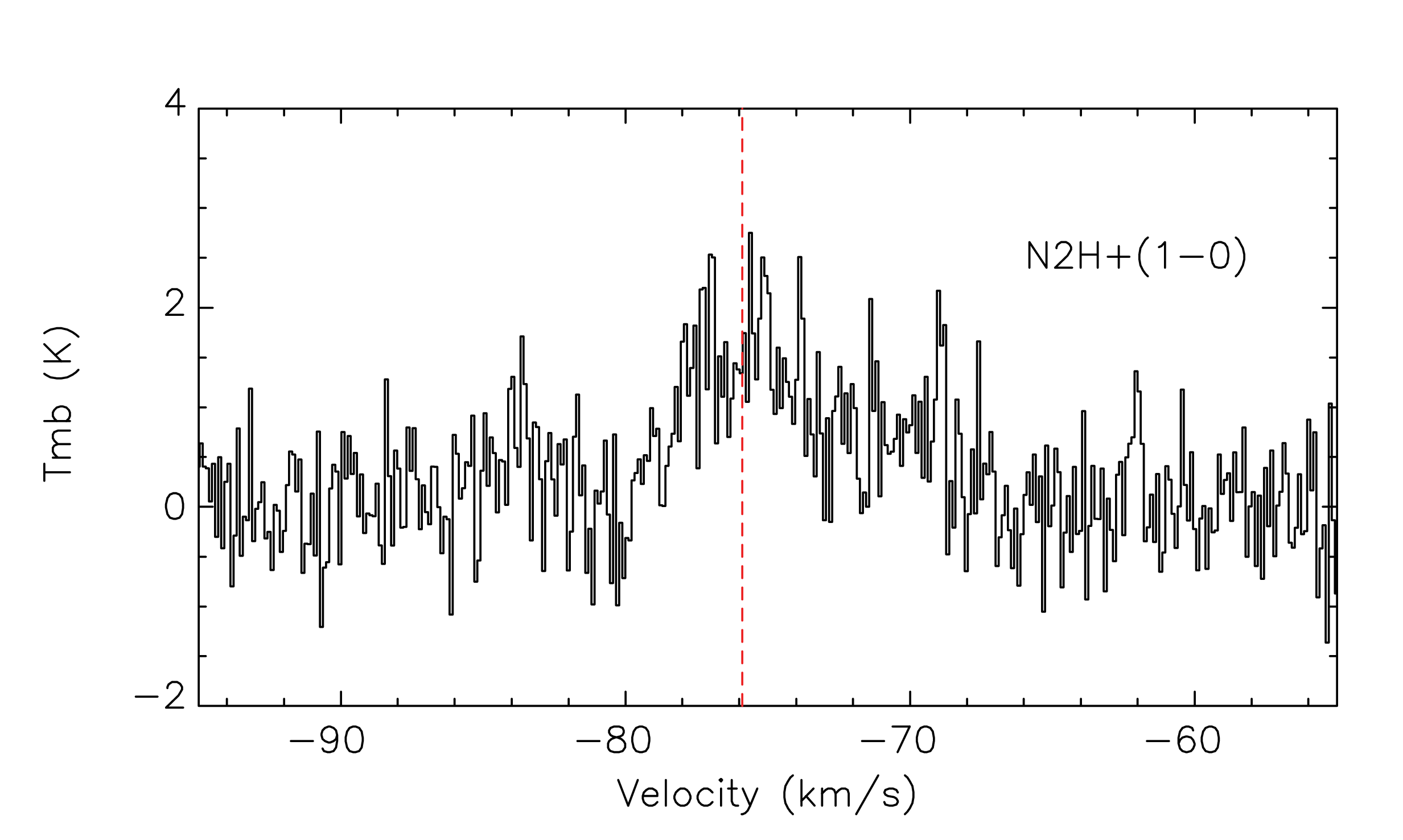,width=2.6in,height=1.8in}}
\centerline{\psfig{file=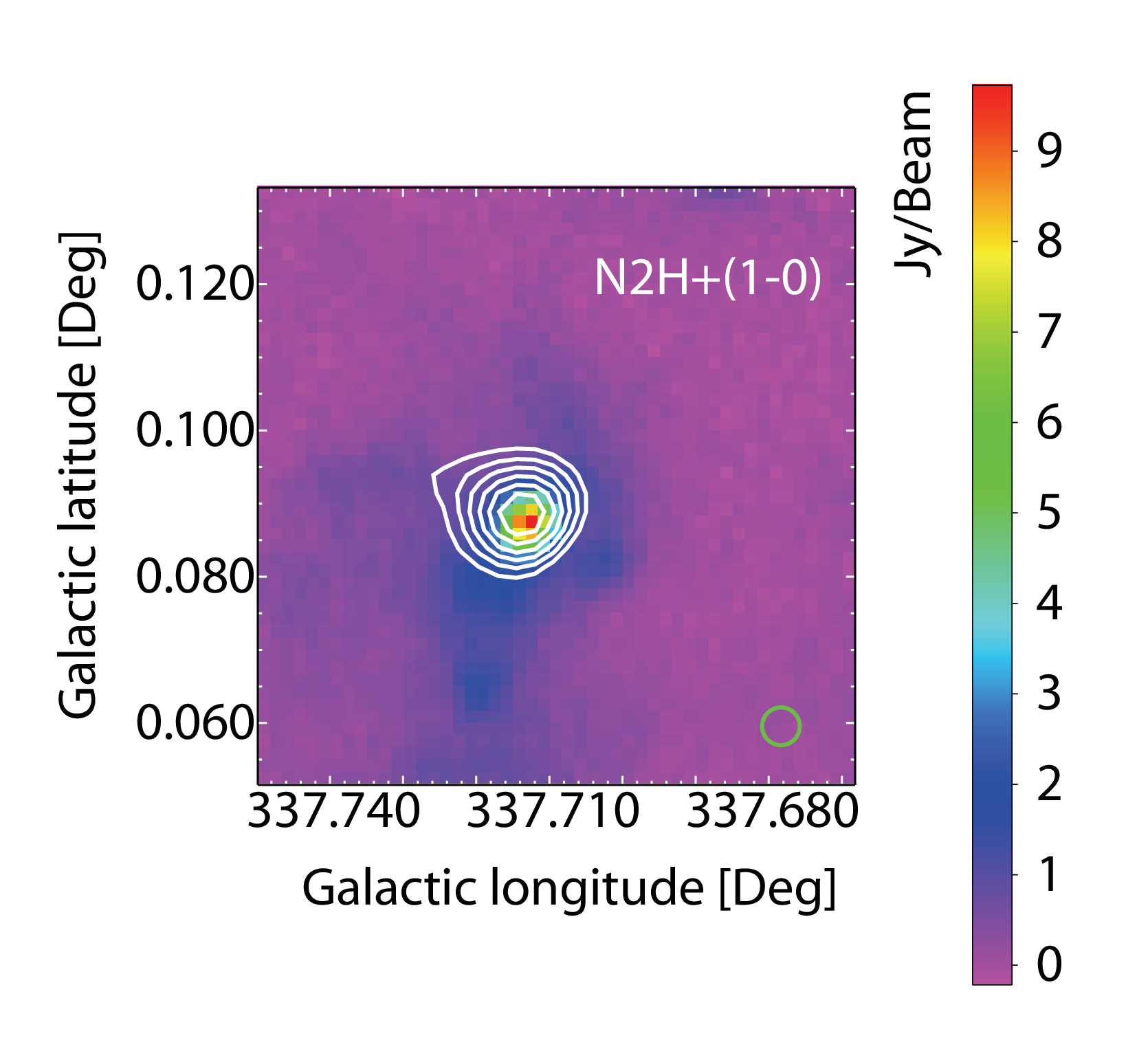,width=2.6in,height=2.3in}}
\caption{Spectra and integrated intensities superimposed on the 870
$\mu$m map in gray scale of G337.7091+00.0932A. The red dash line
represents the V$_{LSR}$ of N$_2$H$^+$ line. Contour levels are
30$\%$, 40$\%$...90$\%$ of the center peak emissions. The angular
resolution of the ATLASGAL survey is indicated by the green circle
shown in the lower right corner. }
\end{figure}
\begin{figure}
\centerline{\psfig{file=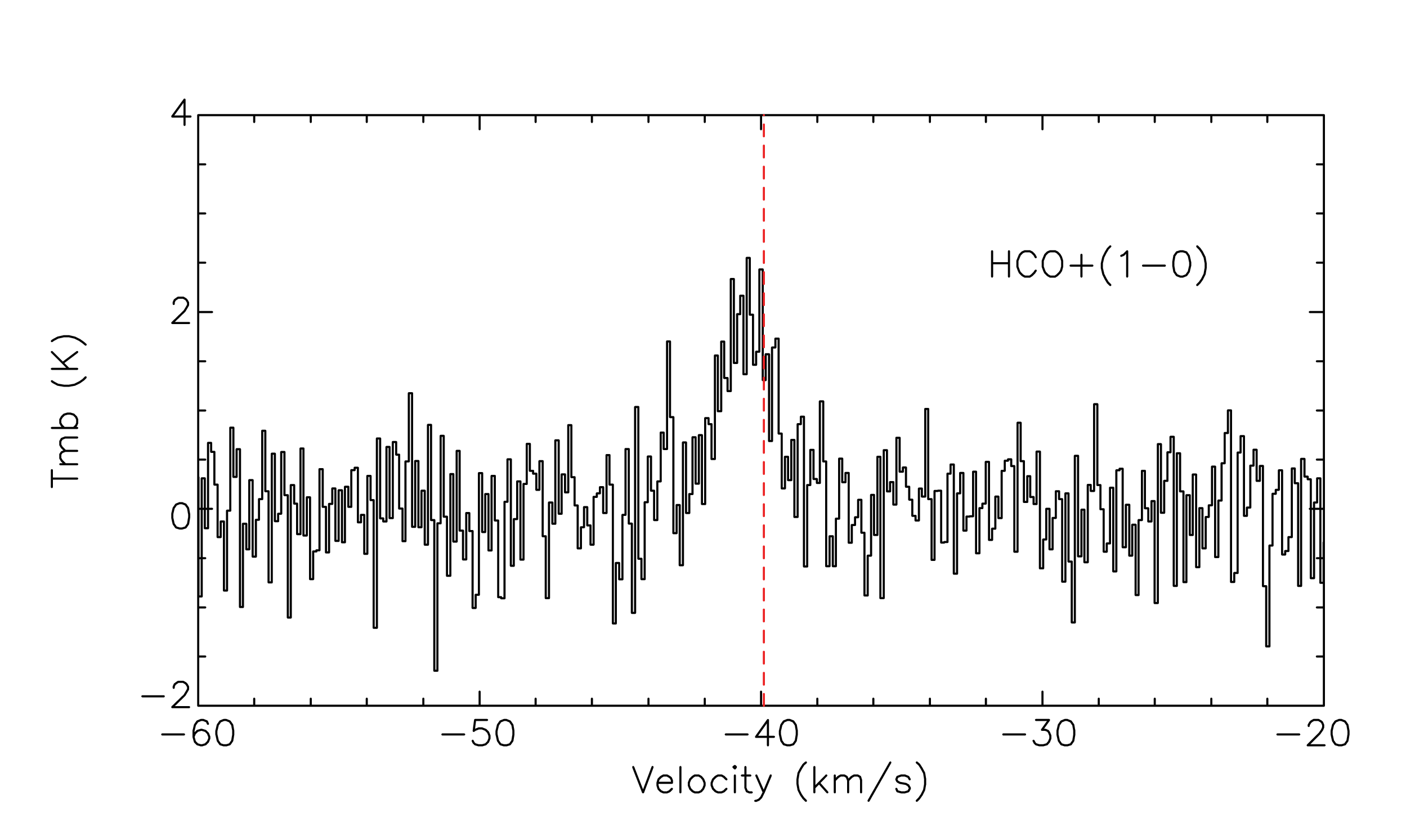,width=2.6in,height=1.8in}}
\centerline{\psfig{file=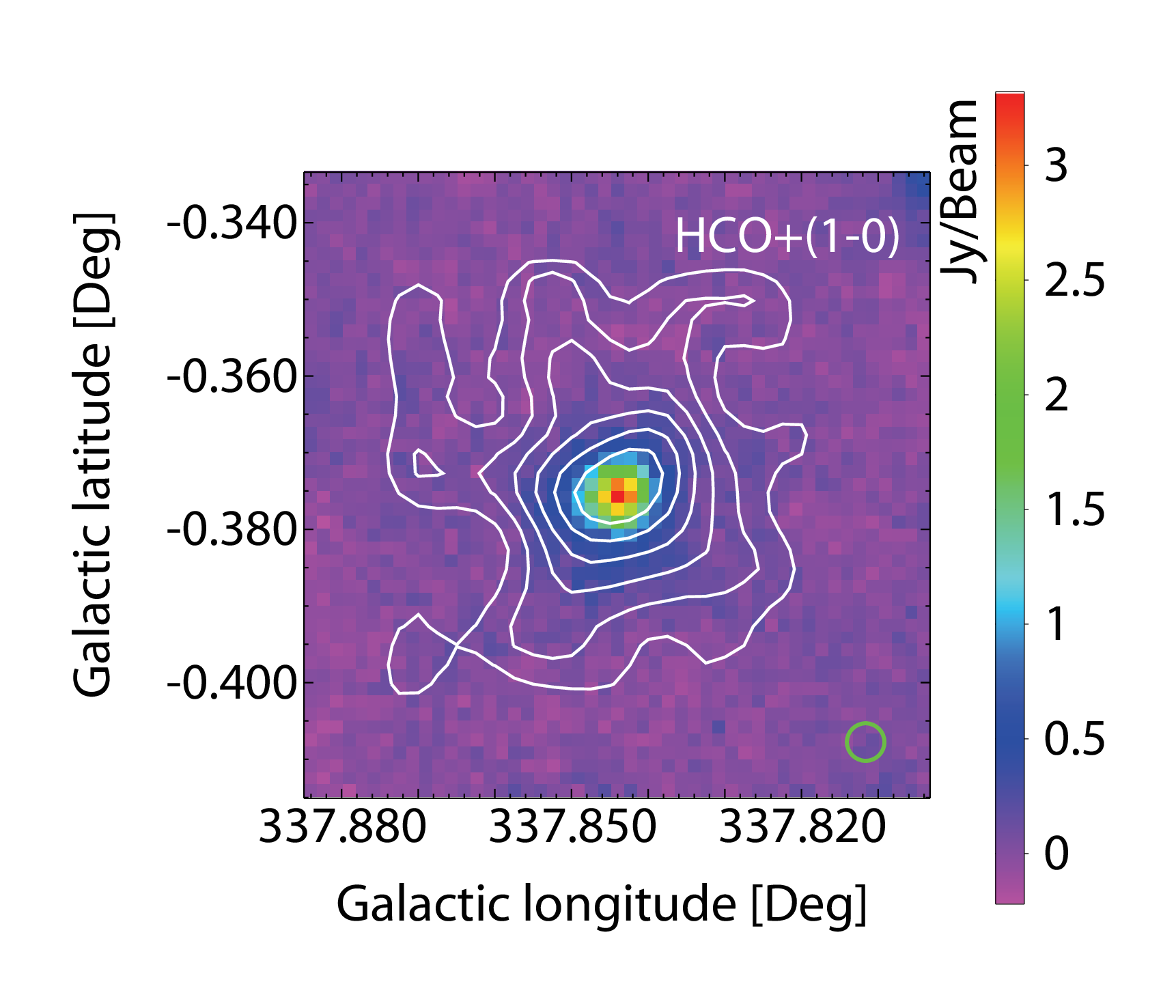,width=2.6in,height=2.3in}}
\centerline{\psfig{file=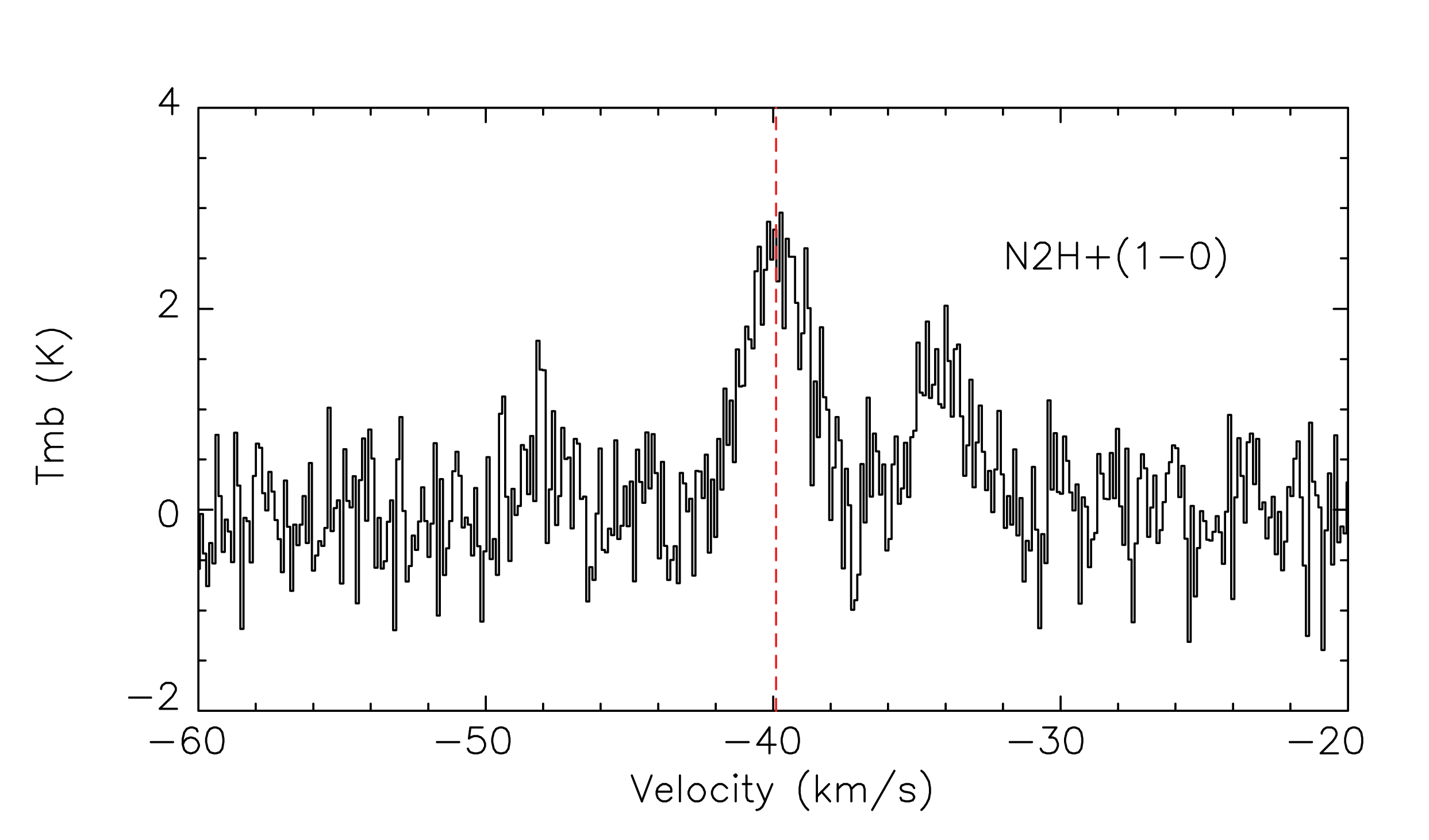,width=2.6in,height=1.8in}}
\centerline{\psfig{file=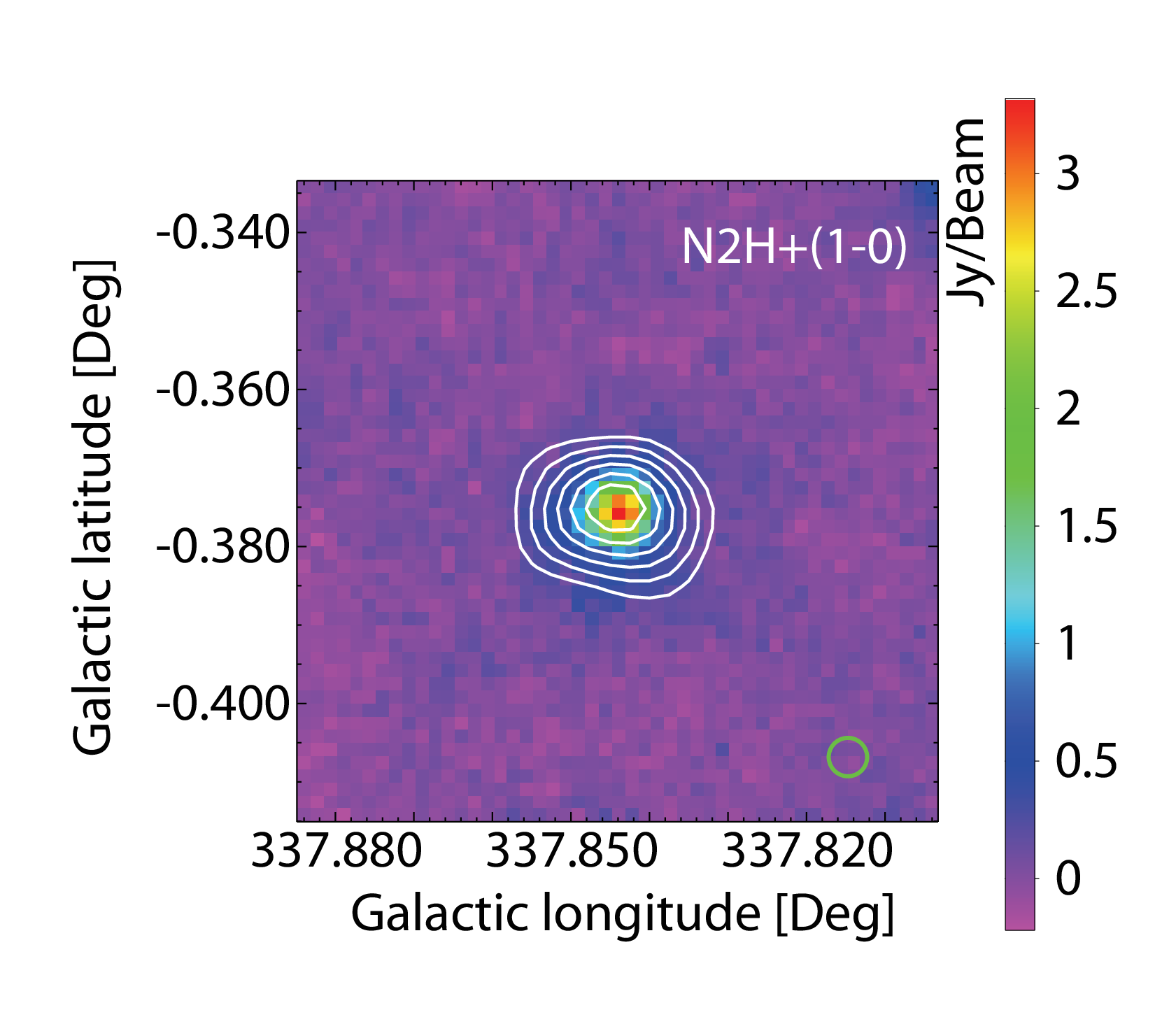,width=2.6in,height=2.3in}}
\caption{Spectra and integrated intensities superimposed on the 870
$\mu$m map in gray scale of G337.8442-00.3748. The red dash line
represents the V$_{LSR}$ of N$_2$H$^+$ line. Contour levels are
30$\%$, 40$\%$...90$\%$ of the center peak emissions. The angular
resolution of the ATLASGAL survey is indicated by the green circle
shown in the lower right corner. }
\end{figure}
\begin{figure}
\centerline{\psfig{file=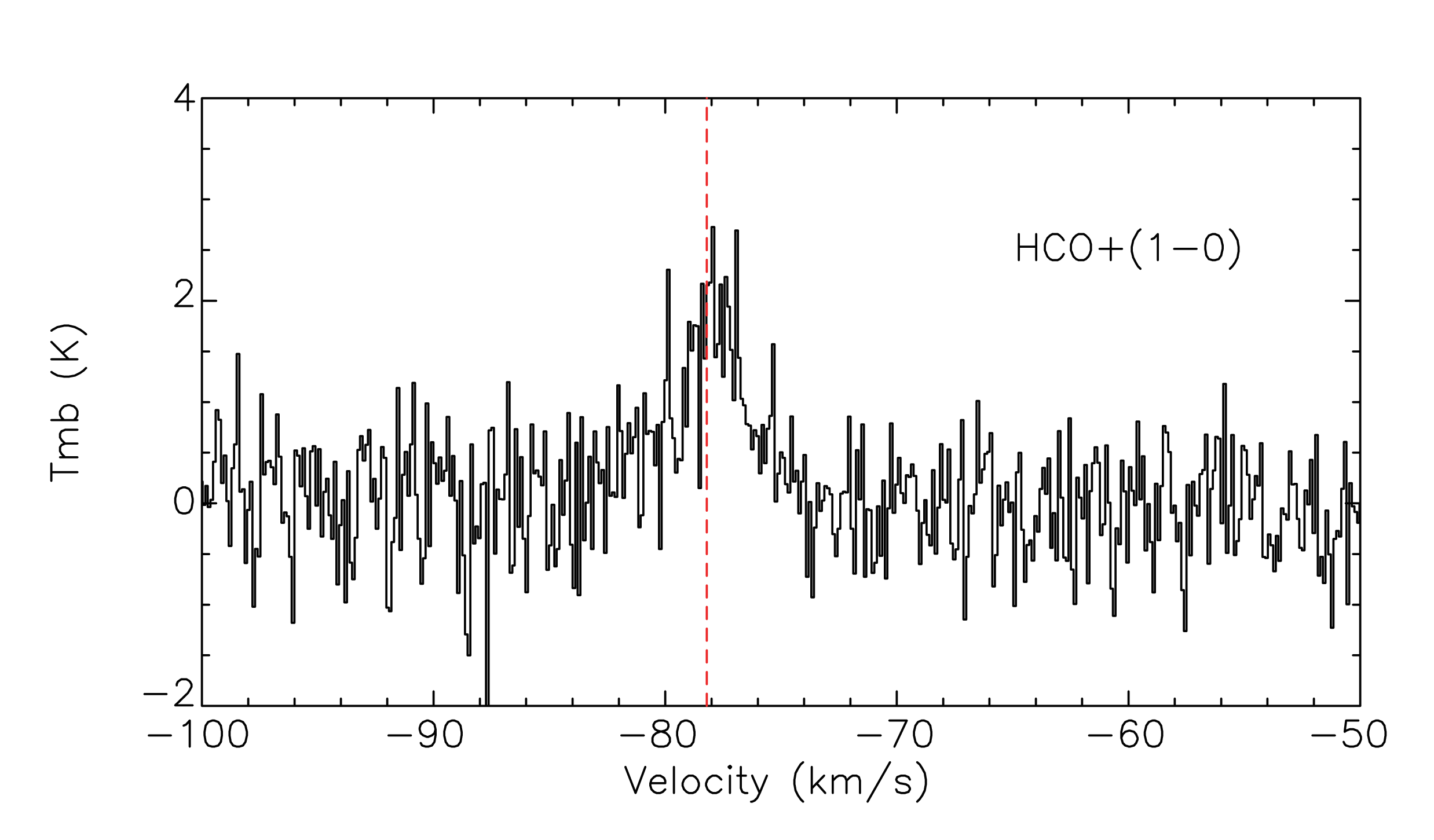,width=2.6in,height=1.8in}}
\centerline{\psfig{file=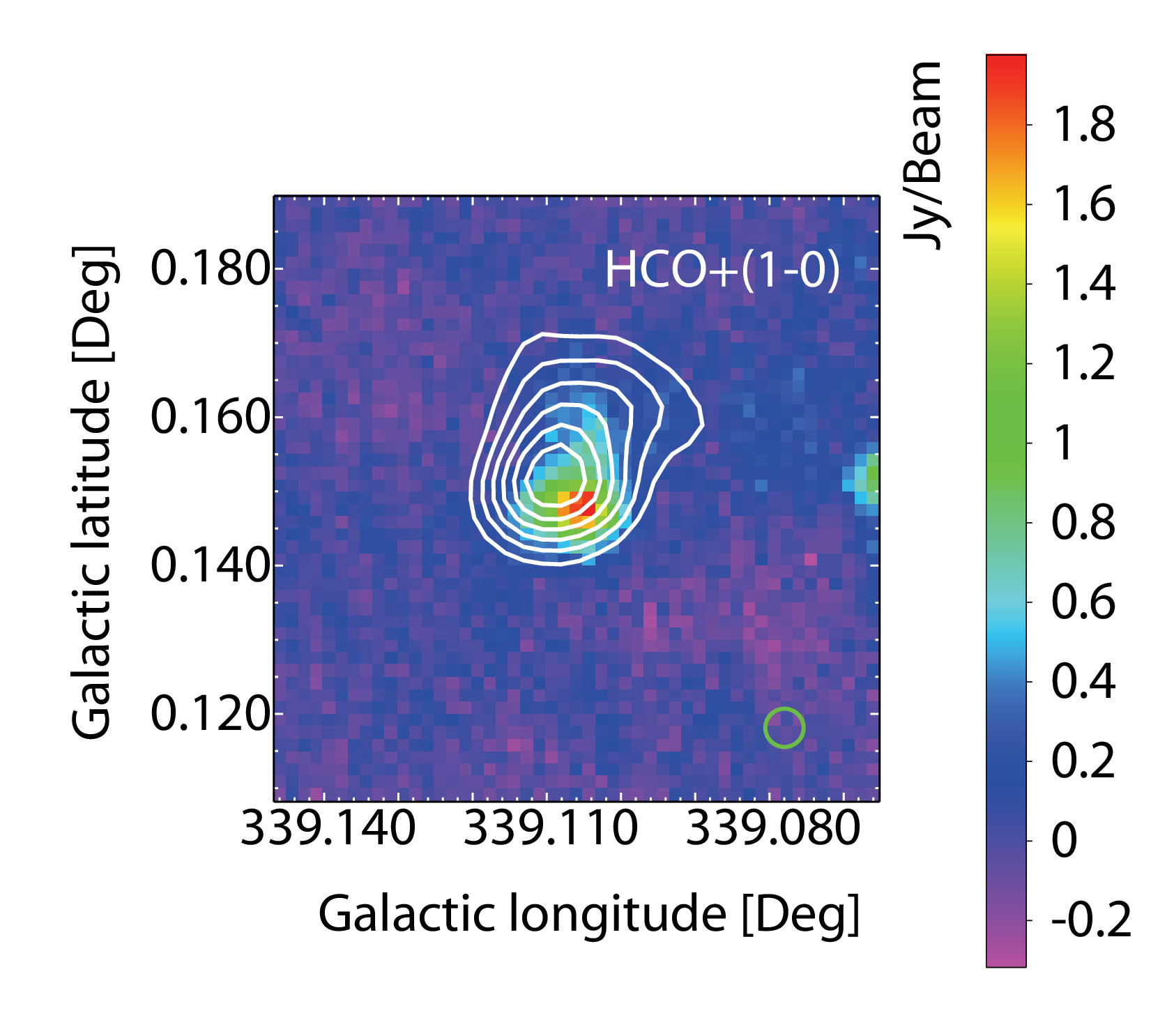,width=2.6in,height=2.3in}}
\centerline{\psfig{file=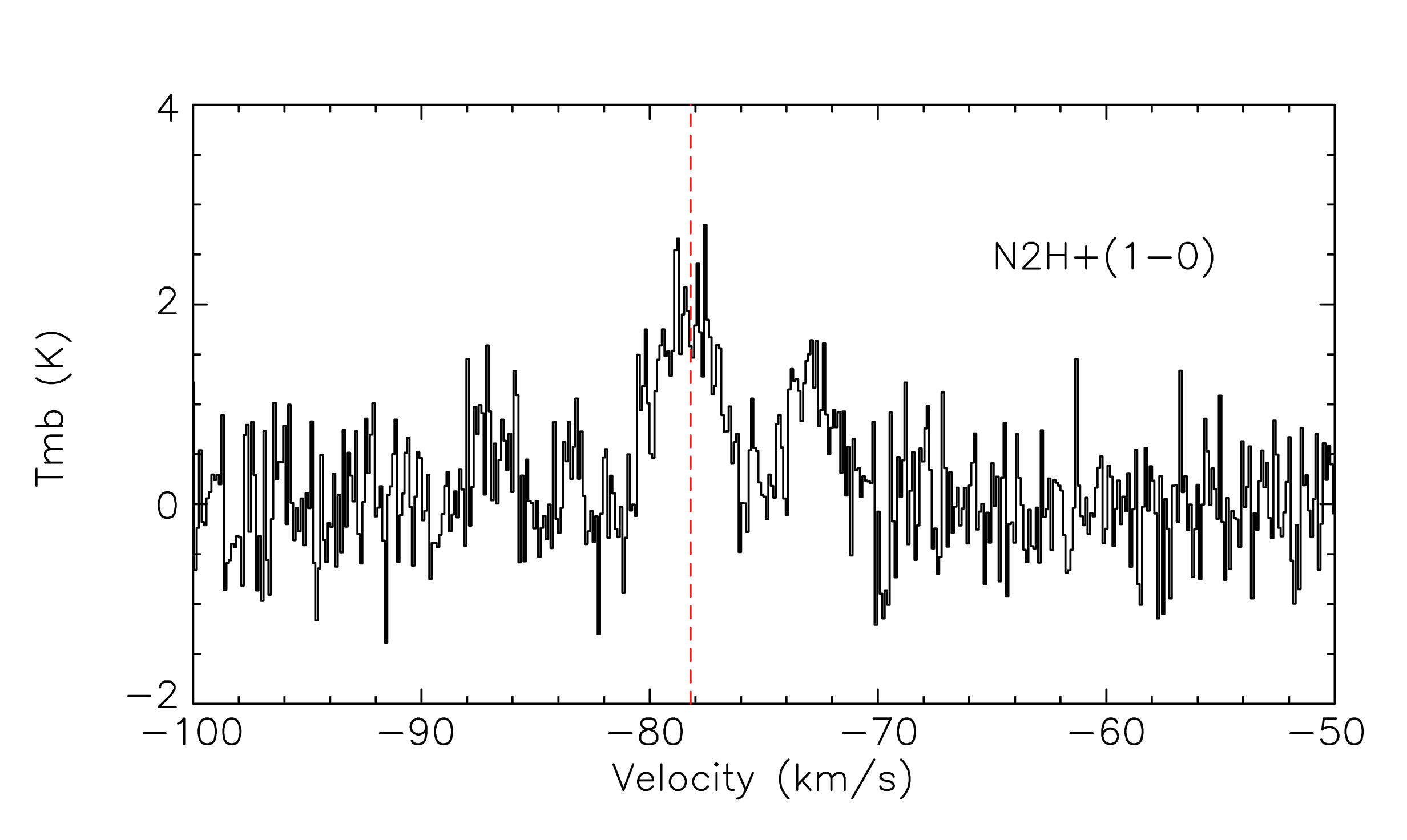,width=2.6in,height=1.8in}}
\centerline{\psfig{file=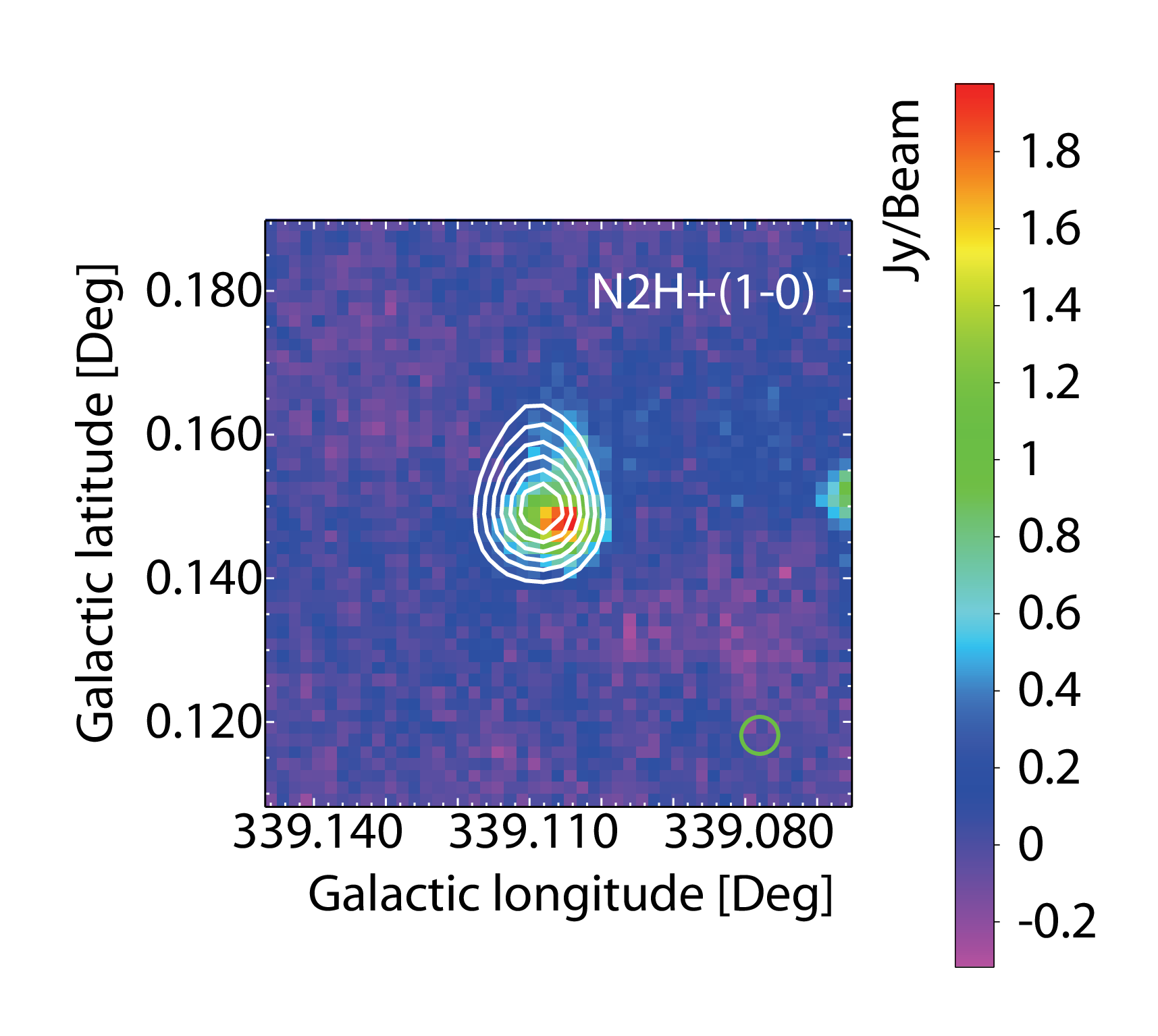,width=2.6in,height=2.3in}}
\caption{Spectra and integrated intensities superimposed on the 870
$\mu$m map in gray scale of G339.1052+00.1490. The red dash line
represents the V$_{LSR}$ of N$_2$H$^+$ line. Contour levels are
30$\%$, 40$\%$...90$\%$ of the center peak emissions. The angular
resolution of the ATLASGAL survey is indicated by the green circle
shown in the lower right corner.  }
\end{figure}
\begin{figure}
\centerline{\psfig{file=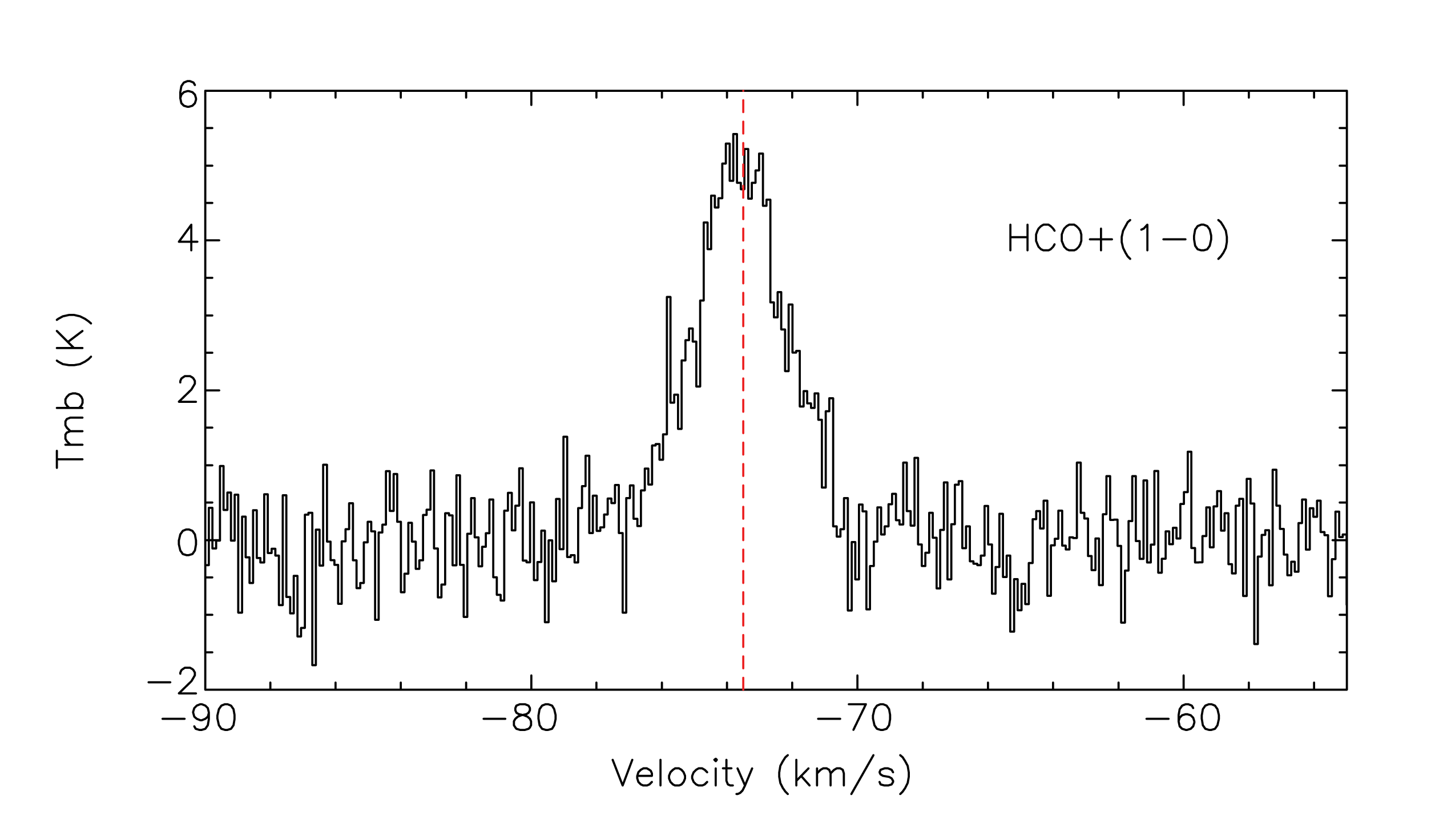,width=2.6in,height=1.8in}}
\centerline{\psfig{file=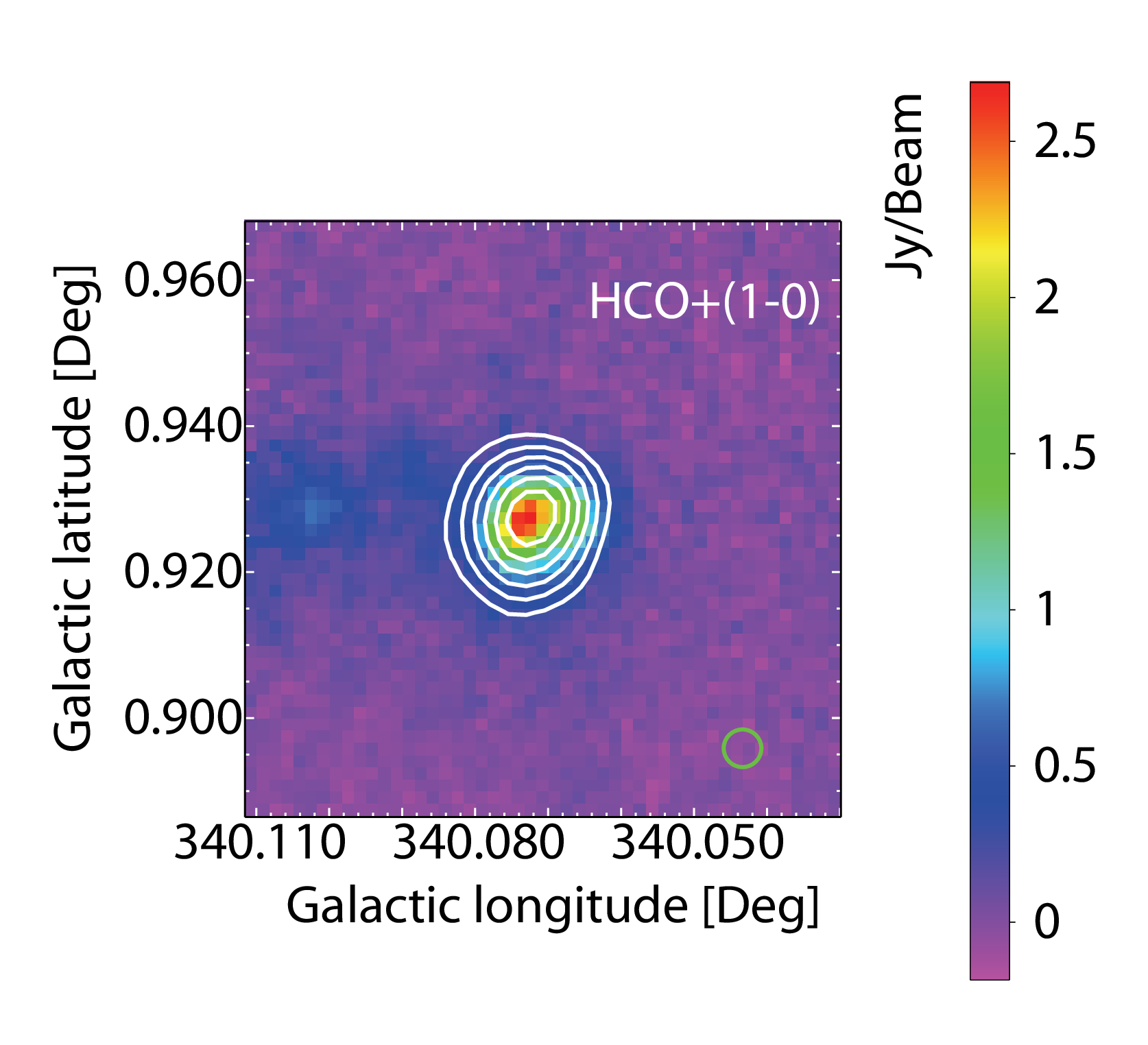,width=2.6in,height=2.3in}}
\centerline{\psfig{file=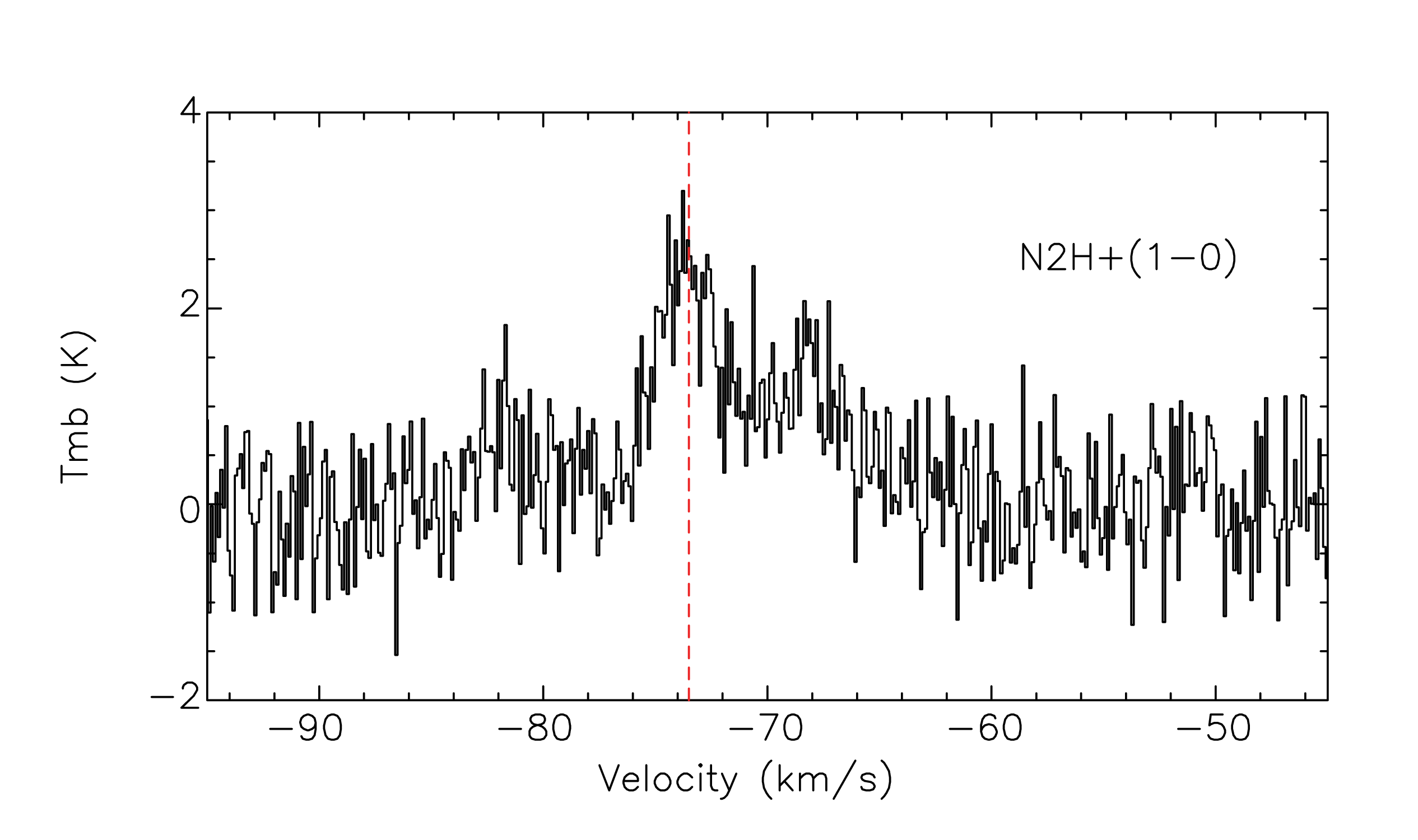,width=2.6in,height=1.8in}}
\centerline{\psfig{file=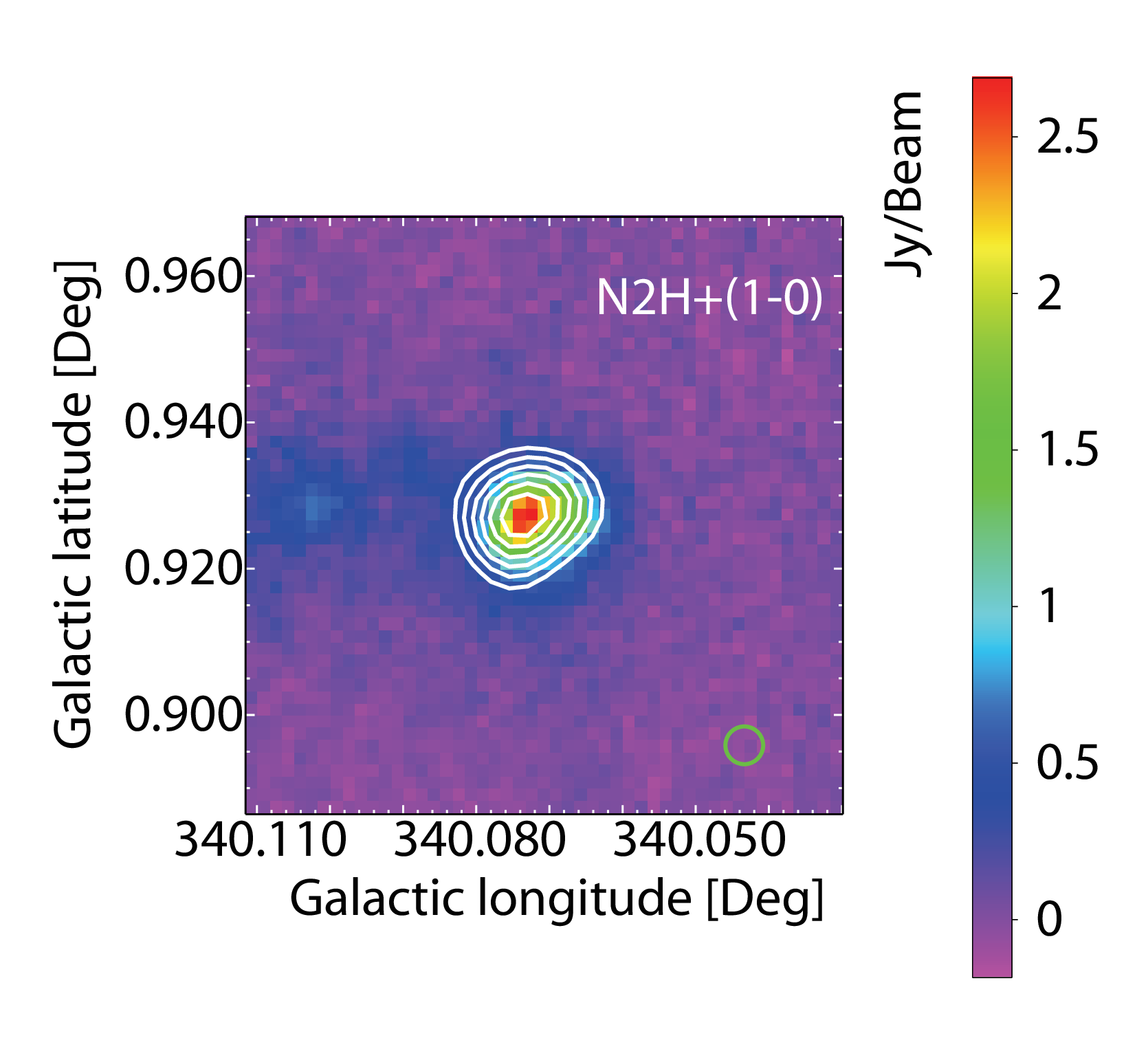,width=2.6in,height=2.3in}}
\caption{Spectra and integrated intensities superimposed on the 870
$\mu$m map in gray scale of G340.0708+00.9267. The red dash line
represents the V$_{LSR}$ of N$_2$H$^+$ line. Contour levels are
30$\%$, 40$\%$...90$\%$ of the center peak emissions. The angular
resolution of the ATLASGAL survey is indicated by the green circle
shown in the lower right corner. }
\end{figure}
\begin{figure}
\centerline{\psfig{file=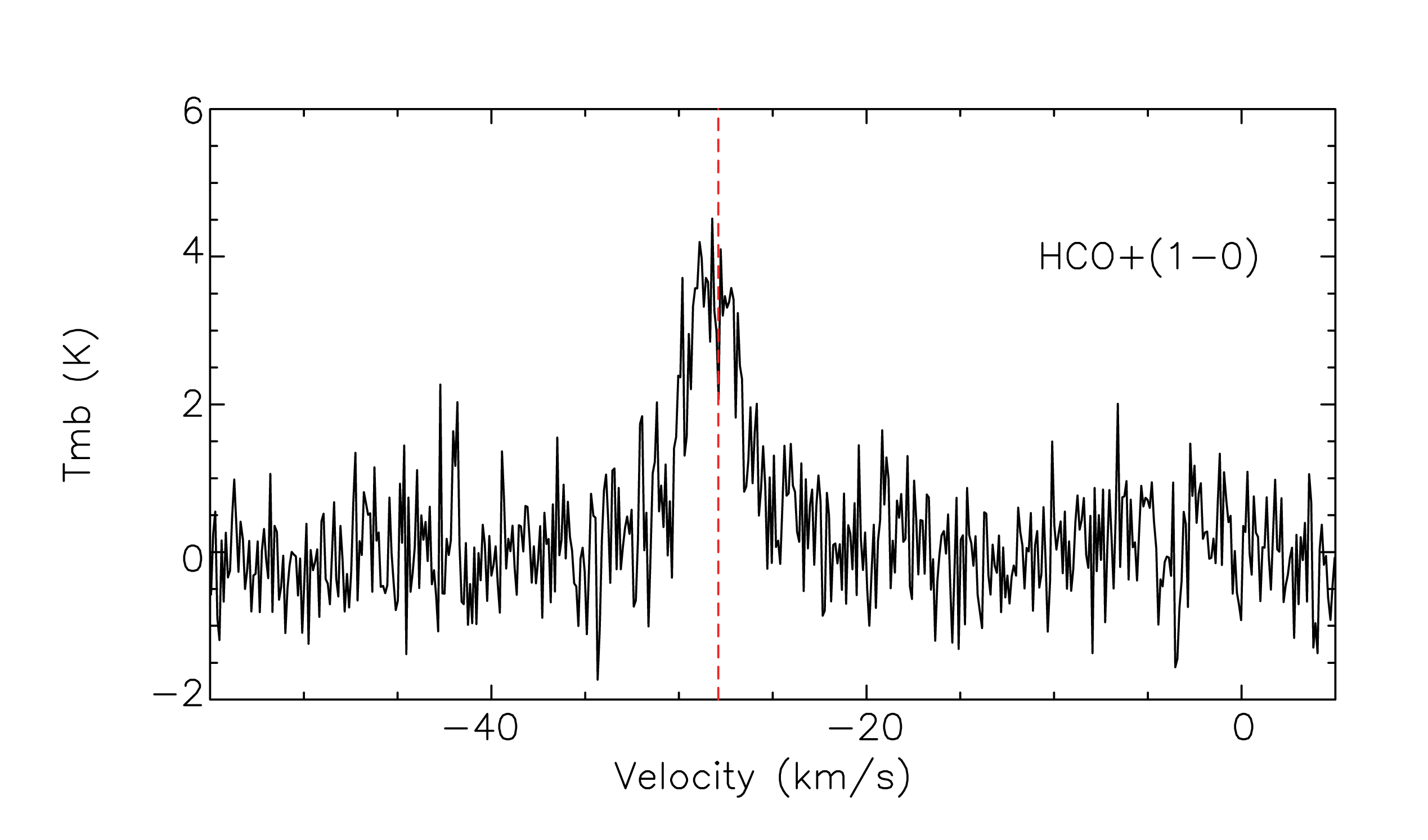,width=2.6in,height=1.8in}}
\centerline{\psfig{file=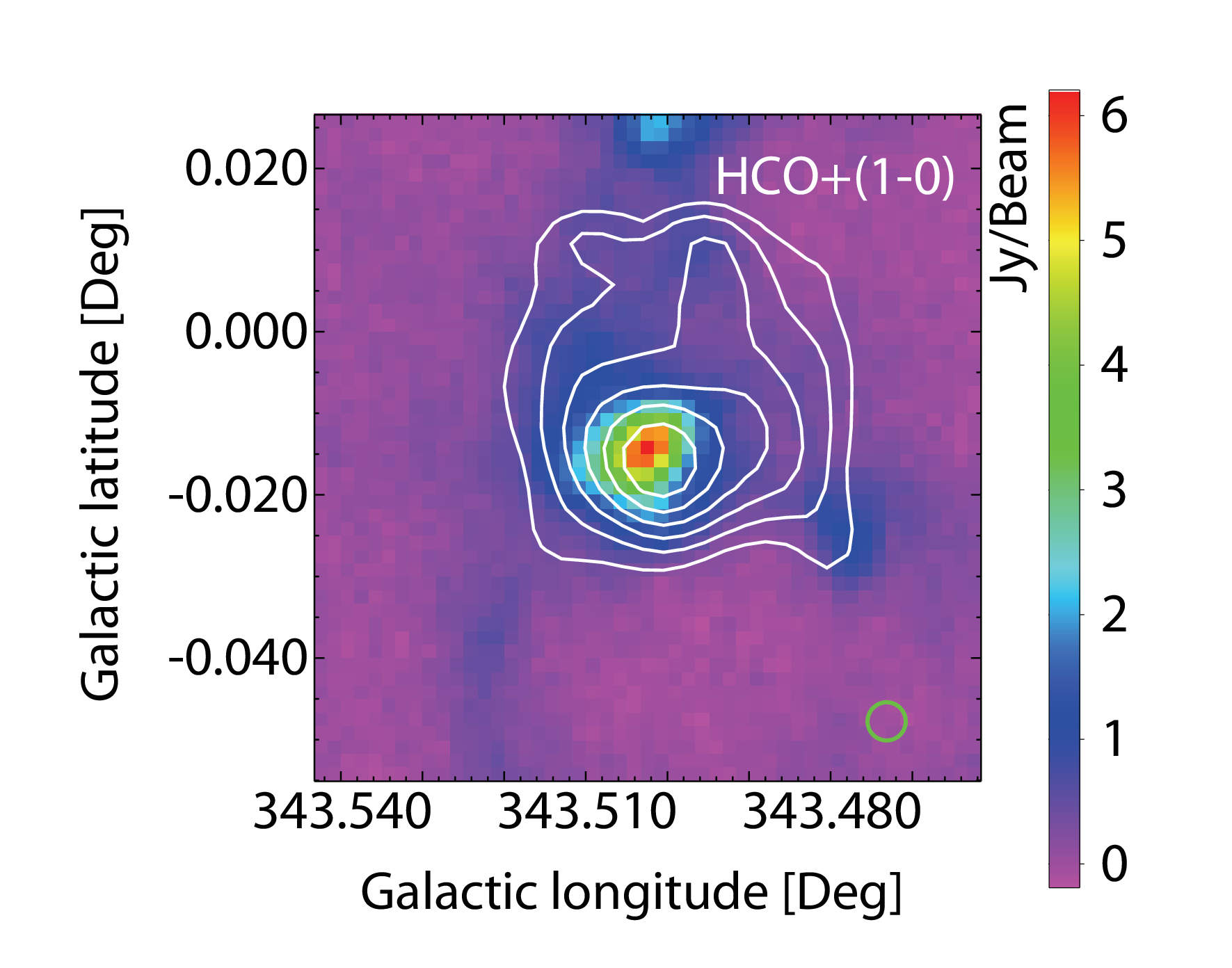,width=2.6in,height=2.3in}}
\centerline{\psfig{file=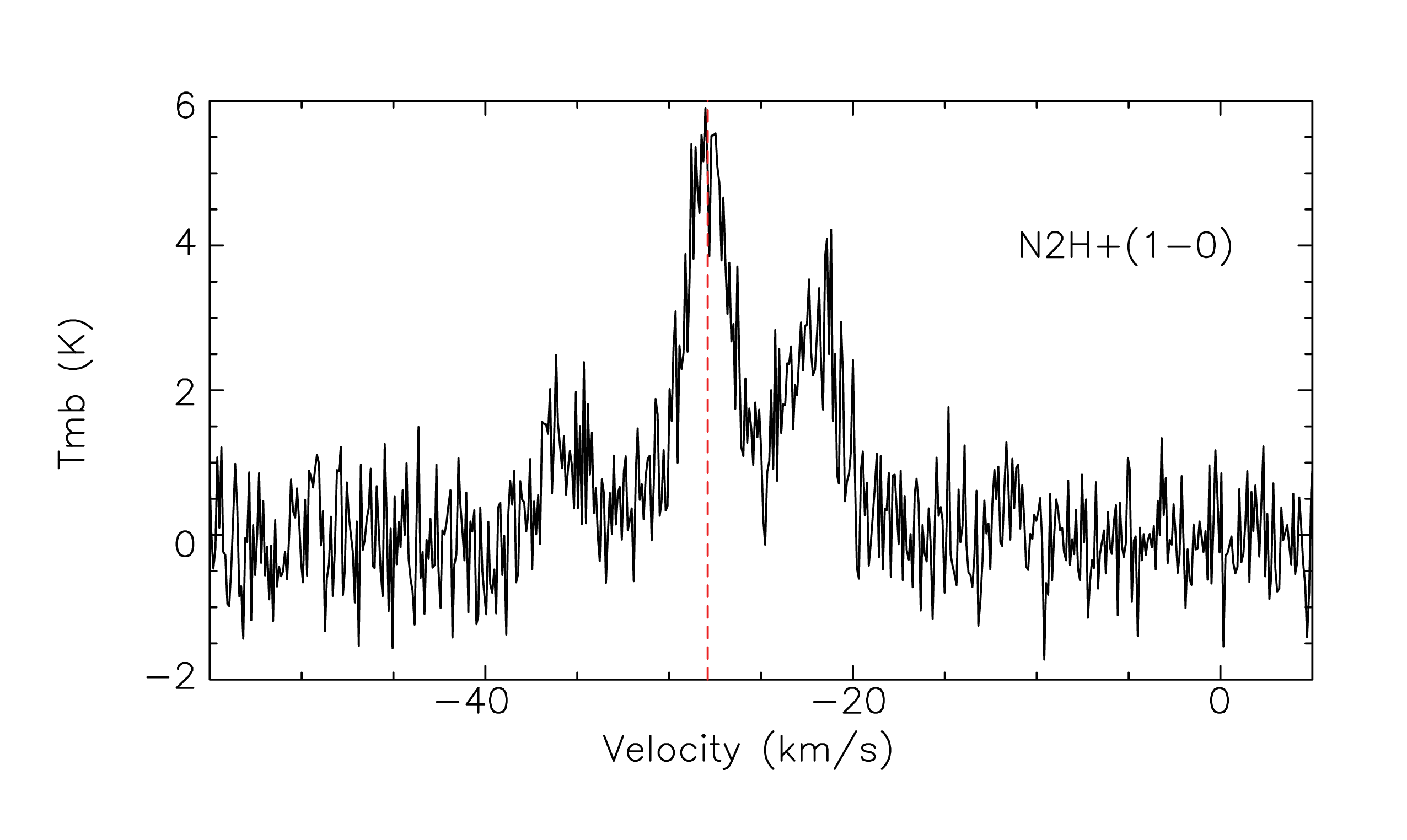,width=2.6in,height=1.8in}}
\centerline{\psfig{file=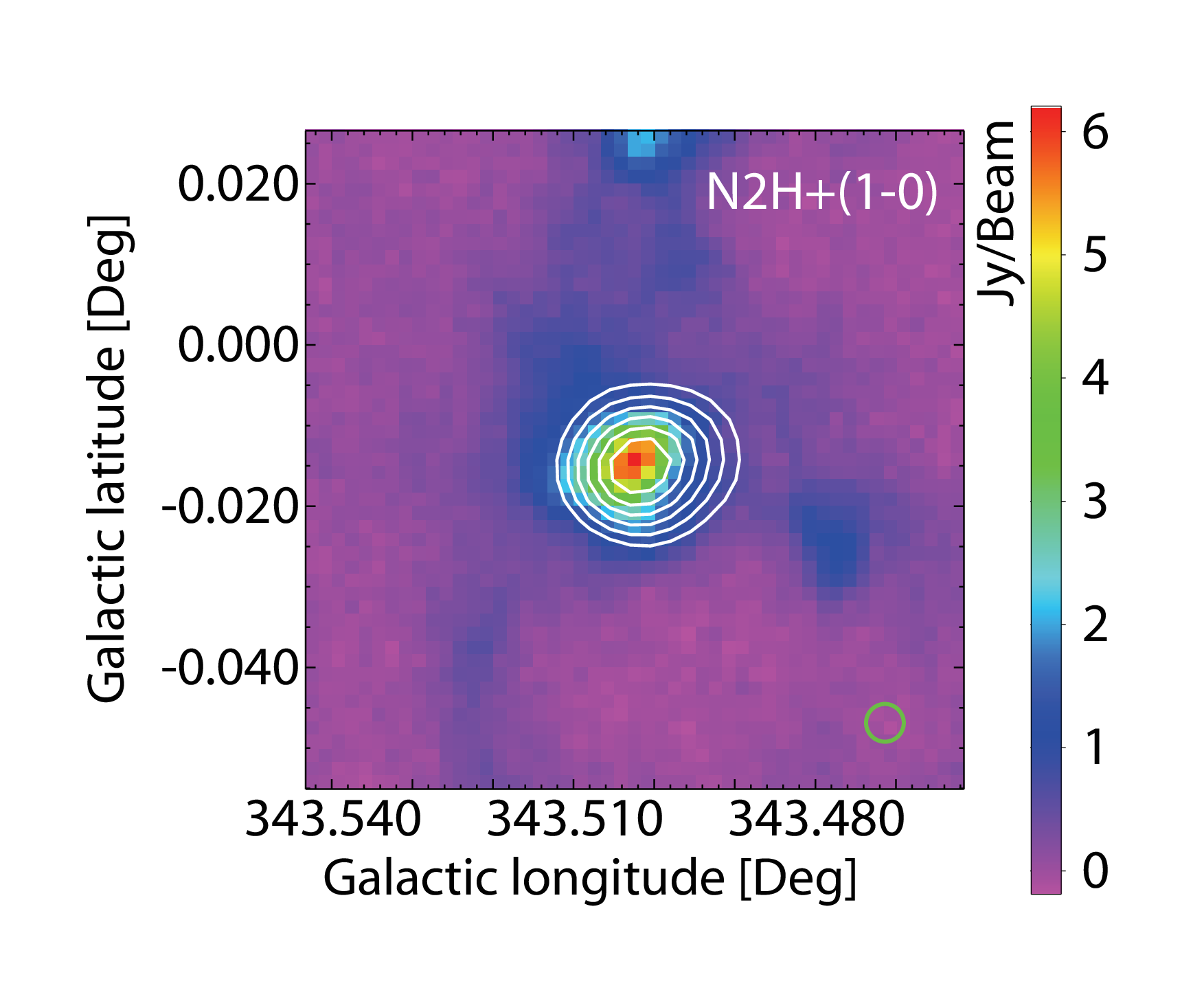,width=2.6in,height=2.3in}}
\caption{Spectra and integrated intensities superimposed on the 870
$\mu$m map in gray scale of G343.5024-00.0145. The red dash line
represents the V$_{LSR}$ of N$_2$H$^+$ line. Contour levels are
30$\%$, 40$\%$...90$\%$ of the center peak emissions. The angular
resolution of the ATLASGAL survey is indicated by the green circle
shown in the lower right corner.  }
\end{figure}
\begin{figure}
\centerline{\psfig{file=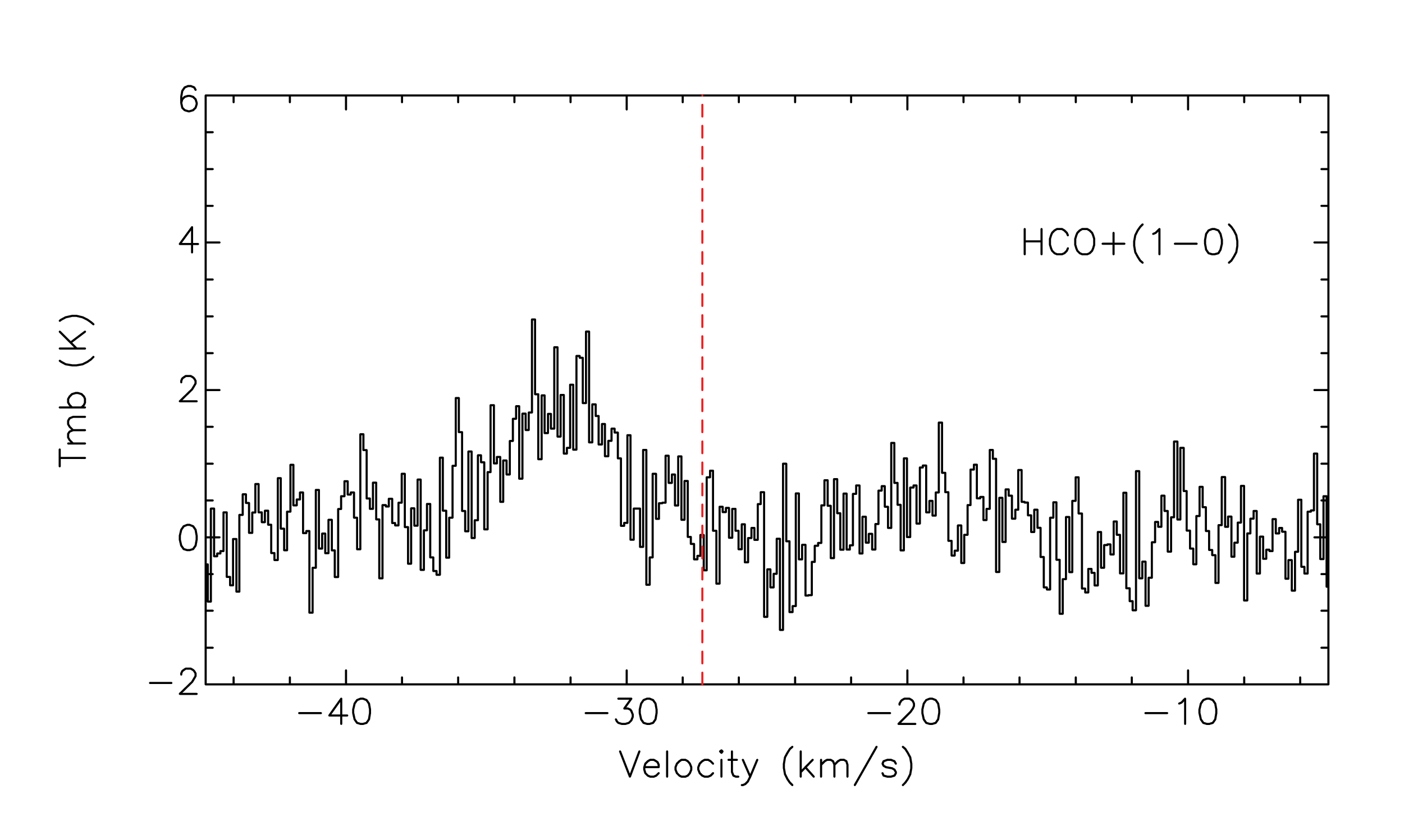,width=2.6in,height=1.8in}}
\centerline{\psfig{file=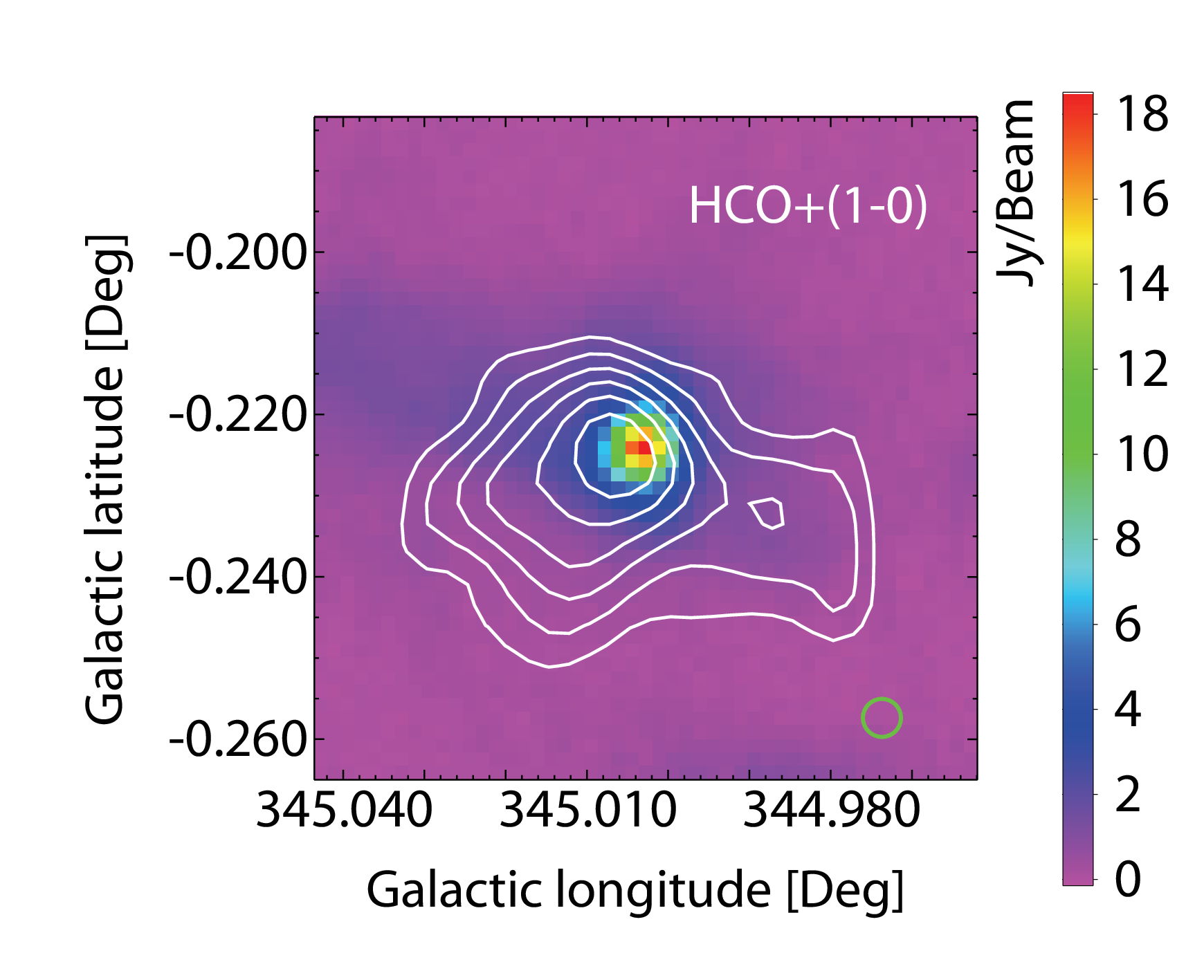,width=2.6in,height=2.3in}}
\centerline{\psfig{file=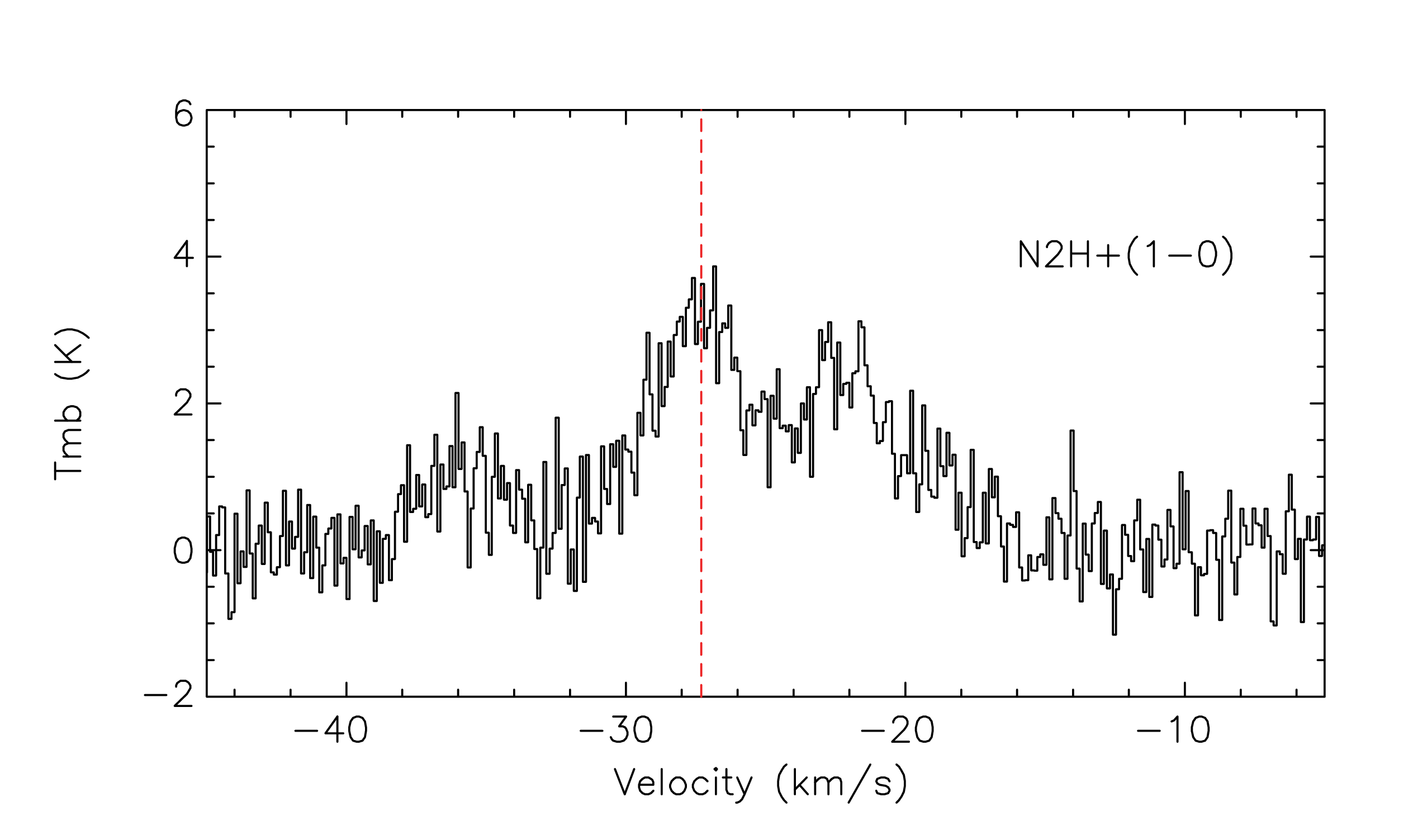,width=2.6in,height=1.8in}}
\centerline{\psfig{file=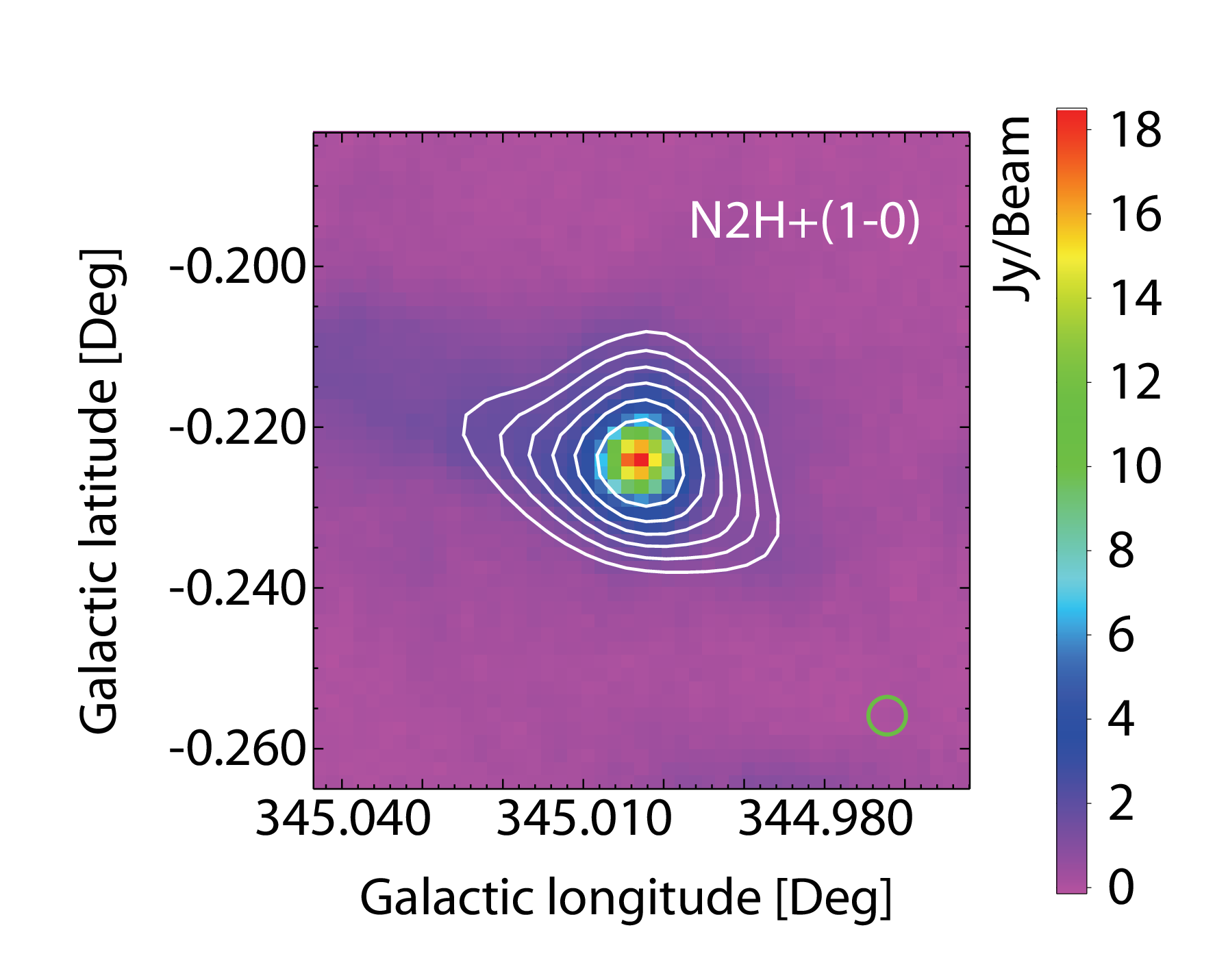,width=2.6in,height=2.3in}}
\caption{Spectra and integrated intensities superimposed on the 870
$\mu$m map in gray scale of G345.0034-00.2240A. The red dash line
represents the V$_{LSR}$ of N$_2$H$^+$ line. Contour levels are
30$\%$, 40$\%$...90$\%$ of the center peak emissions. The angular
resolution of the ATLASGAL survey is indicated by the green circle
shown in the lower right corner. }
\end{figure}
\begin{figure}
\centerline{\psfig{file=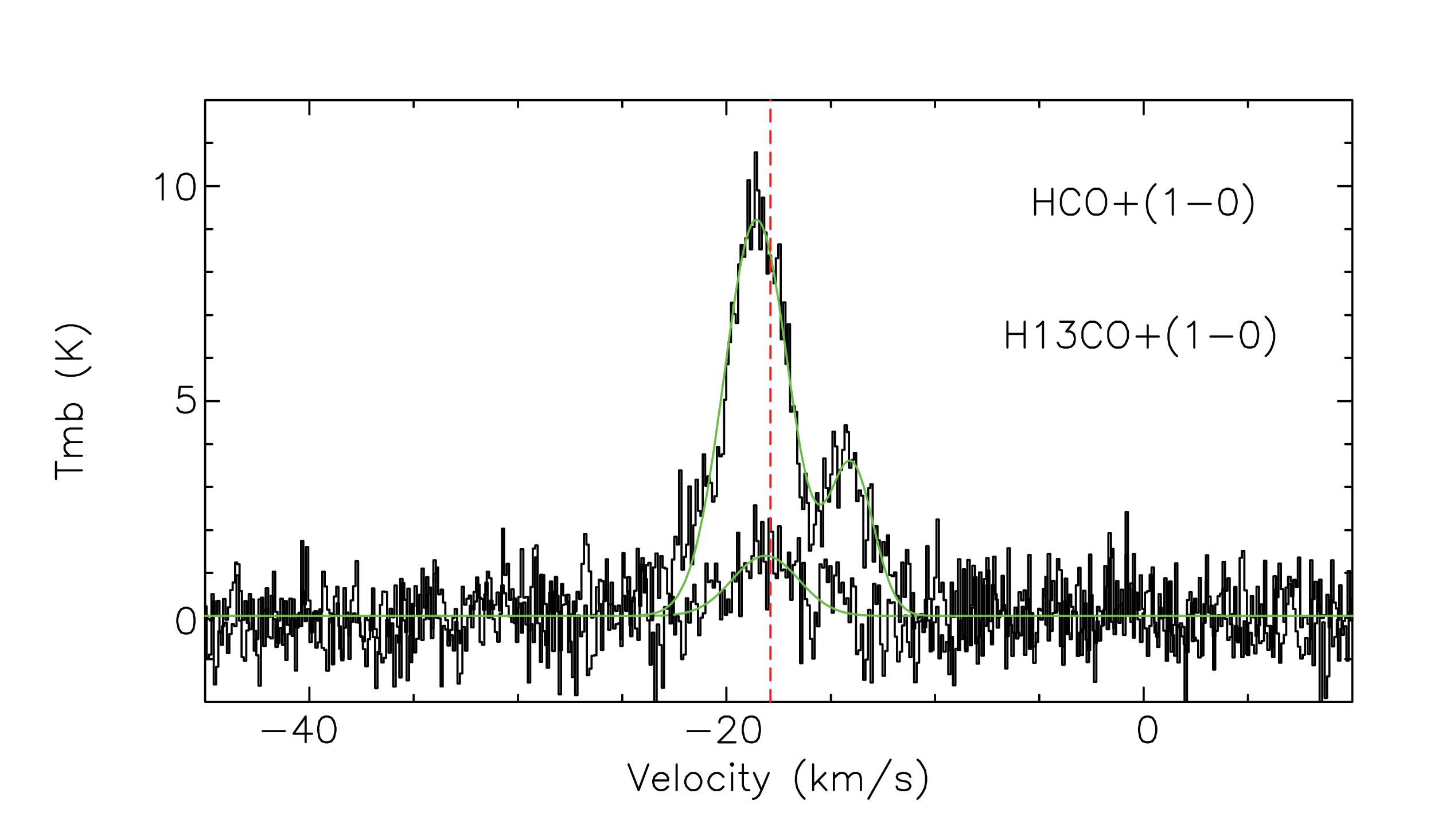,width=2.6in,height=1.8in}}
\centerline{\psfig{file=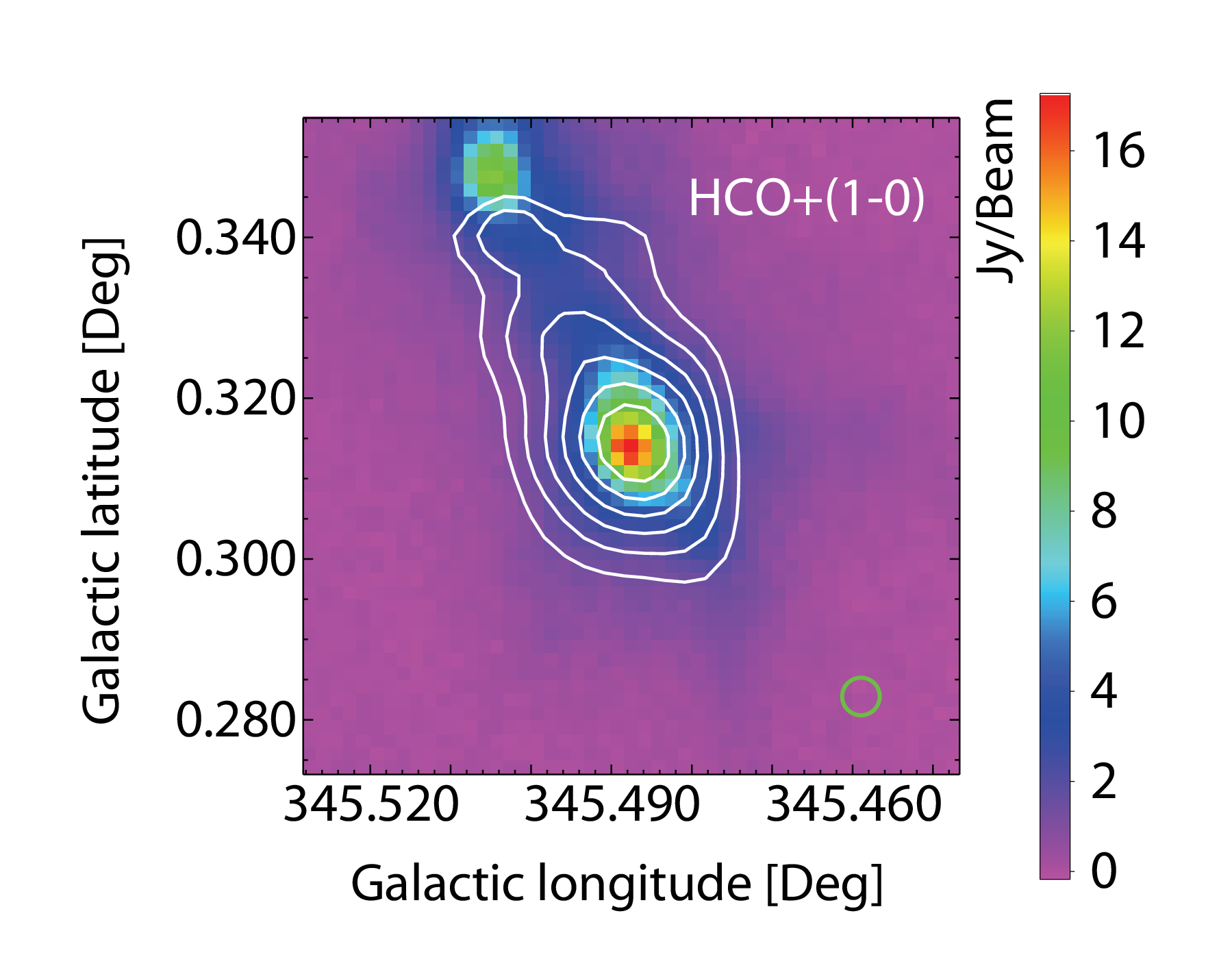,width=2.6in,height=2.3in}}
\centerline{\psfig{file=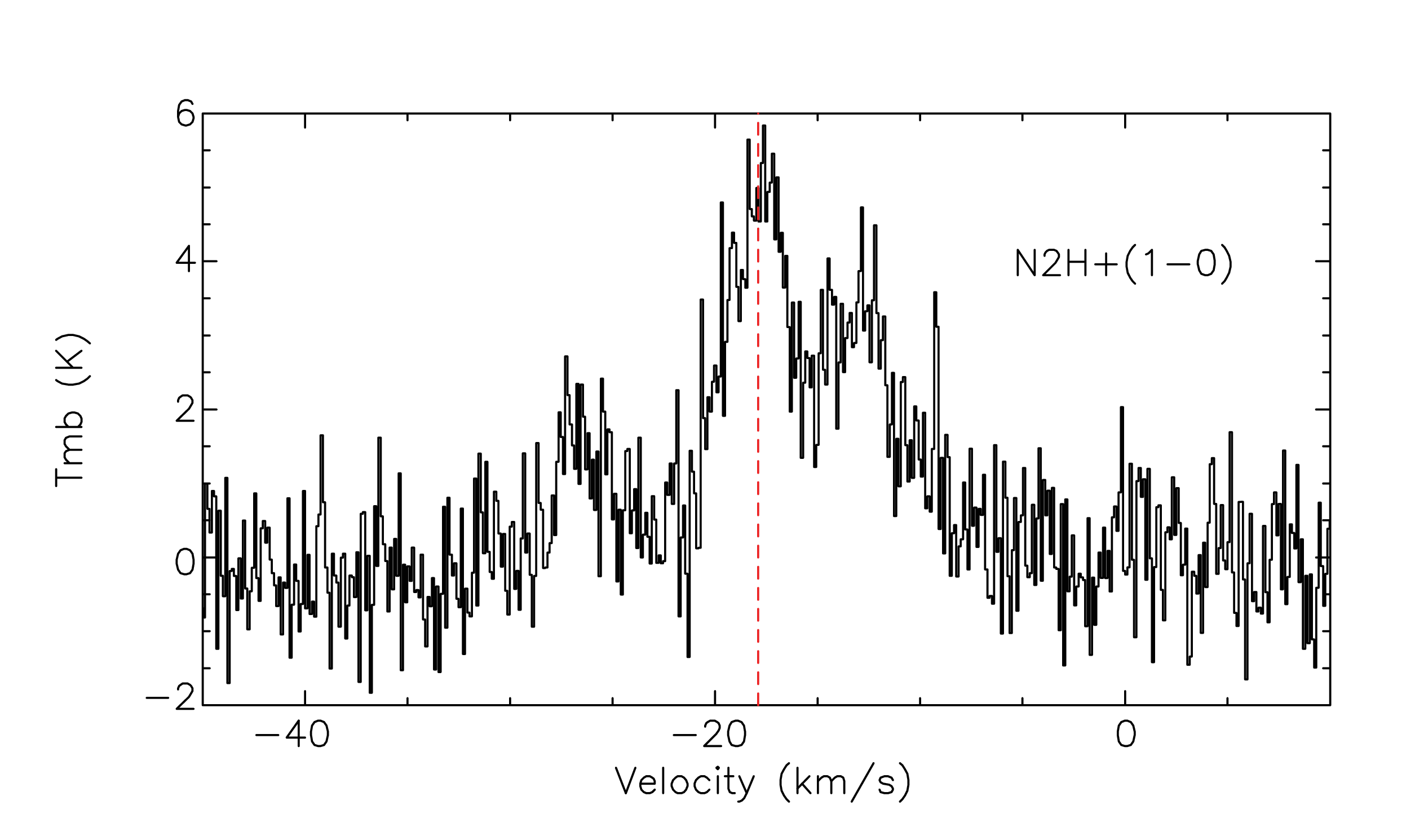,width=2.6in,height=1.8in}}
\centerline{\psfig{file=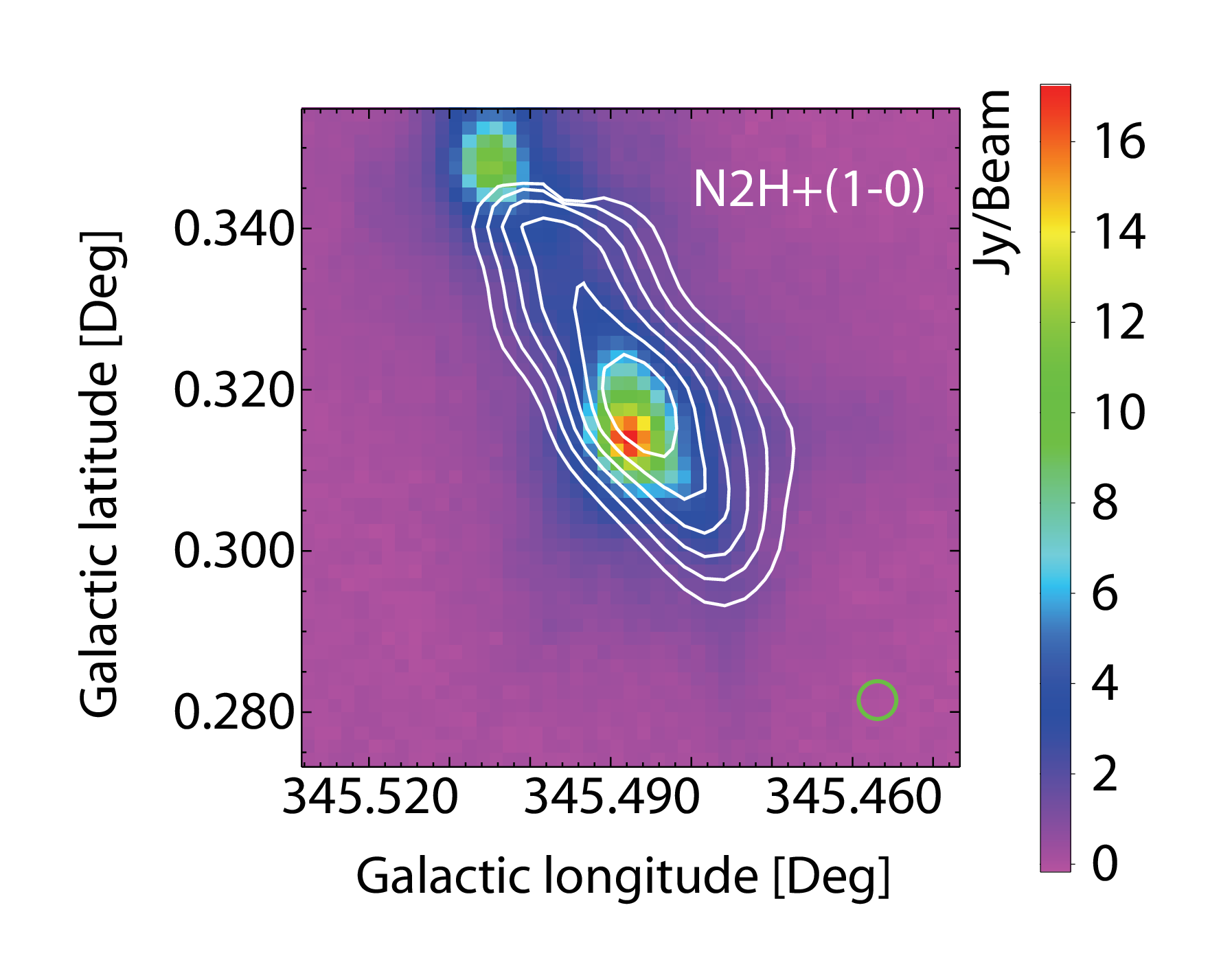,width=2.6in,height=2.3in}}
\caption{Spectra and integrated intensities superimposed on the 870
$\mu$m map in gray scale of G345.4881+00.3148. The red dash line
represents the V$_{LSR}$ of N$_2$H$^+$ line. The fits of the Myers
model are shown with green lines. Contour levels are 30$\%$,
40$\%$...90$\%$ of the center peak emissions. The angular resolution
of the ATLASGAL survey is indicated by the green circle shown in the
lower right corner.  }
\end{figure}
\begin{figure}
\centerline{\psfig{file=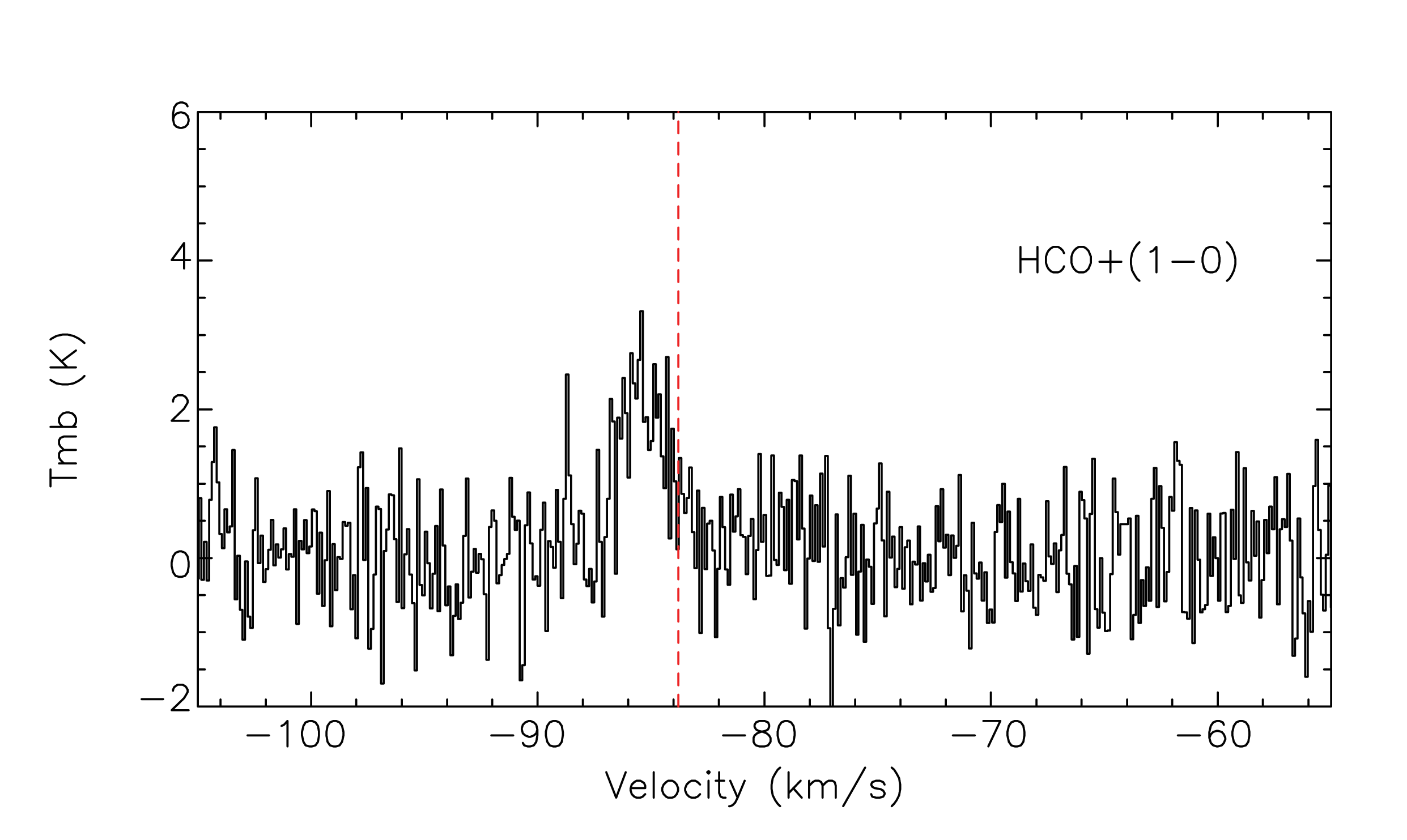,width=2.6in,height=1.8in}}
\centerline{\psfig{file=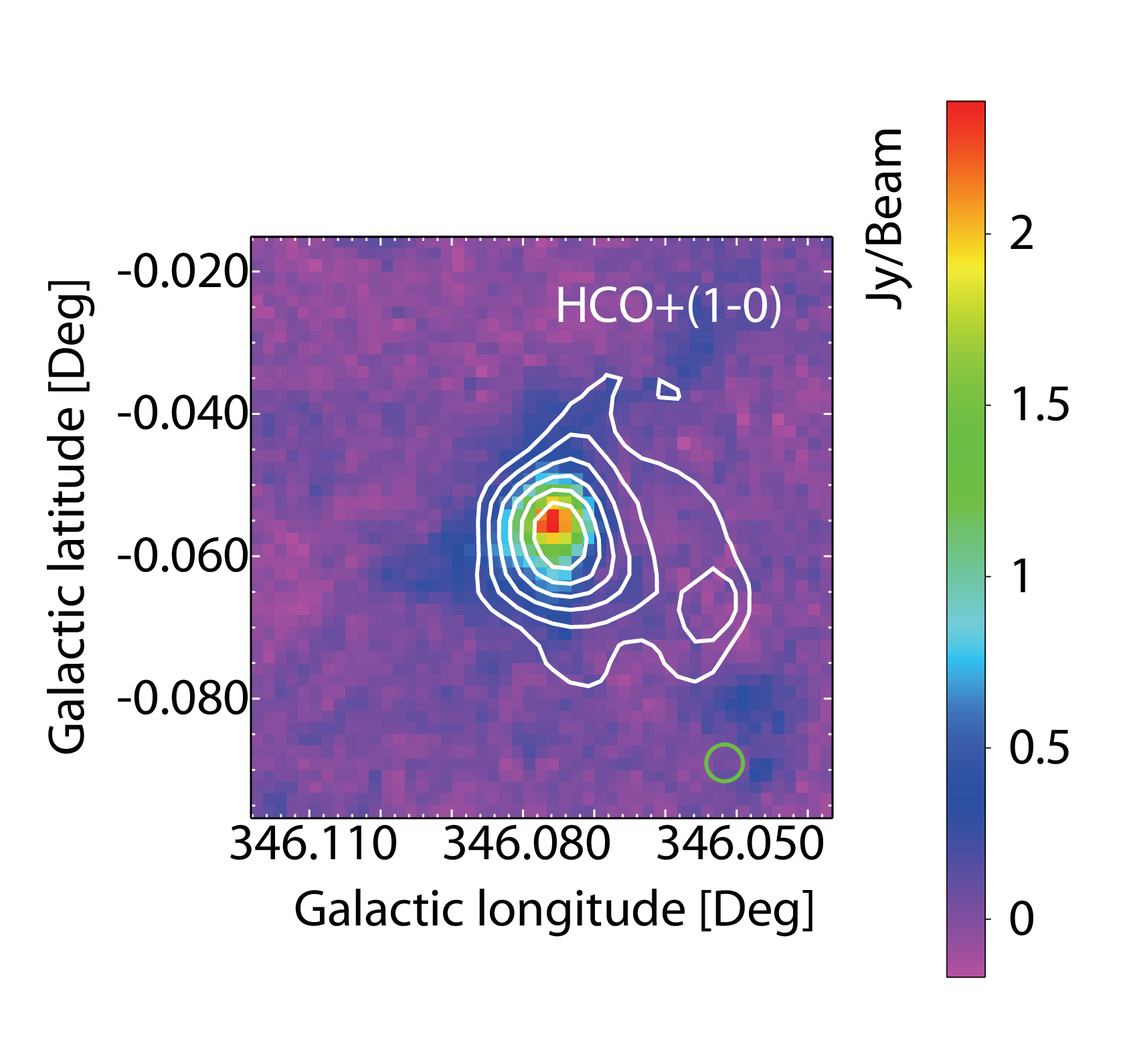,width=2.6in,height=2.3in}}
\centerline{\psfig{file=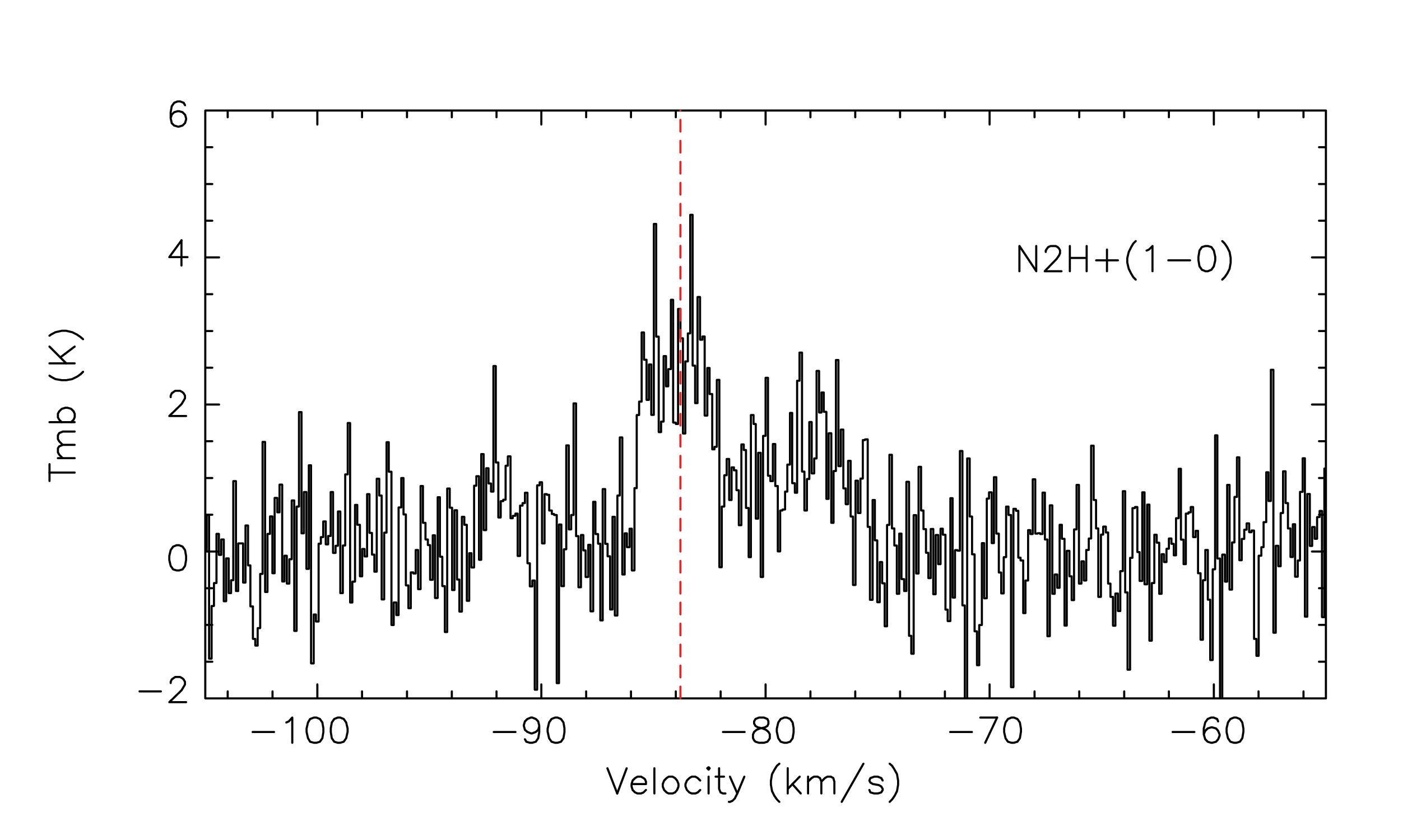,width=2.6in,height=1.8in}}
\centerline{\psfig{file=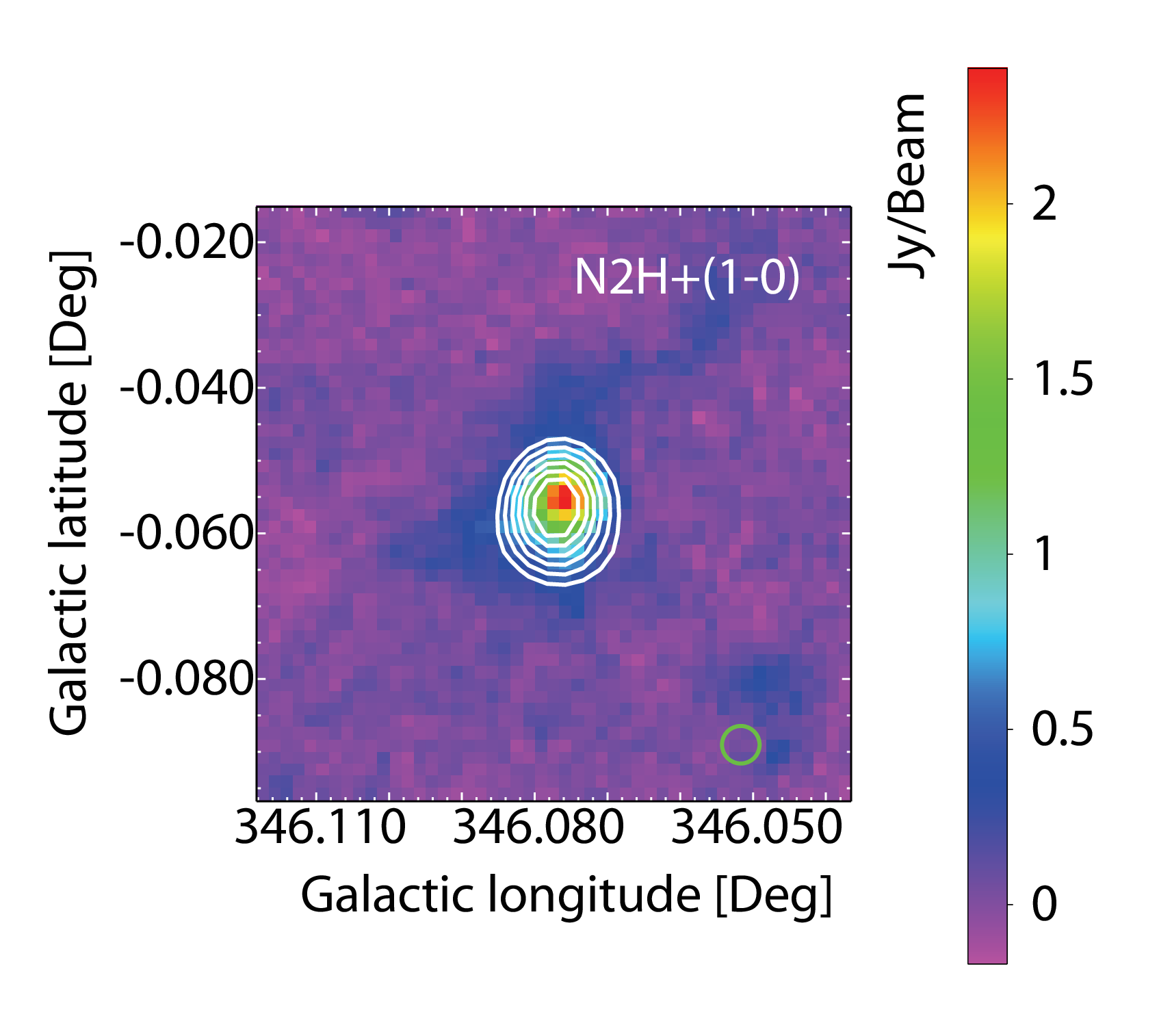,width=2.6in,height=2.3in}}
\caption{Spectra and integrated intensities superimposed on the 870
$\mu$m map in gray scale of G346.0774-00.0562. The red dash line
represents the V$_{LSR}$ of N$_2$H$^+$ line. Contour levels are
30$\%$, 40$\%$...90$\%$ of the center peak emissions. The angular
resolution of the ATLASGAL survey is indicated by the green circle
shown in the lower right corner.  }
\end{figure}
\begin{figure}
\centerline{\psfig{file=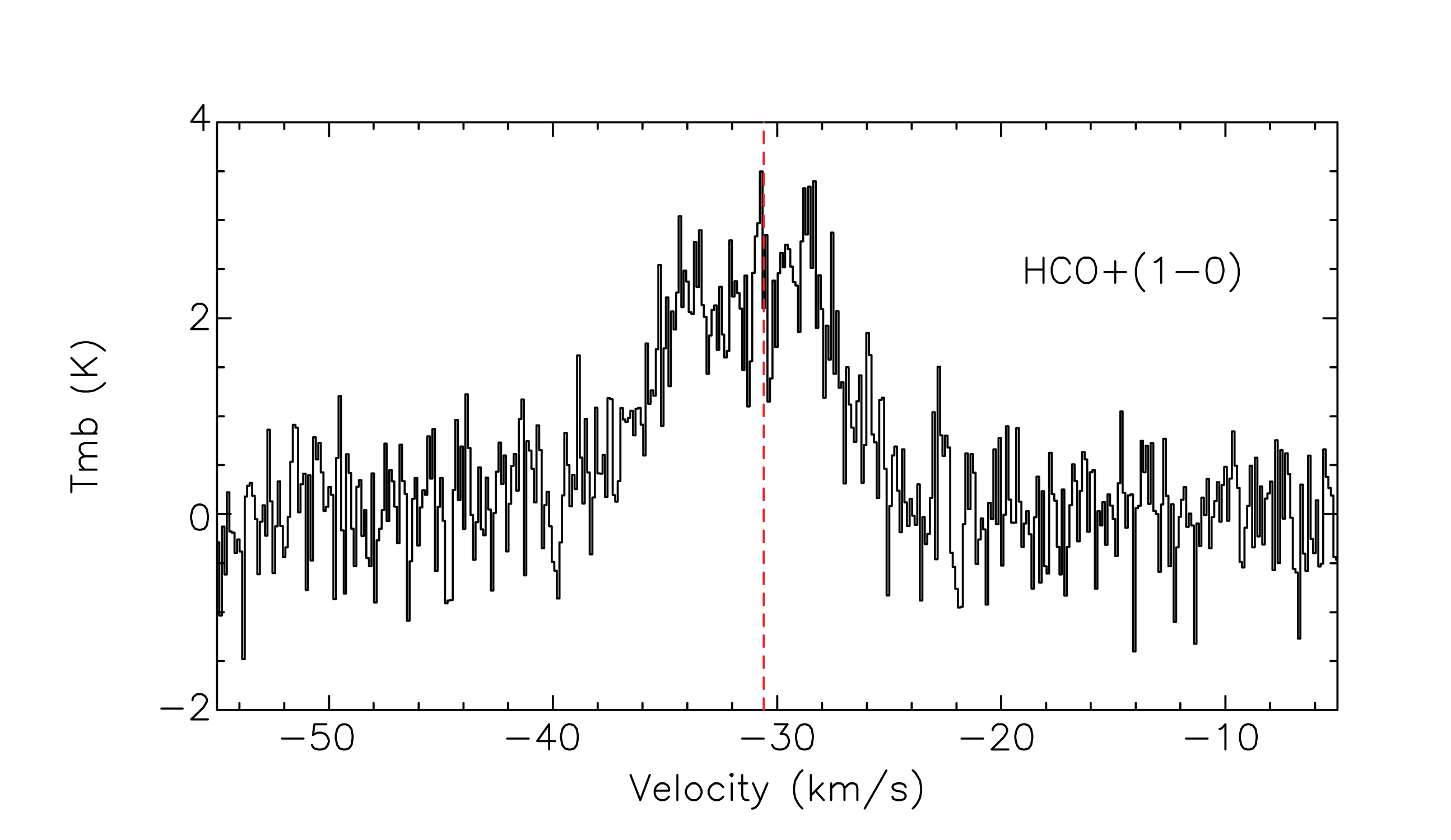,width=2.6in,height=1.8in}}
\centerline{\psfig{file=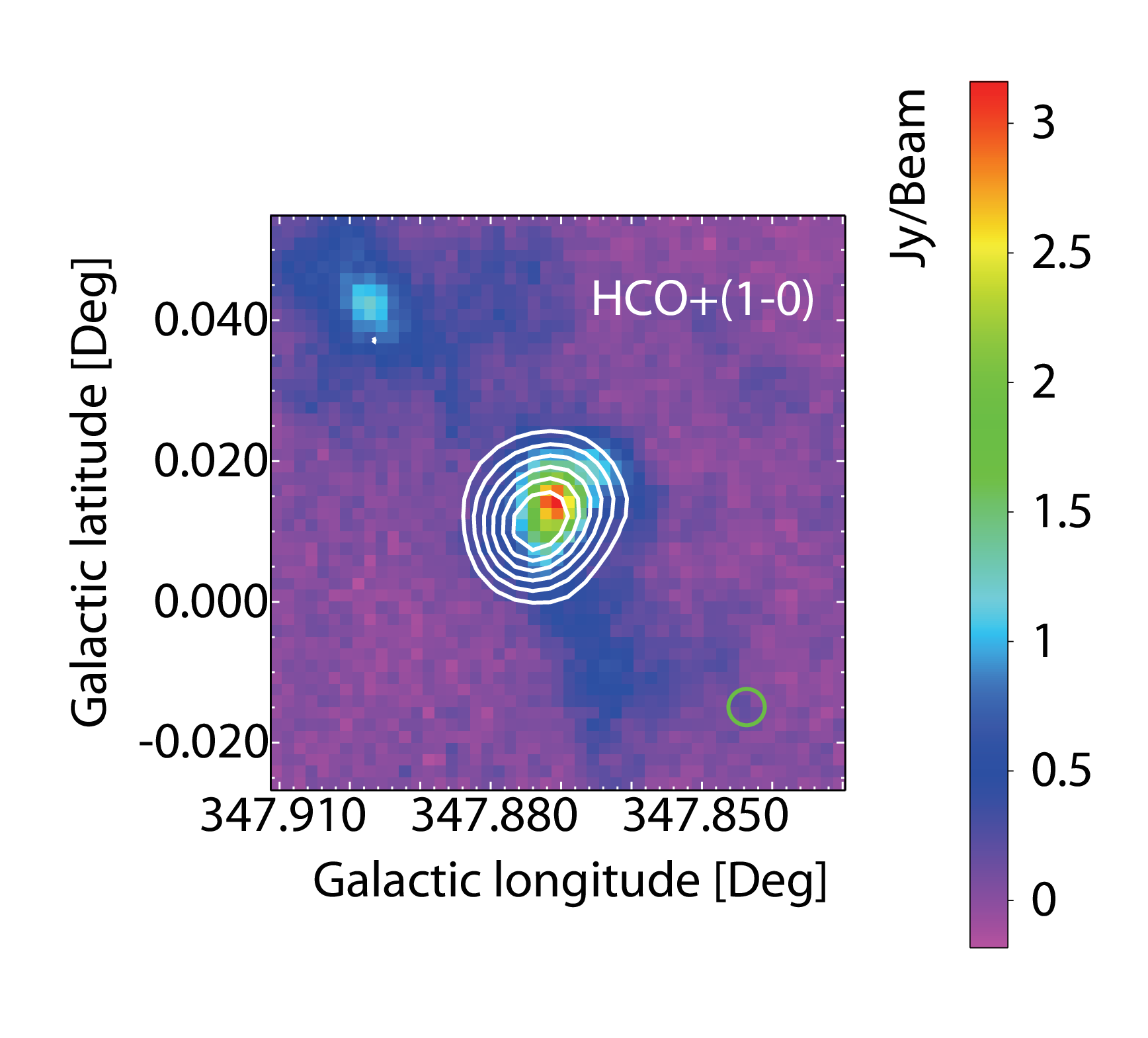,width=2.6in,height=2.3in}}
\centerline{\psfig{file=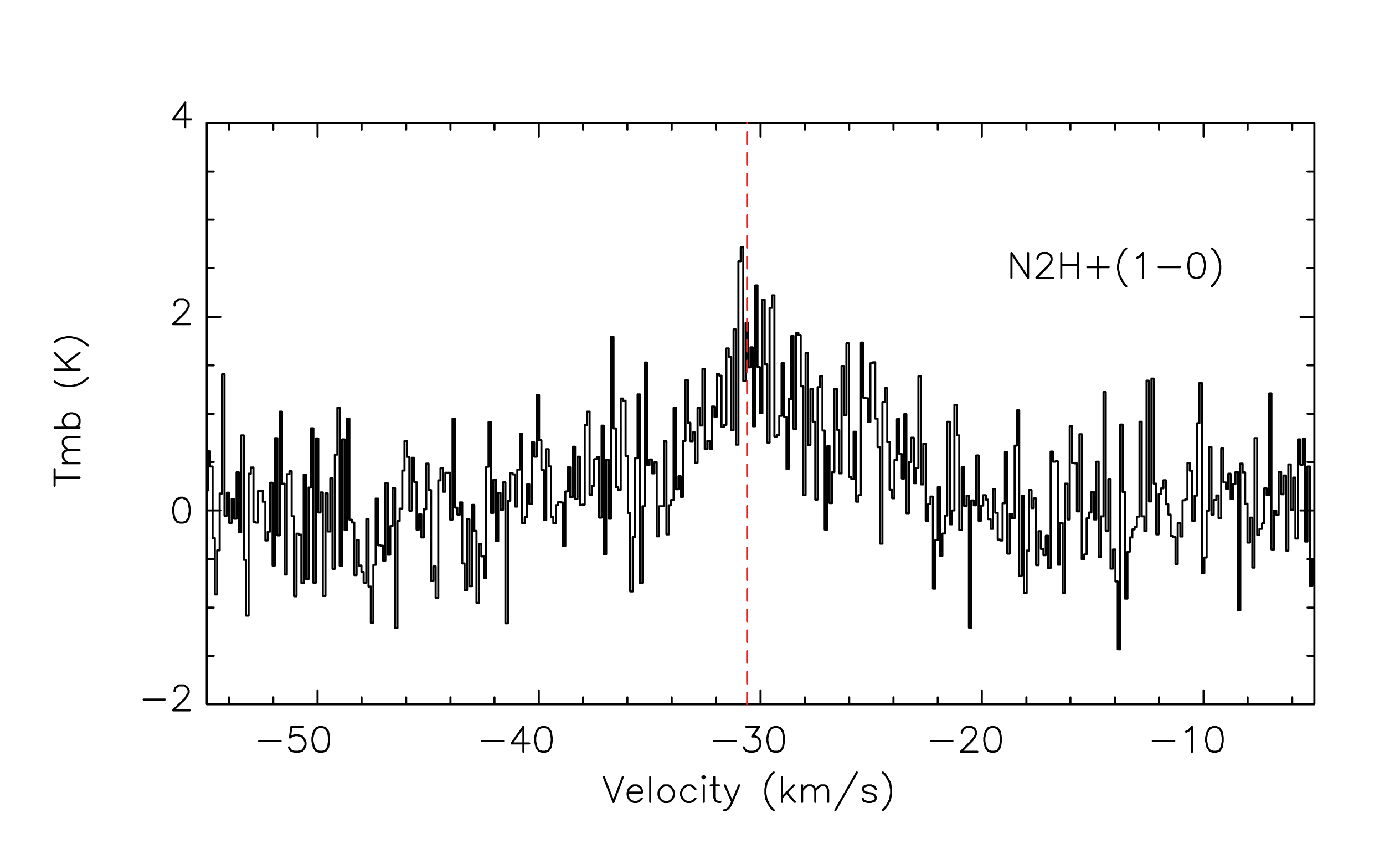,width=2.6in,height=1.8in}}
\centerline{\psfig{file=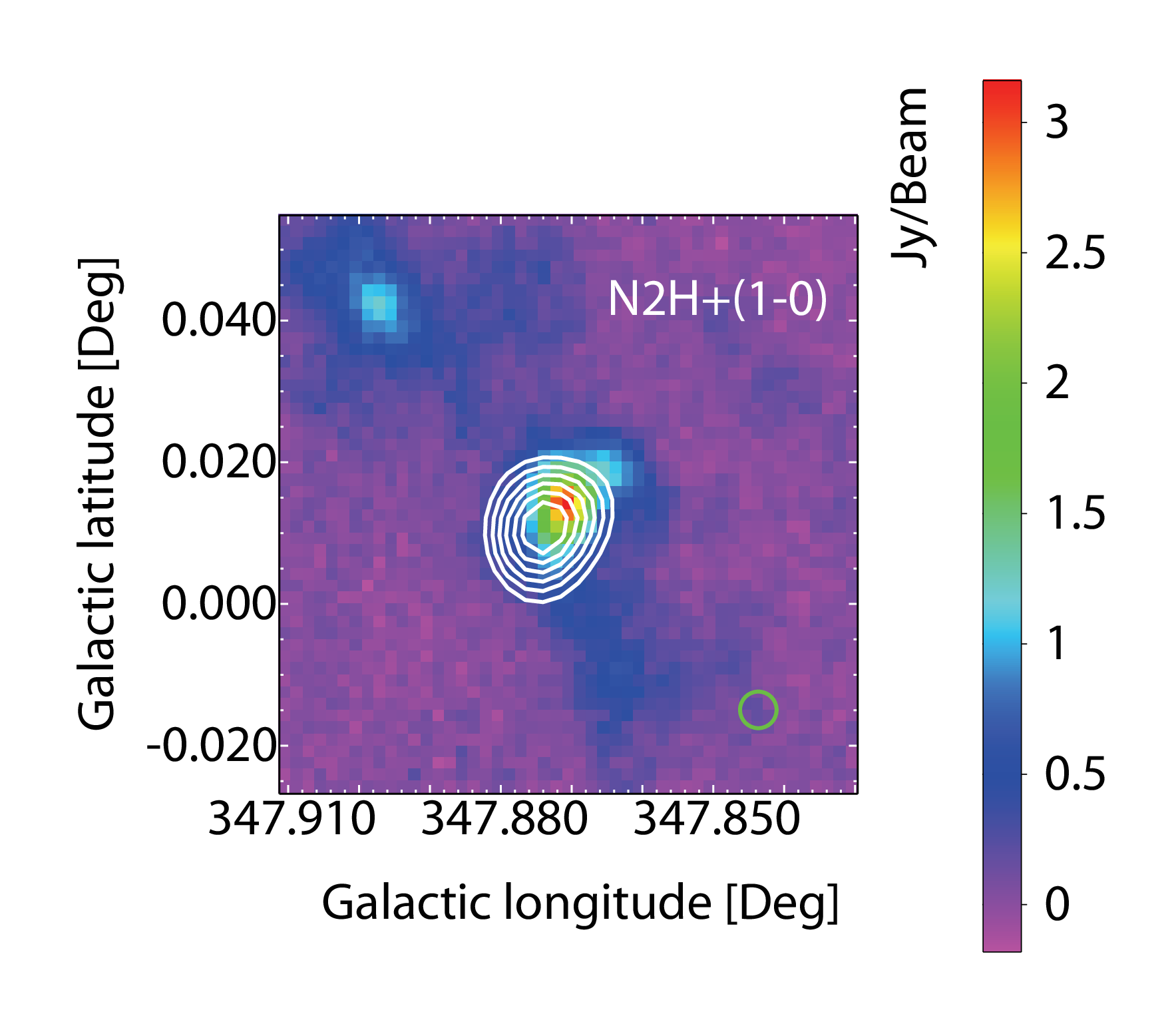,width=2.6in,height=2.3in}}
\caption{Spectra and integrated intensities superimposed on the 870
$\mu$m map in gray scale of G347.8707+00.0146. The red dash line
represents the V$_{LSR}$ of N$_2$H$^+$ line. Contour levels are
30$\%$, 40$\%$...90$\%$ of the center peak emissions. The angular
resolution of the ATLASGAL survey is indicated by the green circle
shown in the lower right corner. }
\end{figure}
\begin{figure}
\centerline{\psfig{file=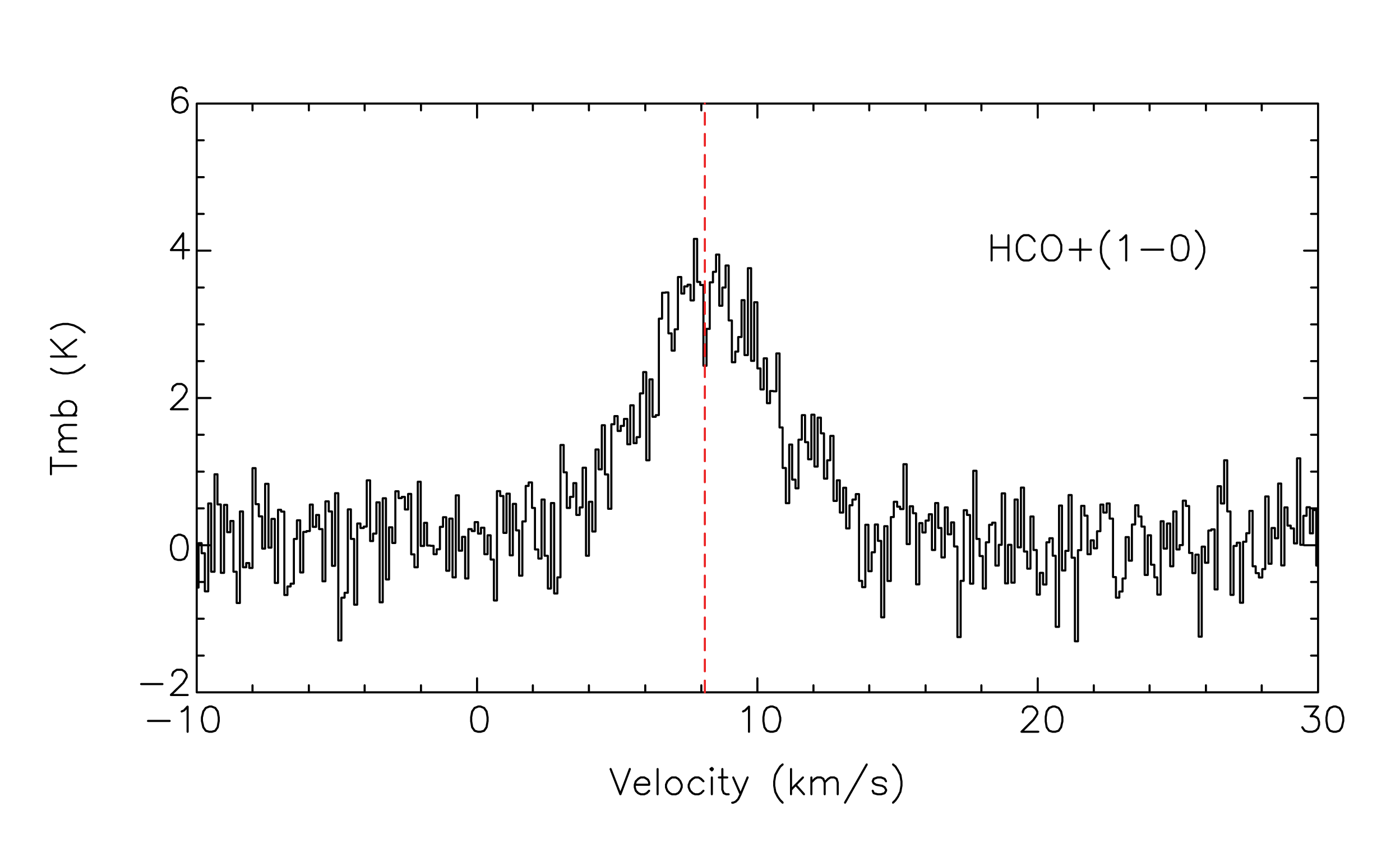,width=2.6in,height=1.8in}}
\centerline{\psfig{file=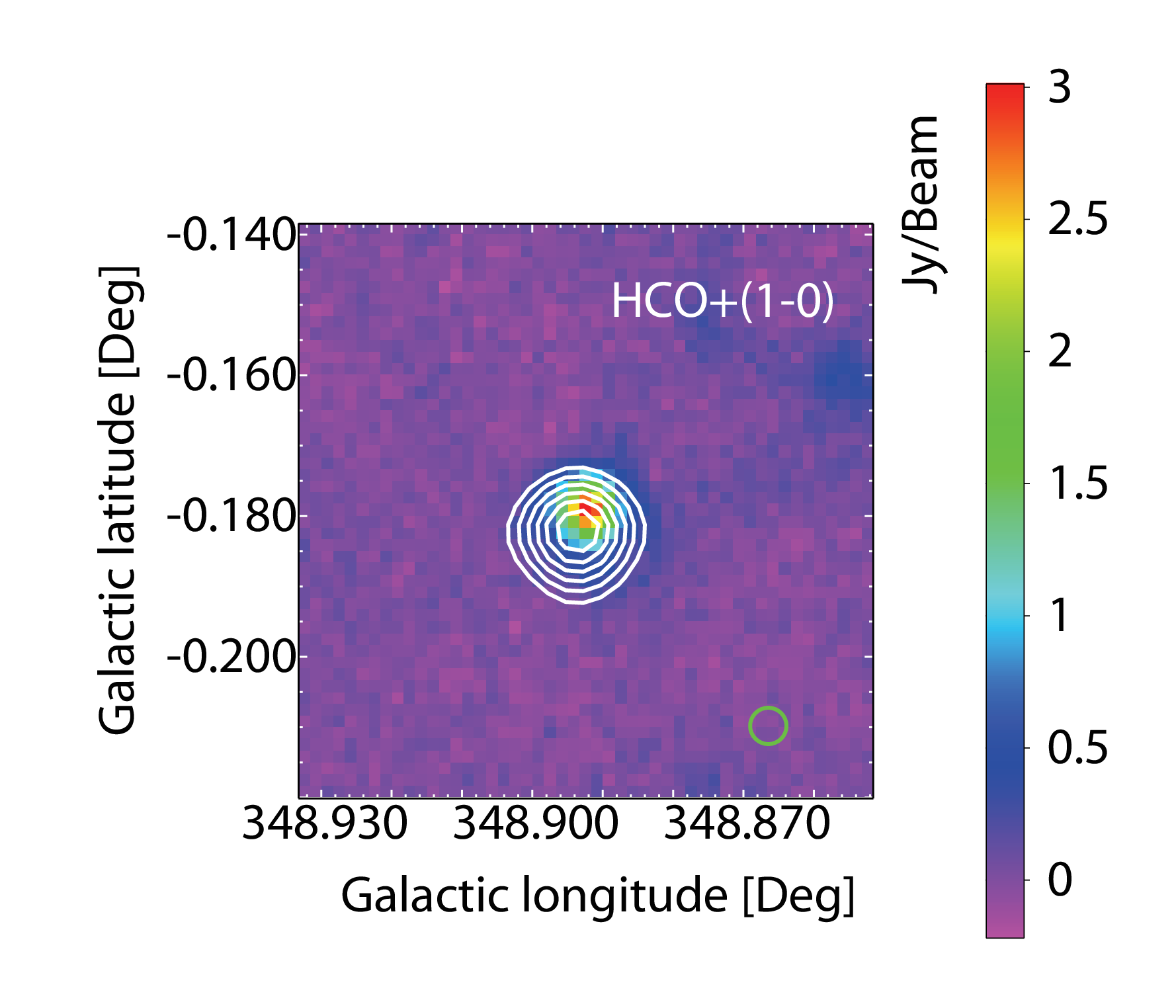,width=2.6in,height=2.3in}}
\centerline{\psfig{file=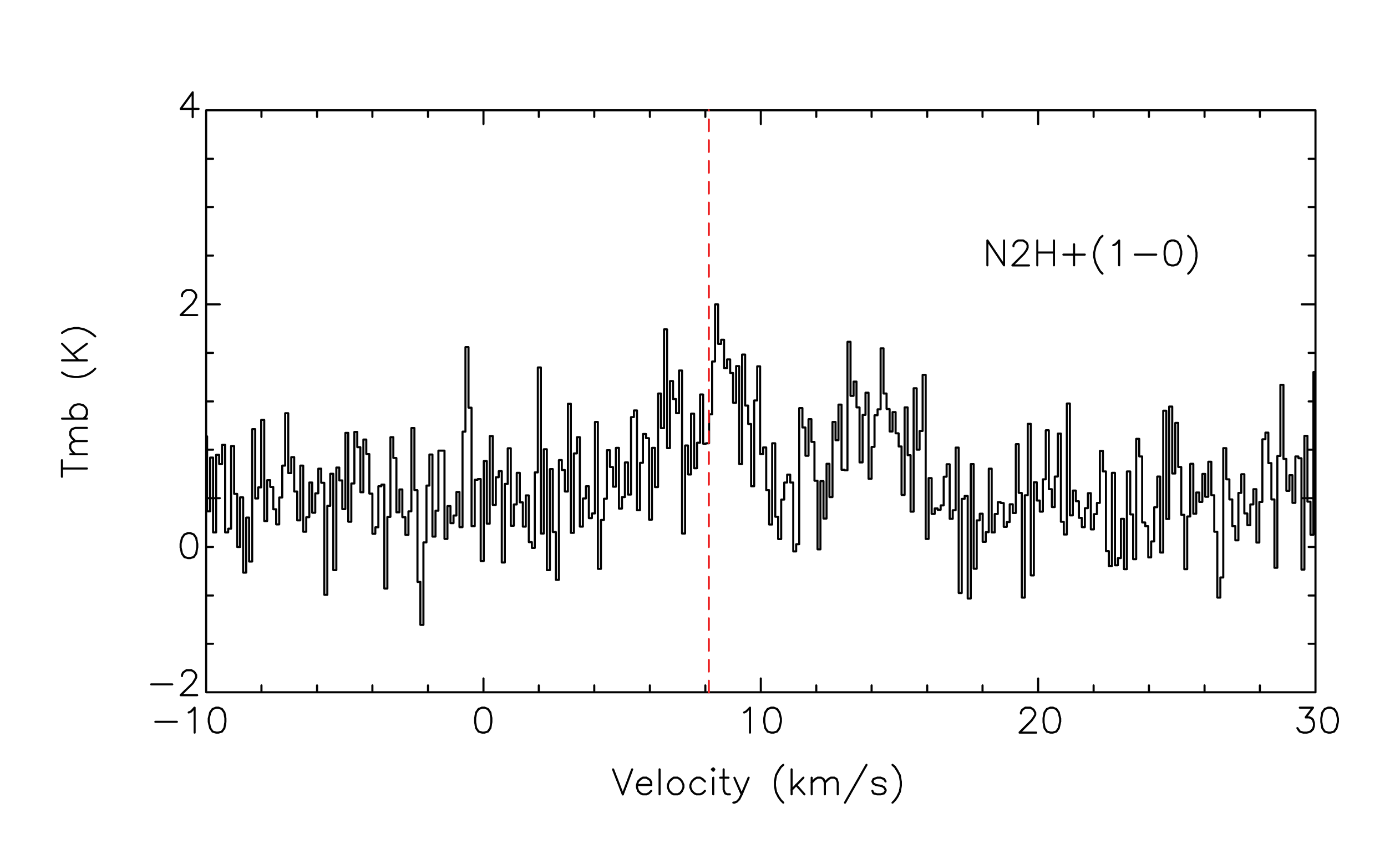,width=2.6in,height=1.8in}}
\centerline{\psfig{file=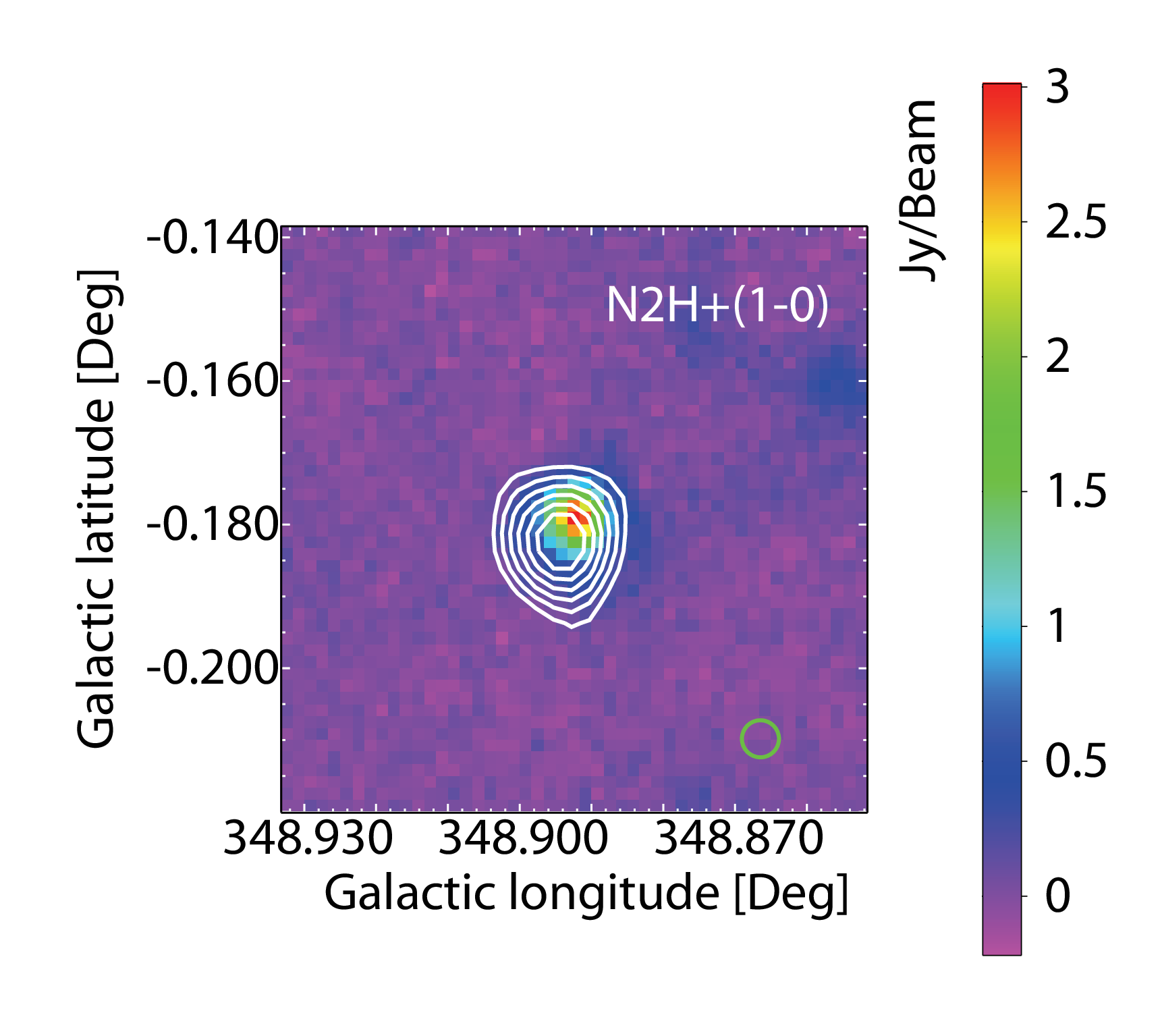,width=2.6in,height=2.3in}}
\caption{Spectra and integrated intensities superimposed on the 870
$\mu$m map in gray scale of G348.8922-00.1787. The red dash line
represents the V$_{LSR}$ of N$_2$H$^+$ line. Contour levels are
30$\%$, 40$\%$...90$\%$ of the center peak emissions. The angular
resolution of the ATLASGAL survey is indicated by the green circle
shown in the lower right corner.}
\end{figure}

\end{document}